\documentclass[twocolumn,a4paper]{article}

\usepackage{amsmath,amsfonts,amssymb}
\usepackage{array}
\usepackage{bm}
\usepackage{cancel}
\usepackage{dcolumn}
\usepackage{graphicx}
\usepackage{picins}
\usepackage[scanall]{psfrag}
\usepackage{times}

\usepackage[
	paperwidth=210mm,
	paperheight=297mm,
	textwidth=170mm,
	textheight=250mm,
	dvips]
{geometry}

\newcommand{\psfragstyle}[1]{\small{#1}}

\newcommand{\figref}[1]{Fig.~\ref{#1}}

\newlength{\tempfboxsep}
\newcommand{\sidebar}[1]{%
	\setlength{\fboxsep}{0pt}%
	\setlength{\temptabcolsep}{\tabcolsep}%
	\setlength{\tabcolsep}{0pt}%
	\begin{tabular}{rcl}%
		\vrule width 0.03\columnwidth &%
		\hspace*{0.05\columnwidth} &%
		\begin{minipage}{0.92\columnwidth}%
			#1%
		\end{minipage}%
	\end{tabular}%
	\setlength{\tabcolsep}{\temptabcolsep}%
	\setlength{\fboxsep}{\tempfboxsep}%
}
\newlength{\temptabcolsep}

\newcommand{\zero}{\ensuremath{\text{0}}}
\newcommand{\one}{\ensuremath{\text{1}}}
\newcommand{\two}{\ensuremath{\text{2}}}

\newcommand{\thend}{\ensuremath{^{\text{nd}}}~}

\newcommand{\theth}{\ensuremath{^{\text{th}}}~}

\newcommand{\trademark}{\ensuremath{^{\text{\tiny{TM}}}}~}

\begin{document}

\newlength{\figurewidth}
\setlength{\figurewidth}{\columnwidth}
\newlength{\halffigurewidth}
\setlength{\halffigurewidth}{0.47\columnwidth}
\newlength{\figureseparation}
\setlength{\figureseparation}{0.2cm}

\setlength{\parindent}{0pt}
\setlength{\parskip}{1ex plus 0.5ex minus 0.2ex}

\title{
  \textbf{Transportation Planning and Traffic Flow Models}
}

\author{
	Sven Maerivoet\footnote{\texttt{sven.maerivoet@esat.kuleuven.be}}
		\quad and \quad
	Bart De Moor\footnote{\texttt{bart.demoor@esat.kuleuven.be}}\\
	\small{\emph{Department of Electrical Engineering ESAT-SCD (SISTA)}}\\
	\small{\emph{Phone: +32 (0) 16 32 17 09 Fax: +32 (0) 16 32 19 70}}\\
	\small{\emph{URL: \texttt{http://www.esat.kuleuven.be/scd}}}\\
	\small{\emph{Katholieke Universiteit Leuven}}\\
	\small{\emph{Kasteelpark Arenberg 10, 3001 Leuven, Belgium}}
}

\date{(\small{Dated: \today})}

\maketitle

\begin{abstract}
	In this paper, we focus on the different traffic flow models that exist in 
	literature. Due to our frequently encountered confusion among traffic 
	engineers and policy makers, this paper goes into more detail about 
	transportation planning models on the one hand, and traffic flow models on the 
	other hand. The former deal with households that make certain decisions which 
	lead to transportation and the use of infrastructure, as opposed to the latter 
	which explicitly describe the physical propagation of traffic flows in a road 
	network. Our goal is not to give a full account (as that would be a 
	dissertation of its own, given the broadness of the field), but rather to 
	impose upon the reader a thorough feeling for the differences between 
	transportation planning and traffic flow models. Because of the high course of 
	progress over the last decade (or even during the last five years), this paper 
	tries to chronicle both past models, as well as some of the latest 
	developments in this area.


\end{abstract}

\setlength{\parskip}{0pt}

\tableofcontents

\setlength{\parskip}{1ex plus 0.5ex minus 0.2ex}

%

Due to our frequently encountered confusion among traffic engineers and policy 
makers when it comes to transportation planning models and the role that traffic 
flow models play therein, this paper strives to alleviate that bewilderment. The 
material elaborated upon in this paper, spans a broad range going from 
transportation planning models that operate on a high level and deal with 
households that make certain decisions which lead to transportation and the use 
of infrastructure, to traffic flow models that explicitly describe the physical 
propagation of traffic flows in a road network.

	\section{Transportation planning models}

Before going into detail about the possible mathematical models that describe 
the physical propagation of traffic flows, it is worthwhile to cast a glance at 
a higher level, where transportation planning models operate. The main rationale 
behind transportation planning systems, is that travellers within these systems 
are motivated by making certain decisions about their wishes to participate in 
social, economical, and cultural activities. The ensemble of these activities is 
called the \emph{activity system}. Because these activities are spatially 
separated (e.g., a person's living versus work area), the need for 
transportation arises. In such a system, the so-called \emph{household activity 
patterns} form the main explanation for what is seen in the transportation 
network.

These models have as their primary intent the performing of impact and 
evaluation studies, and conducting `before and after' analyses. The fact that 
such transportation studies are necessary, follows from a counter-intuitive 
example whereby improving the transportation system (e.g., by making extra 
infrastructure available), can result in an \emph{increase} of the travel times. 
This phenomenon, i.e., allowing more flexible routing that results in more 
congestion, is known as Brae\ss' paradox, after Dietrich 
Brae\ss~\cite{BRAESS:69}. The underlying reason for this counter-intuitive 
behaviour, is that people generally only \emph{selfishly} try to minimise their 
own travel times, instead of considering the effects they have on other people's 
travel times as well \cite{PAS:97}.

As transportation is inherently a temporal and spatial phenomenon, we first take 
a look at the concept of land-use models and their relation to the 
socio-economical behaviour of individual people. In the two subsequent sections, 
we consider two types of transportation planning models, i.e., the classic 
trip-based models, and the class of activity-based models, respectively. The 
section concludes with a brief reflection on the economist's view on 
transportation systems.

		\subsection{Land use and socio-economical behaviour}

As already stated, transportation demand arises because of the desire to 
participate in a set of activities (e.g., social, economical, cultural, \ldots). 
In order to deduce this \emph{derived} transportation demand, it is necessary to 
map the activity system and its spatial separations. This process is commonly 
referred to as \emph{land use}, mainly playing the role of forging a relation 
between economical and geographical sciences. In general, land-use models seek 
to explain the growth and layout of urban areas (which is not strictly 
determined by economical activities alone, i.e., ethnic considerations et cetera 
can be taken into account),

Because transportation has spatial interactions with land use and vice versa, it 
can lead to a kind of chicken-and-egg problem \cite{RODRIGUE:05}. For example, 
building a new road will attract some economical activity (e.g., shopping malls 
et cetera), which can lead to a possible increase of the travel demand. This in 
turn, can lead to an increase of extra economical activity (because of the 
well-suited location), and so on, resulting in a local reorganisation of the 
spatial structure. Resolving this chicken-and-egg paradox, is typically done by 
means of feedback and iterations between land-use and transportation models, 
whereby the former provide the basic starting conditions for the latter models 
(with sometimes a reversal of the models' roles).

In the following two sections, we first shed some light on several of the 
archetypical land-use models, after which we take a look at some of the more 
modern models for land use in the context of geosimulation.

		\subsubsection{Classic land-use models}

The discussion given in this section, talks about several kinds of land-use 
models that --- at their time --- were considered as landmark studies. That 
said, the models presented here should be judged as being general in that they 
deal with (pre-)industrial American societies in the first part of the 20\theth 
century. They are devised to gain insight into the general patterns that govern 
the growth and evolution of a city. As such, they almost never `fit' perfectly, 
leading to the obvious criticism that they are more applicable to American 
cities than elsewhere. Notwithstanding these objections, the models remain very 
useful as explanations for the mechanisms underpinning the socio-economical 
development of cities.

One of the oldest known models describing the relation between economic markets 
and spatial distances, is that of Johann Heinrich von Th\"unen 
\cite{VONTHUNEN:26}. As the model was published in 1826, it presents a rather 
`pre-industrial' approach: the main economical ingredients are based on 
agricultural goods (e.g., tomatoes, apples, wheat, \ldots), whereas the 
transportation system is composed of roads on which carts pulled by horses, 
mules, or oxen ride. The spatial layout of the model, assumes an \emph{isolated 
state} (self-sustaining and free of external influences), in which a central 
city location is surrounded by concentric regions of respectively farmers, 
wilderness, field crops, and meadows for grazing animals. All farmers aim for 
maximum profits, with transportation costs proportionally with distance, thus 
determining the land use around the city centre.

\begin{figure}[!htb]
	\centering
	\psfrag{CBD}[][]{\psfragstyle{\emph{CBD}}}
	\psfrag{I}[][]{\psfragstyle{\emph{I}}}
	\psfrag{L}[][]{\psfragstyle{\emph{L}}}
	\psfrag{M}[][]{\psfragstyle{\emph{M}}}
	\psfrag{H}[][]{\psfragstyle{\emph{H}}}
	\psfrag{C}[][]{\psfragstyle{\emph{C}}}
	\includegraphics[width=\figurewidth]{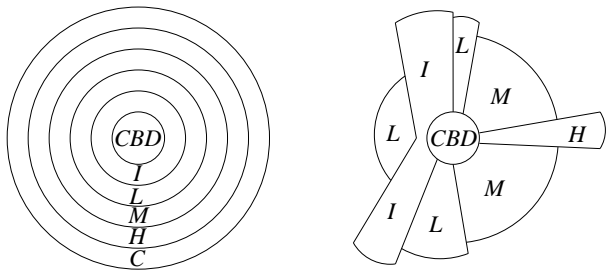}
	\caption{
		Typical examples of two models relaying the evolution of land use. 
		\emph{Left:} the concentric zone model of Burgess. \emph{Right:} the sector 
		model of Hoyt. In both figures, \emph{CBD} corresponds to the central 
		business district, \emph{I} to the industrial factories, \emph{L}, \emph{M}, 
		and \emph{H} to the low-, middle-, and high-class residents respectively. In 
		the Burgess model at the left, \emph{C} denotes the commuter zone.
	}
	\label{fig:TFM:LandUseModels}
\end{figure}

Some 100 years later, inspired by von Th\"unen's simple and elegant model, 
Ernest W. Burgess developed what is known as the \emph{concentric zone model} 
\cite{BURGESS:25}. It was based on observations of the city of Chicago at the 
beginning of the 20\theth century. As can be seen in the left part of 
\figref{fig:TFM:LandUseModels}, Burgess considered the city as growing around a 
central business district (CBD), with concentric zones of respectively the 
industrial factories and the low-, middle-, and high-class residents. The 
outermost ring denotes the commuter zone, connecting the CBD with other cities. 
As time progresses, the city develops and the radii of these concentric zones 
would grow by processes of `invasion' and `succession': an inner ring will 
expand, invading an outer ring that in turn has to grow, in order to make space.

Fifteen years after Burgess' theory, Homer Hoyt introduced refinements, 
resulting in the \emph{sector model} \cite{HOYT:39}. One of the main incentives, 
was the observation that low-income residents were typically located in the 
vicinity of railroads. His model accommodates this kind of observation, in that 
it assumes that a city expands around major transportation lines, resulting in 
wedge-shaped patterns (i.e., sectors), stretching outward from the CBD. A 
typical example of this development, can be seen in the right part of 
\figref{fig:TFM:LandUseModels}.\\

Halfway the previous century, Chauncy D. Harris and Edward L. Ullman were 
convinced that the previous types of models did not correspond to many of the 
encountered cities. The main reason for this discrepancy, was to be found in the 
stringent condition of a central area being surrounded by different zones. As a 
solution to this shortcoming, Harris and Ullman presented their \emph{multiple 
nuclei model} \cite{HARRIS:45}. Their theory assumed that in larger cities, 
small suburban areas could develop into fully fledged business districts. And 
although Harris and Ullman did not dispose of the CBD as the most important city 
centre, their smaller `nuclei' would take on roles of being areas for 
specialised socio-economical activities.

To end our discussion of classic land-use models, we highlight the work of Peter 
Mann in 1965 \cite{MANN:65}, who considered a \emph{hybrid model} for land-use 
representation. He combined both Burgess's and Hoyt's models, when deriving a 
model that described a typical British city. In his model that studied the 
cities of Huddersfield , Nottingham, and Sheffield, the CBD still remained the 
central location, surrounded by zones of pre- and post 1918 housing 
respectively. Dispersed around the outer concentric zone, the low-, middle-, and 
high-class residents would live. A most notable feature of Mann's model, is the 
fact that he considered the industrial factories to be on one side of the city, 
with the high-class residents diametrically opposed (the rationale being that 
high-class residents would prefer to stay upwind of the factories' smoke 
plumes).

		\subsubsection{The modern approach to land-use models}

In the current time of living, most modern citizens have a different behaviour 
than their former counterparts at the beginning of the 20\theth century. It 
seems there is an increased trend towards expansion, as people are feeling more 
comfortable about covering larger distances, e.g., working in a busy city centre 
or at a remote industrial facility, coupled with living on the countryside). The 
activities related to working, living, and recreation appear to occur at 
substantially different spatial locations. Furthermore, several urban regions 
are composed of unique ethnic concentrations, among other things leading to the 
conclusion that the emphasis on the geographical aspect of a city gets less 
important during its evolution.

Recognising these radical changes in the development, modern land-use models 
approach the integration of an activity system from a completely different 
perspective. The growth of a city is represented as the evolution of a 
\emph{multi-agent system}, in which a whole population of individual households 
is simulated. Due to the tremendous increase in computational power over the 
last two decades, these large-scale simulations are now possible. As an example, 
it is feasible to consider residential segregation in urban environments: within 
these environments (e.g., the city and housing market), individual agents (i.e., 
households) interact locally in a well-defined manner, leading to emergent 
structures, i.e., the evolving city. Besides data surveys that try to capture 
the households' behaviour, the basic landscape and mapping data is fed into 
\emph{geographical information systems} (GIS) that is coupled with a 
\emph{computer aided design} (CAD) representational model of the real world 
(although the difference between the traditional GIS and CAD concepts is slowly 
fading away) \cite{WADELL:04}. A recent example of such an all-encompassing 
approach, is the work related to the UrbanSim project, where researchers try to 
interface existing travel models with new land use forecasting and analysis 
capabilities \cite{WADELL:02}. It is being developed and improved by the Center 
for Urban Simulation and Policy Analysis at the University of Washington.

To conclude this section, we refer to the work of Benenson and Torrens, who 
adopted the terminology of \emph{geosimulation} \cite{BENENSON:04}. Their 
methodology is based on what they call the `collective dynamics of interacting 
objects'. As such, geosimulation hinges on the representation of what we would 
call a \emph{socio-economy} that is simulated, taking into account hitherto 
neglected dynamic effects (e.g., demographic changes, shifts of the economic 
activities, \ldots).

		\subsection{Trip-based transportation models}

The relation between activity patterns and the transportation system has a long 
history, starting around 1954 with the seminal work of Robert B. Mitchell and 
Chester Rapkin \cite{MITCHELL:54}. They provided the first integrated study, 
establishing a link that introduced a framework for transportation analysis, 
primarily intended for studying large scale infrastructure projects 
\cite{MCNALLY:00b}. Their methodology was based on four consecutive steps (i.e., 
submodels), collectively called the \emph{four step model} (4SM). In 1979, 
Manheim casted the model's structure into a larger framework of transportation 
systems analysis, encapsulating both activity and transportation systems 
\cite{MANHEIM:79}. Central to this framework, was the notion of `demand and 
performance procedures', which we can validly call \emph{demand} and 
\emph{supply procedures}. In a typical setup, they respectively represent the 
traffic that wants to use this infrastructure and the road infrastructure. For a 
more historically tinted recollection of the trip-based approach, we refer the 
reader to the outstanding overview of Boyce \cite{BOYCE:04b}.\\

\sidebar{
	With respect the 4SM's history, a subtle --- almost forgotten --- fact is that 
	the classic four step model was actually conceived independently from the 
	integrated network equilibrium model proposed by Beckmann, McGuire, and 
	Winsten in the mid-fifties; the 4SM can actually be perceived as a 
	trimmed-down version of this latter model \cite{BOYCE:04b}. Intriguingly, over 
	the years, the work of the `BMW trio' has had profound impacts on the 
	mathematical aspects of determining network equilibria, optimal toll policies, 
	algorithms for variational inequalities, stability analyses, supply chains, 
	\ldots \cite{ALTMAN:03,NAGURNEY:03,BOYCE:04b,BOYCE:04}.
}\\

In the next four sections, we consider the basic entities and assumptions of the 
four step model, followed by a brief overview of the four individual submodels 
with some more detail on the fourth step (traffic assignment), concluding with 
some remarks on the criticisms often expressed against the four step model. For 
a more extensive survey of the four step model, we refer the reader to the books 
of Sheffi \cite{SHEFFI:85} and Ortuzar and Willumsen \cite{ORTUZAR:01}.

			\subsubsection{Basic entities and assumptions}

The basic ingredients on which the four step model is rooted, are the 
\emph{trips}. These trips are typically considered at the household level, and 
relate to aggregate information (individuals are no longer explicitly 
considered). This level of detail, essentially collapses the whole tempo-spatial 
structure of transportation planning based on individual travellers into bundles 
of trips, going from one point in the transportation network to another.

In the four step model, one of the most rigid assumptions is that all trips 
describe departure and arrival within the planning period (e.g., the morning 
commute). Furthermore, the usage of the model's structure is intended for 
large-scale planning purposes, excluding small infrastructural studies at e.g., 
a single intersection of urban roads. Another assumption is based on the fact 
that an entity within the four step model has to make certain decisions, e.g., 
what is the departure time, which destination is picked, what kind of 
transportation (private or public) will be used, which route will be followed, 
\ldots In many cases, these decisions are considered concurrently, but the four 
step model assumes they are made independently of each other. And finally, as 
each submodel needs input, most of the data is aggregated into \emph{spatial 
zones} (often presumed to be distinguished by socio-economic characteristics) in 
order to make the model computationally feasible. These zones are typically 
represented by their centrally located points, called \emph{centroids}.

			\subsubsection{The four steps}
			\label{sec:TFM:4SM}

Within the four step model, the first three steps (I) -- (III) can collectively 
be seen as a methodology for setting up the travel demand, based e.g., on land 
use and other socio-economical activities. This travel demand is expressed as 
\emph{origin-destination} (OD) pairs (by some respectively called `sources' and 
`sinks'), reflecting the amount of traffic that \emph{wants} to travel from a 
certain origin to a certain destination (these are typically the zones mentioned 
in the previous section). The last step (IV) then consists of loading this 
travel demand onto the network, thereby assigning the \emph{routes} that 
correspond to the trips.\\

\textbf{(I) Trip generation}\\
In an essential first step, transportation engineers look at all the trips that 
on the one hand originate in certain zones, and on the other hand arrive in 
these zones. As such, the first step comprises what are called the 
\emph{productions} and \emph{attractions}. Central to the notion of a trip, is 
the \emph{motive} that instigated the trip. An example of such a motive is a 
home-based work trip, i.e., a trip that originates in a household's residential 
area, and arrives in that household's work area. Other examples include 
recreational and social motives, shopping, \ldots and the \emph{chaining of 
activities}. Based on these intentions, productions and attractions consist of 
absolute counts, denoting the number of trips that depart from and arrive in 
each zone. Because of this, productions and attractions are in fact \emph{trip 
ends}. Both of them are derived using techniques based on regression analysis, 
category analysis, or even logit models. As different models can be used for the 
derivations of the number of productions and attractions, an a posteriori 
balancing is performed that equalises both results. In the end, step (I) gives 
the magnitude of the total travel demand on the network. Note that all 
activities (i.e., the original motives) are at this point in effect transformed 
and \emph{aggregated} into trips. More importantly, these trips are \emph{only 
considered for a specific time} period (e.g., the morning rush hour).\\

\textbf{(II) Trip distribution}\\
Once the total number of productions and attractions for all zones in the 
transportation network is known, the next step then consists of deriving how 
many trips, originating in a certain zone, arrive at another zone. In other 
words, step (II) \emph{connects} trip \emph{origins} to their 
\emph{destinations} by distributing the trips. The result of step (II) is then 
the construction of a complete \emph{origin-destination} table (\emph{OD 
table}). In such an OD table (or \emph{OD matrix} as some people say), an 
element at a row $i$ and a column $j$ denotes the total number of trips 
departing from origin zone $O_{i}$ and arriving in destination zone $D_{j}$. 
Diagonal elements denote intra-zonal trips. Note that step (II) does not state 
anything about the different \emph{routes} that can be taken between two such 
zones; this is something that is derived in the final step (IV). Because of the 
implicit assumption in step (I), namely that all trips are considered for a 
specific time period, the same premise holds for all the derived OD tables. 
Consequently, the four step model is applied for different time periods, e.g., 
during rush hours or off-peak periods. In this context, we advise to use the 
nomenclature of \emph{time-dependent} or \emph{dynamic} OD tables, denoting OD 
tables that are specified for a certain period, e.g., from 07:00 until 08:00 (or 
even tables given for consecutive quarter-hours).

Considering the fact that an OD table contains a large amount of unknown 
variables (it is a considerably under determined system of equations), several 
techniques have been introduced to deal with this problem by introducing 
additional constraints. If an OD table for a previous period (called a 
\emph{base table}) is known, then a new OD table can be derived by using a 
so-called \emph{growth factor model}. Another method is by using \emph{gravity 
models} (also known as \emph{entropy models}, see e.g., the discussion in by 
Helbing and Nagel \cite{HELBING:04}), which are based on \emph{travel impedance 
functions}. These functions reflect the relative attractiveness of a certain 
trip e.g., based on information retrieved from household travel surveys. In most 
cases, they are calibrated as power or exponential functions. One of the harder 
problems that still remains to be solved, is how to deal with so-called 
\emph{through trips}, i.e., trips that originate or end outside of the study 
area. Horowitz and Patel for example, directly incorporate rudimentary 
geographical information and measured link flows into a model that allows to 
derive through-trip tables, using a notion of external stations located in an 
external territory. Application of his methodology to regions in Wisonsin and 
Florida, result in reasonable estimates of link flows that are comparable with 
empirically obtained data \cite{HOROWITZ:99}.

Besides using results from productions and attractions, gathering the necessary 
information for construction of OD tables can also be done using other 
techniques. An equivalent methodology is based on the consideration of 
\emph{turning fractions} at intersections. The process can be largely automated 
when using video cameras coupled with image recognition software. Furthermore, 
there literally exist thousands of papers devoted to the estimation of 
origin-destination matrices, mostly applicable to small-scale vehicular 
transportation networks and local road intersections. Some past methodologies 
used are the work of Nihan and Davis who developed a recursive estimation scheme 
\cite{NIHAN:87}, the review Cascetta and Nguyen who casted most earlier methods 
into a unified framework \cite{CASCETTA:88}, and Bell who estimated OD tables 
based on constrained generalised least squares \cite{BELL:91}. An example of a 
more recent technique is the work of Li and De Moor who deal with incomplete 
observations \cite{LI:02}.\\

\textbf{(III) Mode choice / modal split}\\
Once the origin-destination table for the given network and time period is 
available, the next step deals with the different \emph{modes of transportation} 
that people choose between. Typical examples are the distinction between private 
and public transportation (both vehicular and railroad traffic). The `split' in 
this step, refers to the fact that the OD table obtained from step (II), is now 
divided over the supported transportation modes. To this end, \emph{discrete 
choice theory} is a popular tool that allows a disaggregation based on the 
choice of individual travellers, e.g., by using utility theory based on a nested 
logit model \cite{BENAKIVA:85}. A modern trend in this context is to work with 
fully \emph{multi-modal transportation networks}; these multi-layered networks 
provide access points for changing from one layer (i.e., mode of transportation) 
to another \cite{VANNES:02}.

Historically, steps (I) --- production and attraction --- and (III) were 
executed simultaneously, but nowadays they are considered separate from step 
(I): the main reason is the fact that the modal choice is not only dependent on 
e.g., a household's income, but also on the type of trip to be undertaken, as 
well as the trip's destination. As a result, the modal split can be intertwined 
with step (II), trip distribution, or it can be executed subsequently after step 
(II). In the former case, the same kind of travel impedance functions are used 
in combination with an adjusted gravity model, whereas in the latter case, a 
hierarchic logit model can be used.\\

\textbf{(IV) Traffic assignment}\\
At this point in the four step model, the total amount of trips undertaken by 
the travellers is known. The fourth and final step then consists of finding out 
\emph{which routes} these travellers follow when going from their origins to 
their destinations, i.e., which sequence of consecutive links they will follow~? 
In a more general setting, this process is known as \emph{traffic assignment}, 
because now the total travel demand (i.e., the trips) are assigned to routes in 
the transportation network. Note that in some approaches, an iteration is done 
between the four steps, e.g, using the traffic assignment procedure to calculate 
link travel times that are fed back as input to steps (II) and (III).

It stands to reason that all travellers will endeavour to take the 
\emph{shortest route} between their respective origins and destinations. To this 
end, a suitable measure of distance should be defined, after which a shortest 
path algorithm, e.g., Dijkstra's algorithm \cite{DIJKSTRA:59}, can calculate the 
possible routes. Such a notion of distance typically includes both spatial and 
temporal components, e.g., the physical length of an individual link and the 
travel time on this link, respectively. The use of the travel time is one of the 
most essential and tangible components in travellers' route choice behaviour. 
Note that in a more general setting, the distance can be considered as a 
\emph{cost}, whereby travellers then choose the \emph{cheapest route} (i.e., the 
quickest route when time is interpreted as a cost). Daganzo calls these 
formulations the \emph{forward shortest path problem}, as opposed to the 
\emph{backward shortest path problem} that tries to find the cheapest route for 
a given arrival time \cite{DAGANZO:02d}.

The basic principles that underlay route choice behaviour of individual 
travellers, were developed by Wardrop in 1952, and are still used today. In his 
famous paper, relating space- to time-mean speed, Wardrop also stated two 
possible criteria governing the distribution of traffic over alternative routes 
\cite{WARDROP:52}:

\begin{quote}
	\textbf{User equilibrium (W1):} \emph{``The journey times on all the routes 
	actually used are equal, and less than those which would be experienced by a 
	single vehicle on any unused route.''}\\

	\textbf{System optimum (W2):} \emph{``The average journey time is a 
	minimum.''}
\end{quote}

The above two criteria are based on what is called the \emph{Nash equilibrium} 
in game theory \cite{NASH:51}, albeit that now a very large number of 
individuals are considered\footnote{The difference between a Wardrop and a Nash 
equilibrium is a subtle but important one. In the Wardrop case, an infinite 
number of individuals is considered, each seeking their own optimum. Note that 
the concept `infinite number of individuals' can practically be approximated by 
'a large amount of individuals'. The Nash case also considers an infinite number 
of individuals, but they are now grouped into a finite number of classes, with 
each class seeking its own optimum. If in this latter case the number of classes 
goes to infinity, then the Nash equilibrium converges to a Wardrop equilibrium 
\cite{HAURIE:85}.}. In the first criterion (W1), it is assumed that all 
individuals' decisions have a negligible effect on the performance of others. 
Two, more important, fundamental principles here are the fact that in the 
equilibrium situation, there is \emph{no cooperation between individuals} 
assumed, and that all individuals make their decisions in an \emph{egoistic and 
rational} way \cite{HAGSTROM:01}. In real-life traffic, everybody is expected to 
follow the first criterion (W1), such that the whole system can settle in an 
equilibrium in which no one is better off by choosing an alternative route. In 
this respect, the work of Roughgarden is interesting because it provides a 
mathematical basis for the quantification of the worst loss of social welfare 
due to \emph{selfish routing}, and the management of networks that limit these 
effects in order to obtain a socially desirable outcome \cite{ROUGHGARDEN:02}. 
In contrast to this user equilibrium situation, the second criterion (W2) is 
unlikely to occur spontaneously. However, when the perceived cost of a route by 
a traveller is changed to a \emph{generalised} or \emph{marginal cost} (i.e., 
including the costs of the effects brought on by adding an extra vehicle to the 
travel demand), then a system optimum is achieved with respect to these latter 
costs. In any case, as some people will be better off, others will be worse off, 
but the transportation system as a whole will be best off.

The above two principles, are a bit idealistic, in the sense that there are many 
exceptions to these behavioural guidelines. For example, in urban city centres, 
a significant part of the congestion can be brought on by vehicles looking for 
parking space. Furthermore, many drivers just follow their \emph{usual route}, 
because this is the route they know best, and they know what to expect with 
respect to travel time. In a broader setting, this make these `standard' routes 
more \emph{appealing} to road users than other unfamiliar alternative routes. In 
some cases however, travellers will opt for these less known routes, thereby 
possibly entering the \emph{risk} of experiencing a higher travel time as has 
been concluded in the work of Chen and Recker \cite{CHEN:01b}. Another fact that 
we expect to have a non-negligible effect on the distribution of traffic flows, 
is that nowadays more and more people use \emph{intelligent route planners} to 
reach their destinations. These planners take into account congestion effects, 
as the trip gets planned both spatially and temporally. This will result in a 
certain percentage of the population that is informed either \emph{pre-route} or 
\emph{en-route}, and these people can consequently change their departure time 
or actual route (e.g., through route guidance), respectively. Another 
interesting research problem arises because transportation infrastructure 
managers should then be able to adapt their policies to the changing travel 
patterns. For example, how should a policy maker optimally control the traffic 
when only 20\% of the population will follow the proposed route guidance~?

Due to the importance of the subject, we have devoted two separate sections in 
this dissertation to the concept of traffic assignment. In these sections, we 
discuss the traffic assignment procedure in a bit more detail, considering two 
prominent methodologies from a historic perspective, namely \emph{static} versus 
\emph{dynamic} traffic assignment.

			\subsubsection{Static traffic assignment}
			\label{sec:TFM:STA}

The classic approach for assigning traffic to a transportation network, assumes 
that all traffic flows on the network are \emph{in equilibrium}. In this 
context, the \emph{static traffic assignment} (STA) procedure can be more 
correctly considered as dealing with \emph{stationary} of \emph{steady-state} 
flows: the travel demand and road infrastructure (i.e., the supply) are supposed 
to be time-independent, meaning that the calculated link flows are the result of 
a constant demand. In a typical setup, this entails the assignment of an hourly 
(or even daily) OD table to the network (e.g., during on- and off-peak periods), 
resulting in \emph{average} flows for the specified observation period. Because 
the STA methodology neglects time varying congestion effects (it assumes 
constant link flows and travel times), various important phenomena such as queue 
spill back effects are not taken into account.

In general, several possible techniques exist for achieving an STA. The first 
one assumes (i) that all drivers will choose the same cheapest route between a 
pair of origins and destinations, (ii) that they all have the same \emph{perfect 
information} about the links' impedances, and (iii) that these impedances are 
considered to be constant, i.e., independent of a link's traffic load (so no 
congestion buildup is taken into account). As the methodology implies, this is 
called an \emph{all-or-nothing assignment} (AON). A second technique refines 
this notion, whereby differences among drivers are introduced (i.e., giving rise 
to imperfect information), resulting in a \emph{stochastic assignment}. In this 
methodology, the link travel impedances are assumed to be probabilistically 
distributed: for each link in the network, an impedance is drawn from the 
distribution after which an AON assignment is performed on the resulting 
network. This Monte Carlo process is repeated until a certain termination 
criterion is met.

Both previous methods carry a significant drawback with respect to link 
capacities, that is to say, no effects are taken into account due to the fact 
that an increased flow on a link will generally result in an increase of the 
travel time (i.e., the link's impedance). To this end, a third method introduces 
\emph{capacity restraints} such that an increase of the travel demand on a link, 
will result in a higher cost (thereby possibly changing the route with the 
cheapest cost). This method is called an \emph{equilibrium assignment}, and just 
like as in the second method, a \emph{stochastic equilibrium assignment} version 
can be derived, taking into account travellers' imperfect knowledge. The 
underlying assumption is that all travellers behave according to Wardrop's user 
equilibrium (W1). Furthermore, the capacity restraints are included in the 
travel impedance functions, as they are now synonymously called \emph{travel 
time (loss) functions}, \emph{congestion functions}, \emph{volume delay 
functions}, \emph{link impedance functions}, or even \emph{link performance 
functions}. A popular form of these functions that express the travel time $T$ 
in function of the flow $q$ on a link, is the \emph{Bureau of Public Roads} 
(BPR) power function \cite{BPR:64}:

\begin{equation}
	T = T_{\text{ff}} \left ( 1 + \alpha \left ( \frac{q}{q_{\text{pc}}} \right )^{\beta} \right ).
\end{equation}

\begin{figure}[!htb]
	\centering
	\psfrag{T}[][]{\psfragstyle{$T$}}
	\psfrag{Tff}[][]{\psfragstyle{$T_{\text{ff}}$}}
	\psfrag{q}[][]{\psfragstyle{$q$}}
	\psfrag{qpc}[][]{\psfragstyle{$q_{\text{pc}}$}}
	\psfrag{0}[][]{\psfragstyle{\zero}}
	\includegraphics[width=\figurewidth]{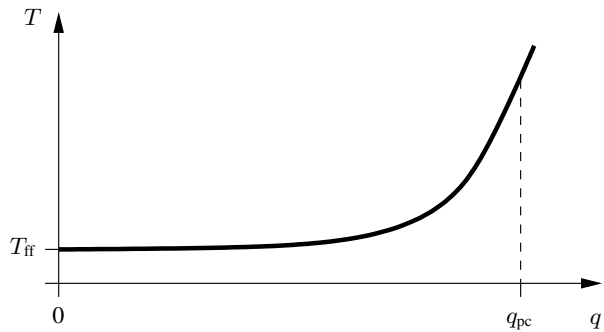}
	\caption{
		The Bureau of Public Roads (BPR) function, relating the travel time to the 
		flow. It is based on the travel time $T_{\text{ff}}$ under free-flow 
		conditions and the practical capacity $q_{\text{pc}}$ of the link under 
		consideration.
	}
	\label{fig:TFM:BPRFunction}
\end{figure}

In this BPR relation, the coefficients $\alpha$ and $\beta$ determine the shape 
of the function. An example of such a function is depicted in 
\figref{fig:TFM:BPRFunction}. For low flows, the BPR function is rather flat and 
the travel time corresponds to the travel time $T_{\text{ff}}$ under free-flow 
conditions. When higher flows occur on the link, the coefficient $\beta$ 
determines the threshold at which the BPR function rises significantly (in some 
formulations it \emph{asymptotically} approaches the capacity flow). The travel 
time will increase with the ratio of the flow $q$ and the so-called 
\emph{practical capacity} $q_{\text{pc}}$. This latter characteristic is derived 
from the value of the travel time under congested conditions. As a result, the 
practical capacity is different from the maximum capacity of a link as defined 
by a fundamental diagram. Finally, note that a serious disadvantage associated 
with these BPR functions in combination with static traffic assignment, is the 
fact that the travel demand on the network at a certain time does not always 
correspond to the actual physical flows that can be sustained. Under congested 
conditions, this implies that the flows in the STA approach can be higher than 
the physically possible link capacities (which are different from the previously 
mentioned practical capacities), leading to an incorrect assignment with faulty 
oversaturated links.

Once the travel time of a link can be related to its current flow using e.g., a 
BPR function, an iterative scheme is adopted to calculate the equilibrium 
traffic assignment. Popular implementations are the \emph{Frank-Wolfe algorithm} 
\cite{FRANK:56}, and the \emph{method of successive averages} (MSA). The former 
method is based on principles of optimisation theory, as demonstrated by 
Beckmann et al. \cite{BECKMANN:55,BOYCE:04} who reformulated the Wardrop 
equilibrium as a convex optimisation problem, consisting of a single objective 
function with linear inequality constraints based on the Karush-Kuhn-Tucker 
(KKT) conditions, thereby resulting in a global minimum. Because travellers do 
not have perfect information, Daganzo and Sheffi formulated a variation on 
Wardrop's first criterion (W1), whereby all traffic distributes itself over the 
network with respect to a \emph{perceived travel time} of the individual drivers 
\cite{DAGANZO:77}. The resulting state of flows on the network is called a 
\emph{stochastic user equilibrium} (SUE), as opposed to the \emph{deterministic 
user equilibrium} (DUE). Note that a further discrimination is possible, as 
proposed by the work of Chen and Recker, who also make a distinction between 
travellers' perception errors on the one hand, and network uncertainty (i.e., 
stochasticity of the travel times) on the other hand \cite{CHEN:01b}. For a 
thorough overview of the STA approach, we refer the reader to the work of 
Patriksson \cite{PATRIKSSON:94}.

Although, as mentioned earlier, time varying congestion effects are not taken 
into account, the STA approach does fit nicely into the concept of long-term 
transportation planning. For short-term analyses however, these effects can have 
a significant impact on the end results, thus requiring a more detailed approach 
to traffic assignment.\\

			\subsubsection{Dynamic traffic assignment}
			\label{sec:TFM:DTA}

As explained in the previous paragraphs, the static traffic assignment heavily 
relies on simple travel time functions (e.g., BPR). An associated problem with 
these is the difficulty in capturing the concept of `capacity of a road'. In 
reality, congestion is a dynamic phenomenon, whereby its temporal character is 
not to be neglected. To tackle theses problems inherent to the STA approach, a 
more dynamic treatment of traffic assignment is necessary \cite{MAERIVOET:04g}. 
A fundamentally important aspect in this \emph{dynamic traffic assignment} (DTA) 
procedure, is the fact that congestion has a temporal character, meaning that 
its buildup and dissolution play an important role: the \emph{history of the 
transportation system} should be taken into account (e.g., congestion that 
occurs due to queue spill back) \cite{DAGANZO:95d}. Neglecting this time 
dependency by assuming that the entry of a vehicle to a link instantaneously 
changes the flow on that link, results in what is called \emph{Smeed's paradox}. 
This leads to incorrect behaviour as a result of a violation of the so-called 
`first-in, first-out' (FIFO) property, because now a vehicle can leave link 
earlier then a vehicle that enters it later (i.e., arriving earlier by departing 
later) \cite{SMEED:67}. The methodology of dynamic traffic assignment was now 
designed to deal with all these particular aspects. The DTA technique is 
composed of two fundamental components:\\

\textbf{Route choice}\\
Just as in the STA approach, each traveller in the transportation network 
follows a certain route based on an instinctive criterion such as e.g., 
Wardrop's (W1). The associated component that takes care of travellers' route 
choices, can be complemented to allow for imperfect information. Another, more 
important, aspect related to the route choice, is a traveller's \emph{choice of 
departure time}. An STA approach assumes that all traffic of a given OD table is 
simultaneously assigned to the network, whereas DTA coupled with departure time 
choice can spread the departures in time (leading to e.g., realistically 
spreading of the morning peak's rush hour).\\

\textbf{Dynamic network loading (DNL)}\\
Instead of using simple travel time functions, a DTA approach typically has a 
component that loads the traffic onto the network. In fact, this step resembles 
the \emph{physical propagation} of all traffic in the network. In order to 
achieve reliable and credible results, a good description of the network's links 
is necessary, as well as the behaviour of traffic at the nodes connecting the 
links within this network (i.e., this is a mandatory requirement to achieve 
correct modelling of queue spill back). The DNL component in the DTA approach 
has been an active field of research during the last decade, and it still 
continues to improve the state-of-the-art. Testimonies include the use of 
\emph{analytic models} that give correct representations of queueing behaviour, 
as well as detailed simulations that describe the propagation of individual 
vehicles in the transportation network. Note that in the case of 
\emph{simulation-based} (also called \emph{heuristic}) traffic assignment, the 
route choice and DNL components can be iteratively executed, whereby the former 
establishes a set of routes to follow, and the latter step feedbacks information 
to the route choice model until a certain termination criterion is met (e.g., a 
relaxation procedure). Furthermore, using simulation-based traffic assignment 
with very large road networks is not always computationally feasible to 
calculate all shortest paths. As a result, it might be beneficial to resort to 
simplifications of either the simulation model (e.g., using faster queueing 
models), or the number of paths to consider (e.g., based on the hierarchy 
inherently present in the road network) \cite{ROSSWOG:01}. Finally, we mention 
the work of Astarita who provides an interesting classification of DNL models, 
based on the discretisation with respect to the spatial and temporal dimensions, 
as well as with respect to the modelling of the traffic demand 
\cite{ASTARITA:02}.

Despite the appealing nature of simulation-based DTA, there is in contrast to 
the STA approach, no unified framework that deals with the convergence and 
stability issues \cite{GAWRON:98b,GAWRON:98,PEETA:03}.

Some examples of these DTA mechanisms are: Gawron who uses a queueing model to 
develop a simulation-based assignment technique that is able to deal with 
large-scale networks and is proven to be empirically stable 
\cite{GAWRON:98b,GAWRON:98}, Bliemer who developed an analytical DTA approach 
(with different user-classes) based on a variational inequality approach 
\cite{BLIEMER:01}, Bliemer's work furthermore culminated in the development of 
\emph{INteractive DYnamic traffic assignment} (INDY) \cite{MALONE:03} which --- 
in combination with the 
OmniTRANS\footnote{\texttt{http://www.omnitrans-international.com}} commercial 
transportation planning software --- can be used as a fully fledged DTA analysis 
tool \cite{VERSTEEGT:03}, Lo and Szeto who developed a DTA formulation based on 
a variational inequality approach leading to a dynamic user equilibrium  
\cite{LO:02}, the group of Mahmassani who is actively engaged in the DTA scene 
with the development of the DYNASMART (\emph{DYnamic Network 
Assignment-Simulation Model for Advanced Roadway Telematics}) simulation suite 
\cite{DYNASMART:03}, \ldots An excellent comprehensive overview of several 
traditional DTA techniques is given by Peeta and Ziliaskopoulos \cite{PEETA:01}.

Another important field of research, is how individual road travellers react to 
the route guidance they are given. In his research, Bottom considered this type 
of \emph{dynamic traffic management} (DTM), providing route guidance to 
travellers whilst taking into account their \emph{anticipated behaviour} during 
e.g., incidents \cite{BOTTOM:00}. Taking this idea one step further, it is 
possible to study the interactions between the behaviour of travellers in a road 
network, and the management of all the traffic controls (e.g., traffic signal 
lights) within this network. An example of such a \emph{dynamic traffic control} 
(DTC) and DTA framework, is the work of Chen who considers the management from a 
theoretic perspective based on a non-cooperative game between road users and the 
traffic authority \cite{CHEN:98}. 

			\subsubsection{Critique on trip-based approaches}
			\label{sec:TFM:CritiqueOnTripBasedApproaches}

Considering its obvious track record of the past several decades, the 
\emph{conventional} use of the trip-based approach is --- to our feeling --- 
running on its last legs. By `conventional' we denote here the fact that the 
current \emph{state-of-the-practice} is still firmly based on the paradigm of 
static traffic assignment, despite the recent (academic) progress on the front 
of dynamic traffic assignment techniques. The four step model still largely 
dominates the commercial business of transportation planning, although its 
structure remained largely unchanged since its original inception. As mentioned 
earlier, in the case of STA, all trips are assumed to depart and arrive within 
the specified planning period. This leads to an unnatural discrepancy between 
models and reality in congested areas during e.g., a morning rush hour: some 
travellers \emph{want} to make a trip and, in the former case, are perfectly 
allowed to achieve this trip, whereas in the latter case they are in fact 
physically \emph{unable} to make the trip due to dynamical congestion effects.

In order to facilitate this disagreement between the balancing of travel demand 
versus supply (i.e., the transportation infrastructure), the DTA approach is 
gaining importance as more features are provided. An example of such a feature 
includes the framework of congestion pricing, where we have an incorporation of 
departure time choice models coupled with the derivation of optimal road tolls. 
Some noteworthy studies that have been carried out in this respect, are the work 
of de Palma and Marchal who present the METROPOLIS toolbox, allowing the 
simulation of large-scale transportation networks \cite{DEPALMA:02}, the work of 
Lago and Daganzo who combined a departure time equilibrium theory with a 
fluid-dynamic model in order to assess congestion policy measures 
\cite{LAGO:03b}, the work of Szeto and Lo who coupled route choice and departure 
time choice with the goal of numerically handling large-scale transportation 
networks \cite{SZETO:04,LO:04}. Closely related to Lago's and Daganzo's work is 
that of Yperman et al., who determined an optimal pricing policy, describing the 
demand side with a bottleneck model and an analytical fluid-dynamic model as the 
DNL component \cite{YPERMAN:05}.

At this point, we should mention some of the complications associated with the 
traditional method of modelling traffic flow propagation using queue-based 
analogies. Historically, there have been two different queueing techniques with 
FIFO discipline that describe this aspect in a trip-based assignment procedure:

\begin{itemize}
	\item Point queues (also called \emph{vertical queues}): this type of queue 
	has no spatial extent. Because vehicles can \emph{always} enter the queue, and 
	leave it after a certain delay time, congestion is incorrectly modelled. A 
	well-known example of a model based on this queueing policy is Vickrey's 
	bottleneck model \cite{VICKREY:69}.
	\item Spatial/physical queues (also called \emph{horizontal queues}): a 
	queue of this type has an associated \emph{finite capacity}, i.e., a buffer 
	storage. Vehicles can only enter the queue when there is enough space for 
	them available.
\end{itemize}

The correct modelling of congestion effects such as queue spill back, is of 
fundamental importance when assessing certain policy measures like e.g., road 
pricing schemes. To this end, the use of vertical queues should be abandoned, in 
favour of horizontal queues. However, even horizontal queues have problems 
associated with them: the buildup and dissolution of congestion in a 
transportation network are flawed, e.g., vehicles that are leaving the front of 
a queue \emph{instantly open up a space at the back of this queue}, thus 
allowing an upstream vehicle to enter. This leads to shorter queue lengths, 
because the physical queueing effect of individual vehicles (i.e., the upstream 
propagation of the empty spot) is absent 
\cite{GAWRON:98,SIMON:99,CETIN:03,LAGO:03b}. In order to alleviate this latter 
issue, a more realistic velocity should be adopted for the backward propagating 
kinematic wave, thus calling for more advanced modelling techniques that 
\emph{explicitly} describe the propagation of traffic (e.g., fluid-dynamic 
approaches, models with dynamical vehicle interactions, \ldots).

As often is the case, a model's criticisms can be found in its underlying 
assumptions. In the case of the four step approach, it is obvious that all 
information regarding individual households is lost because of its aggregation 
to a trip level. As was already recognised from the start, the individual itself 
loses value during this conversion process. This opened the door towards another 
approach to transportation planning, more precisely \emph{activity-based 
modelling} (ABM) which is discussed in the next section.

A final complaint that is more common around many of these grotesque models, is 
their requirement of a vast amount of specific data. In many cases, a diverse 
range of national studies are carried out, having the goal of gathering as much 
data as possible. Regardless of this optimism, some of the key problems remain, 
e.g., it is still not always straightforward to properly interpret and adapt 
this data so it can be used as input to a transportation planning tool.

		\subsection{Activity-based transportation models}

As it was widely accepted that the rationale for travel demand can be found in 
people's motives for participating in social, economical, and cultural 
activities, the classic trip-based approach nevertheless kept a strong foothold 
in the transportation planning community. Instead of focussing attention 
elsewhere, the typical institutional policy was to ameliorate the existing four 
step models \cite{MCNALLY:00}. However, some problems persistently evaded a 
solution with the trip-based approach, e.g., shops that remain open late, 
employers who introduce flexible working hours, the consideration of joint 
activities by members of a household, \ldots

In the next few sections, we illustrate how all this changed with the upcoming 
field of activity-based transportation planning. We first describe its historic 
origins, after which we move on to several of the approaches taken in 
activity-based modelling. The concluding section gives a concise overview of 
some of the next-generation modelling techniques, i.e., large-scale agent-based 
simulations.

			\subsubsection{Historic origins}

The historic roots of the activity-based approach can probably be traced back to 
1970, with the querulous work of Torsten H\"agerstrand \cite{HAGERSTRAND:70}. He 
asserted that researchers in regional sciences should focus more on the 
intertwining of both disaggregate spatial and temporal aspects of human 
activities, as opposed to the more aggregate models in which the temporal 
dimension was neglected. This scientific field got commonly termed as \emph{time 
geography}; it encompasses all time scales (i.e., from daily operations to 
lifetime goals), and focusses on the constraints that individuals face rather 
than predicting their choices \cite{MILLER:04}.

\begin{figure}[!htb]
	\centering
	\includegraphics[width=\figurewidth]{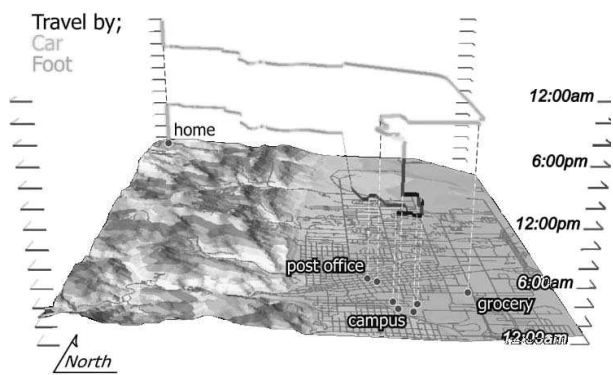}
	\caption{
		An example of a space-time path showing an individuals' activity and travel 
		behaviour in the space-time volume: the two spatial dimensions make up the 
		physical world plane, with the vertical axis denoting the temporal 
		dimension. In this case, we can see a woman going from her home in Boulder 
		(Colorado, USA), to the university's campus, followed by a visit to the post 
		office and grocery store, and finally returning home (image reproduced after 
		\cite{DETTLOFF:01}).
	} 
	\label{fig:TFM:HagerstrandSpaceTimePath}
\end{figure}

Central to H\"agerstrand's work was the notion of so-called \emph{space-time 
paths} of individuals' activity and travel behaviour. In a three-dimensional 
space-time volume, two spatial dimensions make up the physical world plane, with 
the temporal dimension as the vertical axis. The journey of an individual is now 
the path traced out in this space-time volume: consecutive visits to certain 
locations are joined by a curve, with vertical segments denoting places where 
the individual remained stationary during a certain time period. The complete 
chain of activities (called a \emph{tour}) is thus joined by individual 
\emph{trip legs}. In this respect, the space-time path represents the 
\emph{revealed outcome} of an unrevealed behavioural process \cite{MCNALLY:00}. 
An example of such a path can be seen in 
\figref{fig:TFM:HagerstrandSpaceTimePath}: we can see a woman going from her 
home in Boulder (Colorado, USA), to the university's campus, followed by a visit 
to the post office and grocery store, and finally returning home 
\cite{DETTLOFF:01}. Note that H\"agerstrand extended his notion in the 
space-time volume to include \emph{space-time prisms} that encapsulate and 
effectively \emph{constrain} all of a person's reachable points (i.e., all 
his/her possible space-time paths), given a certain maximum travel speed as well 
as both starting and ending points within the volume \cite{CORBETT:05}. This 
environment is sometimes also referred to as a person's \emph{action space}, 
enveloping that person's \emph{time budget}.

Contrary to the belief that the field of activity-based transportation planning 
found its crux with the dissatisfaction of trip-based modelling, it grew and 
emerged spontaneously as a separate research study into human behaviour 
\cite{MCNALLY:00}. The underlying idea however remained the same as in the 
trip-based approach, namely that travel decisions arise from a need to 
participate in social, economical, and cultural activities. But as opposed to 
the more aggregated trip-based view, the basic units here are individual 
activity patterns, commonly referred to as \emph{household activity patterns}. 
In this context, the activity-based approach then studies the interactions 
between members of a household, and the relation to their induced travel 
behaviour \cite{AXHAUSEN:00}.

			\subsubsection{Approaches to activity-based modelling}

Departing from H\"agerstrand's initial comments, activity-based research 
progress has been slowly but steadily. In contrast with the development of the 
trip-based approach that culminated in the four step model, \emph{there is no 
explicit general framework that encapsulates the activity-based modelling 
scheme}. There were however early comprehensive studies into human activities 
and their related travel behaviour, e.g., the synopsis provided by Jones et al. 
\cite{JONES:83}. As the field began to mature, certain ingredients could be 
recognised, e.g. \cite{AXHAUSEN:00}:

\begin{itemize}
	\item the \emph{generation of activities}, which can be regarded as the 
	equivalent of the production/attraction step in trip-based modelling,
	\item the modelling of \emph{household choices}, i.e., with respect to their 
	activity chains; this includes choosing starting time and duration of the 
	activity, its location as well as a modal choice,
	\item the \emph{scheduling of activities}, outlining how a household plans and 
	executes the tasks of its members for long-, middle-, and short-term 
	activities, going from year- and lifetime-long commitments, to daily 
	operations.
\end{itemize}

During the last three decades, many research models that encompass activity 
scheduling behaviour have been developed. An excellent overview is given by 
Timmermans, who makes a distinction between \emph{simultaneous} and 
\emph{sequential models} \cite{TIMMERMANS:01}. The former class is based on full 
activity patterns (e.g., for one whole day), whereas the latter is based on an 
explicit modelling of the activity scheduling process. Simultaneous models 
comprise utility-maximisation models and mathematical programming models (e.g., 
Recker's household activity pattern problem -- HAPP, \cite{RECKER:95}). 
Sequential models are frequently implemented as so-called computational process 
models (CPM), acknowledging the belief that individuals do not arrive at optimal 
choices, but rather employ context-dependent heuristics.

As an example, we illustrate the seminal STARCHILD model, which was originally a 
simultaneous model based on the maximisation of individuals' utilities. Based on 
H\"agerstrand's notion and derivatives thereof, i.e., the central idea that an 
individual's travel behaviour is constrained by its space-time prism, Recker et 
al developed the STARCHILD research tool addressing activity-based modelling 
\cite{RECKER:86,RECKER:86b}. The model hinges on three interdependent 
consecutive steps: (i) the generation of household activities, (ii) constructing 
choice sets for these activities, as well as scheduling them, and (iii) 
constraining these choices within the boundaries of the space-time prism 
\cite{MCNALLY:00}. Note that the principal critique to the model's operation, 
was --- and today still is --- its need for an extensive amount of specifically 
tailored data that encompasses H\"agerstrand's concepts. Just as with the four 
step model, these data are arduous to come by. In short, most of the data are 
based on and transformed from e.g., conventional trip-based surveys, 
\emph{travel diaries} (e.g., the MOBEL (Belgium) and MOBIDRIVE (Germany) surveys 
of Cirillo and Axhausen \cite{CIRILLO:02}) and the like, although more recently 
passive GPS based information is collected 
\cite{AXHAUSEN:00,MCNALLY:00,MILLER:04}.

In the future, a complete integration of activity generation, scheduling, and 
route choice (DTA) is expected to take place, on the condition that suitable 
empirical data will become available. We must however be careful not to be too 
optimistic, as e.g., Axhausen states that depending on the `research-political' 
adoption of the activity-based approach, \emph{``both a virtuous circle of 
progress or a vicious circle of stagnation are a possibility for the future''} 
\cite{AXHAUSEN:00}. An even more harsh argumentation was voiced by Timmermans, 
who looked back at the development of the integration between land-use models 
and transportation planning \cite{TIMMERMANS:03}. In his overview, he identified 
three waves, i.e., (i) aggregate spatial interaction-based models, (ii) 
utility-maximising multinomial logit-based models, and (iii) activity-based 
detailed microsimulation models. His final conclusion states that, despite the 
advances in finer levels of spatial detail, the scientific field has not 
undergone any significant theoretical progress. And although there exists a 
pronounced need for better behavioural models, the critique remains that this 
implies a tremendous complexity, hence the insinuation that many of the 
approaches are in fact based on black-box models.

			\subsubsection{Towards elaborate agent-based simulations}
			\label{sec:TFM:TowardsElaborateAgentBasedSimulations}

One of the most notable critiques often expressed against classic trip-based 
approaches such as the four step model, is the fact that all eye for detail at 
the level of the individual traveller is lost in the trip aggregation process. 
Activity-based modelling schemes try to circumvent this disadvantage by starting 
from a fundamentally different basis, namely individual household activity 
patterns. To this end, it is necessary to retain all information regarding these 
individual households during the planning process.

As hinted at earlier, an upcoming technique that fits nicely in this concept, is 
the methodology of \emph{multi-agent simulations}. In such models, the 
individual households are represented as \emph{agents}; the models then allow 
these agents to make independent decisions about their actions. These actions 
span from long-term lifetime residential housing decisions, the mid-term 
planning of daily activities, to even short-term decisions about an individual's 
driving behaviour in traffic. The following description of such a simulation 
system is based on the work of the group of Nagel et al. 
\cite{NAGEL:02b,BALMER:04c,BALMER:04,BALMER:04b,NAGEL:04,HELBING:04}:

\begin{itemize}
	\item As a first step, a \emph{synthetic population} of agents is generated. 
	There is a close relation with the common land-use models, as these agents 
	come from populations that should be correctly seeded, i.e., they should 
	entail a correct demographic representation of the real world. Once the 
	synthetic population is available to the model, the next step is to 
	\emph{generate activity patterns} (i.e., activity chains), generate these 
	activities' \emph{locations}, and finally the \emph{scheduling} of the 
	activities, as described in the previous section. Finally, \emph{mode} and 
	\emph{route choice} form the bridge between the activity-based model and the 
	transportation layer. As a consequence, it is beneficial to deal with agents' 
	plans directly, rather than to rely on the information contained in OD tables 
	\cite{BALMER:04c}.
	\item The component that represents the \emph{physical propagation of agents} 
	throughout e.g., the road network, sits at the lowest level of the model. In 
	this case, the necessary ingredients constitute the physical propagation of 
	individual vehicles in the traffic streams. Popular models are \emph{traffic 
	cellular automata} and/or \emph{queueing models}, allowing a fast and 
	efficient simulation of individual agents in a network. Higher level models 
	such as e.g., pure fluid-dynamic models are inherently not suitable because 
	they operate on a more aggregated basis and consequently ignore the 
	individuality of each agent in the system. Note that this latter type of model 
	can be deemed appropriate, on the condition that they can incorporate the 
	tracking of individual particles by e.g., a smoothed particle hydrodynamics 
	method \cite{BALMER:04}.
	\item An important issue that revolves around the two previous aspects, is the 
	\emph{clear absence of a rigidly defined direction of causality}, i.e., when 
	exactly do people choose their travel mode, is it before the planning of 
	activities, or is it rather a result from the planning process~? This problem 
	can be dealt with in a broader context, wherein agents make certain plans 
	about their activities, and \emph{iteratively learn} by replanning and 
	rescheduling (either on a \emph{day-to-day} or \emph{within-day} basis). This 
	process of \emph{systematic relaxation} continues until e.g., a Wardrop 
	equilibrium (W1) is reached (see step (IV) in section \ref{sec:TFM:4SM} for 
	more details). However, note that the question of whether or not people in 
	reality strive to reach such an equilibrium, and whether or not such an 
	equilibrium is even reached, remains an open debate. At this stage in the 
	model, we are clearly dealing with aspects from evolutionary game theory, be 
	it cooperative or non-cooperative. In this context, the concept of within-day 
	replanning by agents is getting more attention, as it constitutes a necessary 
	prerequisite for intelligent transportation systems, i.e., when and how do 
	travellers react (e.g., \emph{en-route choice}) to changes (e.g., control 
	signals, incidents, \ldots) in their environment \cite{BALMER:04}~?
\end{itemize}

The above description of a multi-agent activity-based simulation system may seem 
straightforward, nevertheless, no complete practical implementation exists 
today. The model suite that comes the closest to reaching the previously stated 
goals, is the \emph{TRansportation ANalysis and SIMulation System} 
(TRANSIMS\footnote{\texttt{http://www.transims.net}}) project. This project is 
one part of the multi-track Travel Model Improvement Program of the U.S. 
Department of Transportation, the Environmental Protection Agency, and the 
Department of Energy in the context of the Intermodal Surface Transportation 
Efficiency Act and the Clean Air Act Amendments of 1990. Its original 
development started at Los Alamos National Laboratory, but a commercial 
implementation was provided by IBM Business Consulting.

\begin{figure}[!htb]
	\centering
	\includegraphics[width=\figurewidth]{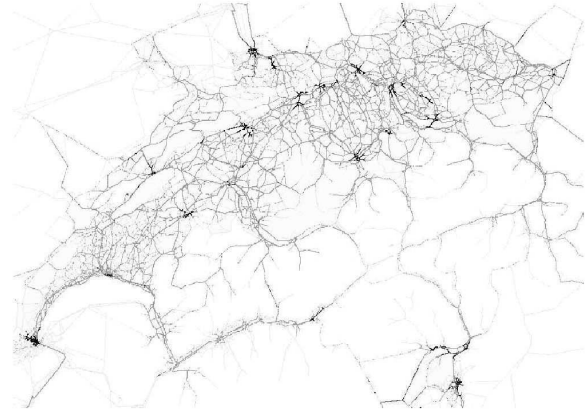}
	\caption{
		An example of a multi-agent simulation of the road network of Switzerland, 
		around 08:00 in the morning: each vehicle is indicated by a single grey 
		pixel, with low-speed vehicles coloured black. The image clearly reveals 
		more vehicular activity (and congestion) in the city centres than elsewhere 
		in the country (image reproduced after \cite{VOELLMY:01}).
	} 
	\label{fig:TFM:SwitzerlandNetwork}
\end{figure}

Since its original inception, TRANSIMS has been applied to a various range of 
case studies. One of the most notable examples, is the truly country-wide 
agent-based detailed microsimulation of travel behaviour in Switzerland (see 
also \figref{fig:TFM:SwitzerlandNetwork}) \cite{VOELLMY:01,RANEY:03}. A similar 
study encompassing the iteration and feedback between a simulation model and a 
route planner, was carried out for the region of Dallas 
\cite{NAGEL:98c,NAGEL:98e}. In the context of large-scale agent-based 
simulations, queueing models were employed as a TRANSIMS component by the work 
of Simon et al. for the city of Portland \cite{SIMON:99}, as well as the work of 
Gawron \cite{GAWRON:98,GAWRON:98b} and Cetin et al. \cite{CETIN:03}. Because of 
the computational complexity involved in dealing with the enormous amount of 
agents in real-world scenarios, a popular approach is the use of parallel 
computations, as described in the work of Nagel and Rickert \cite{NAGEL:01b}. 
Another example of this last type of simulations, is the work of Chopard and 
Dupuis who applied the methodology of large-scale simulations to the city of 
Geneva \cite{CHOPARD:95,CHOPARD:97,DUPUIS:98}.

As a final remark, we would like to draw attention to some more 
\emph{control-oriented aspects of multi-agent simulations}. In this respect, the 
transportation system is considered as a whole, whereby the agents are now the 
local controllers within the system (e.g., traffic lights, variable message 
signs, \ldots), instead of individual households as was previously assumed. 
Using a coordinated control approach, is it then possible to achieve a system 
optimum. An example of such a control system is the \emph{Advanced Multi-agent 
Information and Control for Integrated multi-class traffic networks} 
(AMICI\footnote{\texttt{http://www.amici.tudelft.nl}}) project from The 
Netherlands. As one of its goals, it strives to provide routing information to 
different classes of road users, as well as controlling them by means of 
computer simulated agents who operate locally but can be steered hierarchically.

		\subsection{Transportation economics}

Most of the work related to traffic flow theory has been considered by 
researchers with roots in engineering, physics, mathematics et cetera. With 
respect to transportation planning, the scene has shifted somewhat during the 
last couple of decades towards policy makers who test and implement certain 
strategies, based on e.g., the four step modelling approach. Around 1960 
however, another branch of scientists entered the field of transportation 
planning: economists developed standard models that viewed transportation as a 
market exchange between demand and supply.

Generally stated, the economics of transportation does not exclusively focus on 
traffic as a purely physical phenomenon, but also takes into account the fact 
that \emph{transportation incurs certain costs}, both to the individual as well 
as to the society as a whole \cite{LINDSEY:00}.

In the following sections, we describe the setting in which economists view 
transportation, after which we discuss the concept of road pricing.

			\subsubsection{The economical setting}
			\label{sec:TFM:TheEconomicalSetting}

In the context of economic theory, a transportation system can be seen as an 
interaction between \emph{demand} (profits) and \emph{supply} (costs). In a 
static setting, both demand and supply are frequently described by means of 
functions: they are expressed as the \emph{price} for a good associated with the 
\emph{quantity} of that good. In transportation economics, quantity is 
frequently identified as the \emph{number of trips made} (e.g., by the 
macroscopic characteristic of traffic flow) \cite{BECKMANN:55}. In the remainder 
of this section, we use the term \emph{travel demand} to denote the demand side, 
as opposed to the supply side which is composed of the transportation 
infrastructure (including changes due to incidents, \ldots). In a more broader 
context, travel demand is typically described as the amount of traffic volume 
that \emph{wants} to use a certain infrastructure (i.e., the supply): when 
demand thus exceeds the infrastructure's capacity under congested conditions 
(implying queueing), this supply effectively acts as a constraint to the present 
volume of traffic flow.

\begin{figure}[!htb]
	\centering
	\psfrag{ACF}[][]{\psfragstyle{ACF}}
	\psfrag{TDF}[][]{\psfragstyle{TDF}}
	\psfrag{congested}[][]{\psfragstyle{congested}}
	\psfrag{hypercongested}[][]{\psfragstyle{hypercongested}}
	\psfrag{C}[][]{\psfragstyle{$C$}}
	\psfrag{q}[][]{\psfragstyle{$q$}}
	\psfrag{qcap}[][]{\psfragstyle{$q_{\text{cap}}$}}
	\psfrag{0}[][]{\psfragstyle{\zero}}
	\includegraphics[width=\figurewidth]{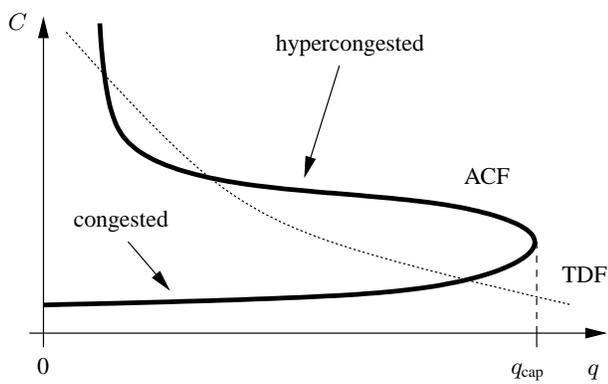}
	\caption{
		A graphical illustration of the economics of transportation operations: the 
		dotted curve represents the demand side, i.e., the travel demand function 
		(TDF), whereas the thick solid curve represents the supply side, i.e., the 
		average cost function (ACF). Both curves express the cost $C$ associated 
		with the number of trips made (e.g., level of traffic flow $q$). The latter 
		curve is said to have two states, namely the congested and the 
		hypercongested area (identified as the backward-bending part of the curve). 
		Points where both demand and supply curves intersect each other denote 
		equilibrium points: given both curves, the transportation system as assumed 
		to settle itself at their intersection(s), with a certain travel cost 
		associated with the equilibrium traffic demand.
	}
	\label{fig:TFM:TDFACF}
\end{figure}

According to the aforementioned conventions, a demand side function is expressed 
as a certain cost associated with a level of flow (i.e., number of trips). We 
call such a curve a \emph{travel demand function} (TDF), and it is typically 
decreasing with increasing flow; an example of such a function can be seen as 
the dotted curve in \figref{fig:TFM:TDFACF}. Note that, to be correct, the 
depicted curve actually represents a \emph{marginal travel demand}: it gives the 
additional profit that is received with the obtaining of one extra unit (the 
total amount of profit is just the area under (i.e., the integral of) the demand 
curve). Translated to a transportation system, this means that the benefits of a 
traveller tend to decrease with increasing travel demand (i.e., congestion).

In similar spirit, we can consider a supply side curve, i.e., price (costs) 
versus quantity (flow). One of the most used approaches for describing traffic 
flow operations at the supply side from an economical point of view, is the use 
of an \emph{average cost function} (ACF) \cite{VERHOEF:98} as proposed by A.A. 
Walters in 1961 \cite{WALTERS:61}. The theory was based on the description of 
traffic flow by means of fundamental diagrams. Consider for example 
Greenshields' simple linear relation between density $k$ and space-mean speed 
$\overline v_{s}$ \cite{GREENSHIELDS:35,MAERIVOET:05d}. The corresponding 
$\overline v_{s_{e}} (q)$ fundamental diagram, consists of a tilted parabola, 
lying on its side \cite{MAERIVOET:05d}. As the travel time $T$ is inversely 
proportional to the space-mean speed $\overline v_{s}$, Walters' idea was to 
assume certain \emph{costs} related to the travel demand. Some examples of these 
costs are those associated with \cite{DEBORGER:01,HAU:05,HAU:05b}:

\begin{itemize}
	\item (i) the construction of the transportation infrastructure,
	\item (ii) vehicle ownership and use,
	\item (iii) taxes,
	\item (iv) travel time, i.e., \emph{value of time} (VOT),
	\item and (v) environmental and social costs.
\end{itemize}

Based on these costs, and using the relation between travel time and travel 
demand, Walters derived a functional relationship for the economical cost $C$ 
associated with the travel demand $q$. This relationship (i.e., the ACF) denotes 
the supply side of transportation economics; an example is the thick solid curve 
in \figref{fig:TFM:TDFACF}.\\

\sidebar{
	Once both travel demand and average cost functions are known, they can be used 
	to determine the \emph{equilibrium points} that occur at their 
	intersection(s): given both curves, the transportation system is assumed to 
	settle itself at these equilibria, with a certain travel cost associated with 
	the equilibrium traffic demand. Note that because of the nature of the 
	analysis procedure, i.e., using static (stationary) curves, the resulting 
	travel costs are \emph{average} costs, hence the name of the average cost 
	function.
}\\

There are some distinct features noticeable in the relation described by the 
average cost function. For starters, the curve does not go through the origin, 
i.e., at low travel demands there is already a \emph{fixed cost} incurred. The 
depicted cost then typically increases with increasing travel demand, mainly due 
to the contribution of the value of time associated with the travel time. The 
most striking feature however, is the fact that the curve contains a 
\emph{backward-bending} upper branch \cite{ELSE:81}. This peculiar branch has an 
asymptotic behaviour, i.e.,

\begin{equation}
	\lim_{q \rightarrow q_{\text{cap}} \rightarrow \zero} C(q) = +\infty,
\end{equation}

where we have denoted the path taken by the limit, i.e., passing through the 
capacity flow $q_{\text{cap}}$ towards the upper branch, which in fact 
corresponds to an increase towards the jam density $k_{\text{jam}}$. Also note 
the presence of an inflection point (for concave $q_{e}(k)$ fundamental 
diagrams), which can be located analytically by differentiating the functional 
relation twice, and solving it with a right hand side equal to zero.\\

\sidebar{
	In contrast to the nomenclature adopted by the traffic engineering community 
	and in this dissertation, economists typically refer to the lower branch of 
	\figref{fig:TFM:TDFACF} as the \emph{congested} state, and to the upper branch 
	as the \emph{hypercongested} state. Their line of reasoning being the 
	conviction that in a certain sense, congestion also occurs when the speed 
	drops below the free-flow speed on the free-flow branch 
	\cite{VERHOEF:98,LINDSEY:00,SMALL:03}.
}\\

With respect to the relevance of this hypercongested state, there has been some 
debate in literature. Among most economists there seems to be a consensus, in 
the sense that the hypercongested branch is actually a transient phenomenon 
\cite{YANG:98,LINDSEY:00}. Walters thought of the branch as just a collection of 
inefficient equilibria, but it was shown by Verhoef that all equilibria obtained 
on the hypercongested branch are inherently unstable \cite{VERHOEF:98,SMALL:03}. 
Another argument, that discards the use of the branch, goes as explained by Yang 
and Huang \cite{YANG:98}: many traditional economical models of transportation 
assume a static (stationary) model of congestion, similarly as in classic static 
traffic assignment described in section \ref{sec:TFM:STA}. Under this premise, 
the relations as described by the fundamental diagrams, should be considered for 
complete links, and not only --- as is usual in traffic flow theory --- at local 
points in space and time. Therefore, a link may contain two different states: a 
free-flow state and a congested state. Hence, the average cost function should 
only describe the properties that are satisfied on the whole link, and as a 
result this excludes the global hypercongested regime.

Several ad hoc solutions exist for dealing with this problem, which is a 
consequence of using cost functions based on stationary equilibria: some of 
these solutions typically entail the use of vertical segments near the capacity 
flow in \figref{fig:TFM:TDFACF}, resulting in finite queueing delays on heavily 
congested links \cite{VERHOEF:98,YANG:98,SMALL:03,VERHOEF:05}. Another much used 
solution that ignores the backward-bending branch, is to directly specify the 
average cost function based on a link's observed capacity, instead of deriving 
it through the fundamental diagram of space-mean speed versus flow 
\cite{LINDSEY:00}. Note that in most cases, \emph{the economic interpretation of 
capacity is different from the one in traffic flow theory}: the capacity 
considered by economists has a lower value as opposed to its engineering 
counterpart. A similar example that specifies the relation between travel demand 
and travel time (e.g., VOT), is the BPR travel impedance function as described 
in section \ref{sec:TFM:STA}. Notwithstanding these proposed specific solutions, 
the mainstream tendency nowadays seems to imply the use of traffic flow models 
that explicitly describe the dynamics of congestion, either by using queueing 
models, or more elaborate models based on fluid dynamics or detailed simulations 
of individual vehicles \cite{YANG:98}.

			\subsubsection{Towards road pricing policies}
			\label{sec:TFM:TowardsRoadPricingPolicies}

In an economical treatment of transportation, road users in general only take 
into account their own \emph{private costs}, such as (ii) vehicle ownership and 
use, (iii) taxes, and (iv) costs related to the travel time. Note that because, 
as mentioned earlier, we are working with \emph{marginal cost functions}, the 
cost (i) related to the transportation infrastructure is not taken into account 
(as this is only a one-time initial cost).

To this end, we consider the average cost function from two different points of 
view: on the one hand, we have the \emph{private costs} borne by an individual 
traveller, and on the other hand, we have the \emph{external costs} that the 
traveller bears to the rest of society. In accordance with economic literature, 
we call the former associated cost function the \emph{marginal private cost 
function} (MPCF), and the latter the \emph{marginal social cost function} 
(MSCF). The extra costs to society brought on by individual travellers, are 
called \emph{negative externalities}.

\begin{figure}[!htb]
	\centering
	\psfrag{TDF}[][]{\psfragstyle{TDF}}
	\psfrag{MPCF}[][]{\psfragstyle{MPCF}}
	\psfrag{MSCF}[][]{\psfragstyle{MSCF}}
	\psfrag{C}[][]{\psfragstyle{$C$}}
	\psfrag{q}[][]{\psfragstyle{$q$}}
	\psfrag{que}[][]{\psfragstyle{$q_{\text{ue}}$}}
	\psfrag{qso}[][]{\psfragstyle{$q_{\text{so}}$}}
	\psfrag{optimal}[][]{\psfragstyle{optimal}}
	\psfrag{toll}[][]{\psfragstyle{toll}}
	\psfrag{welfarebenefit}[][]{\psfragstyle{welfare benefit}}
	\psfrag{0}[][]{\psfragstyle{\zero}}
	\includegraphics[width=\figurewidth]{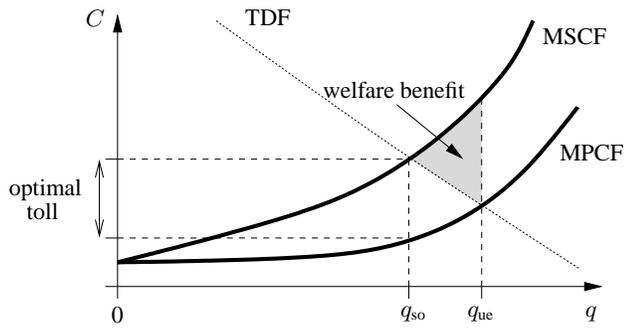}
	\caption{
		An economical equilibrium analysis based on a travel demand function (TDF) 
		represented by the dotted curve, and marginal private and social cost 
		functions (MPCF and MSCF) represented by the thick solid curves. The user 
		equilibrium is located at a demand of $q_{\text{ue}}$, whereas the system 
		optimum is located at a lower demand of $q_{\text{so}}$. The welfare benefit 
		(indicated as the grey triangular region) can be gained by levying a 
		congestion toll equal to the difference between the marginal social and 
		private cost function defined at the system optimum demand level 
		$q_{\text{so}}$.
	}
	\label{fig:TFM:MarginalCostPricing}
\end{figure}

In \figref{fig:TFM:MarginalCostPricing}, we have depicted the resulting 
equilibria that arise from the intersections of the travel demand function with 
both marginal private and social cost functions (note that we disregard the 
upper backward-bending branch of the average cost function as was shown in 
\figref{fig:TFM:TDFACF}). In an unmanaged society, i.e., where no measures are 
taken to change individual travellers' behaviour, the resulting equilibrium will 
be found at $q_{\text{ue}}$, which is in fact a \emph{user equilibrium} 
corresponding to a cost as dictated by the marginal private cost function (MPCF) 
\cite{ARNOTT:94}. As travellers handle selfishly, not considering the costs 
inflicted upon other travellers (e.g., more road users imply more congestion 
\emph{for everybody}), this pricing method is termed \emph{average cost 
pricing}. At this equilibrium, the unpaid external cost to society equals the 
difference between the MSCF and MPCF curves at a demand level of 
$q_{\text{ue}}$.

As early as in 1920, Arthur Cecil Pigou noted that road users do not take into 
account the costs they inflict upon other travellers. In order to rectify this 
situation, he proposed to levy governmental taxes on road use. Pigou actually 
discussed his idea in a broader economic setting, by making a distinction 
between the private and the social costs. Charging of a suitable governmental 
tax could change the balance so the negative externalities would be included, 
resulting in a new equilibrium \cite{PIGOU:20}. This process is called 
\emph{internalising the external costs}.

Some years later in 1924, Frank Hyneman Knight further explored Pigou's 
ideas\footnote{Note that Knight apparently was clued in his research by an error 
made on Pigou's behalf in his study of a two-route road network 
\cite{BOYCE:04b}. Even more intriguing, is the fact that this type of problem 
was already considered as far back as 1841, with the work of the German 
economist Johann Georg Kohl \cite{KOHL:41}.}: Knight fully acknowledged the fact 
that congestion justified the levying of taxes. In contrast to Pigou however, 
Knight raised some criticism in the sense that not governmental taxes were 
necessary, but instead private ownership of the roads would take care of tax 
levying and consequently resulting in a reduction of congestion 
\cite{KNIGHT:24}.

In 1927, Frank Plumpton Ramsey cast this methodology --- called \emph{marginal 
cost pricing} --- in the light of \emph{social welfare economics}. This latter 
type employs techniques from a branch that is called micro-economics, which is 
an economical treatment of society based on the behaviour of individual 
producers and consumers. Welfare economics embraces two important concepts:

\begin{description} 
	\item[Efficiency:] a measure for assessing how much benefit society gains from 
	a certain policy rule. It can be considered with the \emph{(strict) Pareto 
	criterion} (invented by Vilfredo Pareto), which states that efficiency 
	improves if a policy rule implies an increase of welfare for at least one 
	individual, but no other individual of society is worse off. Nicholas Kaldor 
	and John Hicks restated Pareto's criterion, but this time from the point of 
	view of those who gain and those who lose, respectively. Their 
	\emph{Kaldor-Hicks criterion} states that society gains welfare, but not 
	everybody receives personal gain, i.e., there will be winners and losers. The 
	crucial assumption on the Kaldor-Hicks criterion is that the winners could 
	fully compensate the losers, in theory; whether or not this happens at all, is 
	not the issue.
	\item[Equity:] if society benefits from a certain policy rule, then its 
	efficiency can be measured using e.g., the Pareto criterion as stated earlier. 
	However, nothing is said about the size of the benefit each individuals of 
	society receives. This is were the concept of equity enters the picture: it 
	refers to a fair distribution of the total benefits over all individuals in 
	society (note that in this case, there typically is a strong correlation with 
	the \emph{income distribution}).
\end{description} 

In this context, Ramsey thus stated a policy rule, implying a \emph{maximisation 
of the social welfare} \cite{RAMSEY:27}. In the field of transportation, this 
can be done by marginal cost pricing, also called \emph{road pricing}, 
\emph{congestion tolls}, \ldots The nature of the measure is that it signifies a 
demand-side strategy, with the goal of inducing a change in travellers' 
behaviour. Road pricing typically entails a shift from on-peak to off-peak 
periods, switching mode (e.g., from private to public transportation), car 
pooling, route change, \ldots Considering again 
\figref{fig:TFM:MarginalCostPricing}, we can see that if users were to consider 
the marginal social cost function, instead of only their marginal private cost 
function, this would shift the resulting equilibrium from $q_{\text{ue}}$ to 
$q_{\text{so}}$, which is a \emph{social optimum}. As said at the beginning of 
this section, travellers do not take into account the negative externalities 
they cause to the rest of society, and as such, they can be charged with an 
\emph{optimal toll} that is defined as the difference between both marginal 
social and private cost functions. Levying the correct congestion toll, would 
remove the original market failure, resulting in a net social welfare benefit 
that is visualised as the grey triangular region in 
\figref{fig:TFM:MarginalCostPricing}. Note that in an ideal world, congestion 
tolls exactly match the caused negative externalities. In practice however, this 
can not be accomplished, resulting in so-called \emph{second-best pricing} 
schemes. Practical real-life examples of this are tolling the beltway around a 
city upon entering it (e.g., London's recent congestion charge), using 
step-tolls, tolling at fixed time periods instead of based on traffic 
conditions, \ldots \cite{LINDSEY:01}

Reconsidering the work of Wardrop with respect to the criteria (W1) and (W2) 
highlighted in section \ref{sec:TFM:4SM}, Beckmann, McGuire, and Winsten found 
that the system optimum $q_{\text{so}}$ can be reached if the standard cost 
(i.e., journey time) is replaced with a \emph{generalised cost}, which is just 
the marginal social cost function as described earlier \cite{BECKMANN:55}. 
Consequently, the total travel time in the system can be minimised (from an 
engineering perspective), by levying a so-called efficiency toll, which 
corresponds to Ramsey's optimal toll.

One of the most notable extensions in the economic treatment of transportation 
and congestion tolls, is the seminal work of the late Nobel prize winner William 
Vickrey \cite{VICKREY:69}. As already stated in section 
\ref{sec:TFM:CritiqueOnTripBasedApproaches}, correct modelling of e.g., queue 
spill back, is of fundamental importance when assessing the effectiveness of 
road pricing schemes for example. Vickrey's \emph{bottleneck model} is one step 
in this direction: it is based on the behaviour of morning commuters, whereby 
the model takes into account the departure times of all travellers. As 
everybody's desire is to arrive at work at the same time, some will arrive 
earlier, others later. Besides the traditional travel time costs, travellers 
therefore also experience so-called \emph{schedule delay costs}. Levying 
suitable tolls that depend on the travellers' arrival times, allows to reach a 
system optimum \cite{ARNOTT:98,LINDSEY:00}. It is important to realise here that 
the levied toll should vary over time and space, in order to correspond to the 
governing traffic conditions.

To most people in society road pricing is a highly unpopular measure, as well as 
a controversial political issue, whereby public acceptance is everything 
\cite{MARCUCCI:98,HARSMAN:01,HAU:05,HAU:05b}. Alternatives to road pricing can 
include upgrading existing roads and/or public transportation services, strict 
control-oriented regulation by means of \emph{advanced traffic management 
systems} (ATMS), issuing elaborate parking systems, fuel taxes, et cetera 
\cite{ARNOTT:94b}. In spirit of second-best pricing schemes, it was a wise thing 
in the UK to connect London's congestion charge to the simultaneous improvement 
of public transportation \cite{SMALL:05}. Similarly, the cordon toll system in 
the city of Oslo, Norway, quickly found acceptance among the population 
\cite{HARSMAN:01}. Nevertheless, road pricing is considered an \emph{unfair} 
policy measure to most people: households (and firms) with higher incomes, can 
more easily afford to pay the charge, hence they will keep the luxury of 
travelling at their own discretion, whilst others might not be able to pay the 
required toll. As a consequence, an inconsiderate internalisation of the 
external costs, does not lead to an equitable Pareto optimal scenario. Despite 
this resistance, there does seem to be a general consensus among members of 
society that congestion caused by traffic induced by recreational activities, is 
not tolerated during peak periods; congestion tolling for these travellers is 
deemed appropriate.

Despite the advances in the methodology underlying road pricing, there is still 
one major gap that has yet to be filled in, i.e., the equity of the principle, 
or otherwise stated: where do the gained social welfare benefits (i.e., tax 
revenues) go in the redistribution~? As Small states, road pricing is more 
acceptable to the broad public, if it is presented as a complete financial 
package \cite{SMALL:92}. As welfare economists debate on whether or not the 
revenues should go back to the transportation sector or rather elsewhere, Small 
asserts that inclusion of the former is mandatory for achieving substantial 
support from both the political side and the investors. Complementary, in order 
to satisfy the global population, it is advisory to use the collected charges in 
order to diminish e.g., labour taxes, as they are perceived as being too high 
\cite{GOFFIN:04,HAU:05,HAU:05b}.\\

\sidebar{
	In the end, we should note that both economists and traffic engineers are 
	essentially talking about the same subject, although by approaching it from 
	different angles. In the field of economics, road pricing policies are 
	introduced based on average cost functions, allowing an optimisation of the 
	social welfare. This effectively corresponds to the engineers' idea of static 
	traffic assignment, based on a system optimum using travel impedance functions 
	(see e.g., section \ref{sec:TFM:STA}). The validity of using these average 
	cost functions (with or without their backward-bending parts as explained in 
	section \ref{sec:TFM:TheEconomicalSetting}), has instigated several debates in 
	road pricing literature, most notably between Else and Nash 
	\cite{ELSE:82,NASH:82}, Evans and Hills \cite{EVANS:92,EVANS:93,HILLS:93}, and 
	Ohta and Verhoef \cite{OHTA:01,OHTA:01b,VERHOEF:01b}.\\

	In continuation, the approach followed by Vickrey's bottleneck model provides 
	a nice, first alternative, using schedule delay costs (see sections 
	\ref{sec:TFM:CritiqueOnTripBasedApproaches} and 
	\ref{sec:TFM:TowardsRoadPricingPolicies}). Although Vickrey's idea introduces 
	a hitherto absent time dependence, it has the disadvantage that spatial 
	extents are neglected through the assumption of point-queues (see section 
	\ref{sec:TFM:CritiqueOnTripBasedApproaches}). Lo and Szeto have rigourously 
	shown that hypercongestion is essentially a spatial phenomenon, and that by 
	neglecting this facet, a road pricing policy might actually worsen traffic 
	conditions \cite{LO:05}. The correct way out of this problem, is by explicitly 
	taking the tempo-spatial characteristics of traffic flows into account. As an 
	engineering analogy, this can be accomplished by introducing dynamic traffic 
	assignment (see section \ref{sec:TFM:DTA}) which uses physical propagation 
	models to describe the buildup and dissolution of congestion (see also some of 
	the models presented in section \ref{sec:TFM:CritiqueOnTripBasedApproaches}, 
	e.g., the work of Lago and Daganzo \cite{LAGO:03b}).
}\\

	\section{Traffic flow propagation models}

In contrast to the previous section, which dealt with high level transportation 
planning models, the current section considers traffic flow models that 
explicitly describe the \emph{physical propagation} of traffic flows. In a 
sense, these models can be seen as being directly applicable for the physical 
description of \emph{traffic streams}. There exist several methods for 
discriminating between the families of models that describe traffic flow 
propagation, i.e., based on whether they operate in continuous or discrete time 
(or even event-based), whether they are purely deterministic or stochastic, or 
depending on the \emph{level of detail} (LOD) that is assumed, \ldots More 
detailed explanations can found in the overview of Hoogendoorn and Bovy 
\cite{HOOGENDOORN:01}. In this dissertation, we present an overview that is 
based on the latter method of discriminating between the level of detail. We 
believe that this classification most satisfactorily describes the discrepancies 
between the different traffic flow models. Thus, depending on the level of 
aggregation, we can classify the propagation models into the following four 
categories:

\begin{itemize}
	\item \textbf{macroscopic} (highest level of aggregation, lowest level of 
	detail, based on continuum mechanics, typically entailing fluid-dynamic 
	models),
	\item \textbf{mesoscopic} (high level of aggregation, low level of detail, 
	typically based on a gas-kinetic analogy in which driver behaviour is
	explicitly considered),
	\item \textbf{microscopic} (low level of aggregation, high level of detail, 
	typically based on models that describe the detailed interactions between 
	vehicles in a traffic stream),
	\item and \textbf{submicroscopic} (lowest level of aggregation, highest level 
	of detail, like microscopic models but extended with detailed descriptions of 
	a vehicles' inner workings).
\end{itemize}

\sidebar{
	Note that some people regard macroscopic models more from the angle of 
	\emph{network models}. In this case, the focus lies on \emph{performance 
	characteristics} such as total travel times (a measure for the quality of 
	service), number of trips, \ldots \cite{GARTNER:97} To this end, several 
	quantitative models were introduced, such as Zahavi's so-called $\alpha$ 
	relation between traffic flow, road density, and space-mean speed 
	\cite{ZAHAVI:72}, and Prigogine and Herman's \emph{two-fluid theory} of town 
	traffic \cite{HERMAN:79}.
}\\

		\subsection{Macroscopic traffic flow models}
		\label{sec:TFM:MacroscopicTrafficFlowModels}

In this section, we take a look at the class of traffic flow models that 
describe traffic streams at an aggregated level. We first introduce the concept 
behind the models (i.e., the continuum approach), after which we discuss the 
classic first-order LWR model. Because of its historical importance, we devote 
several sections to the model's analytical and numerical solutions, as well as 
to some extensions that have been proposed by researchers. We conclude our 
discussion of macroscopic models with a description of several higher-order 
models, and shed some light on the problems associated with both first-order and 
higher-order models.

			\subsubsection{The continuum approach}
			\label{sec:TFM:TheContinuumApproach}

Among the physics disciplines, exists the field of \emph{continuum mechanics} 
that is concerned with the behaviour of solids and fluids (both liquids and 
gasses). Considering the class of \emph{fluid dynamics}, it has spawned a rich 
variety of branches such as aerodynamics, hydrodynamics, hydraulics, \ldots

Underlying these scientific fields, is the \emph{continuity assumption} (also 
called the \emph{continuum hypothesis}) that they all have in common. In a 
nutshell, this assumption states that fluids are to be treated as continuous 
media (in contrast to e.g., molecular gasses, which consist of distinct 
particles). Stated more rigourously, the macroscopic spatial (i.e., the length) 
and temporal scales are considerably larger than the largest molecular 
corresponding scales \cite{CRAMER:04}. As a consequence, all quantities can be 
treated as being continuous (in the infinitesimal limit). The decision to use 
either a liquid-like or a gas-like treatment, is based on the \emph{Knudsen 
number} of the fluid: a low value (i.e., smaller than unity) indicates a 
fluid-dynamic treatment, whereas a high value is indicative of a more granular 
medium. In this section, we consider the former approach. In the latter case, we 
enter the realm of statistical mechanics that deals with e.g., kinetic gasses, 
requiring the use of the Boltzmann equation (as will be explained in section 
\ref{sec:TFM:MesoscopicTFM} on mesoscopic traffic flow models).

Historically, the fluid-dynamic approach found its roots in the seminal work of 
Claude Louis Navier (1822), Adh\'emar de Saint-Venant (1843), and George Gabriel 
Stokes (1845) \cite{GIRVAN:03}. This gave rise to what we know as the 
\emph{Navier-Stokes equations} (NSE), formulated as a set of \emph{non-linear 
partial differential equations} (PDEs). For our overview, the most relevant 
equation is actually the local \emph{conservation law}, stating that the net 
flux is accompanied by an increase or decrease of material (i.e., fluid). In 
general, we can discern two subtypes: \emph{compressible} or 
\emph{incompressible} fluids, and \emph{viscous} or \emph{inviscid} fluids. 
Incompressibility assumes a constant density (and a high \emph{Mach number}), 
whereas inviscid fluids have a zero viscosity (with a corresponding high 
\emph{Reynolds number}) and are typically represented as the \emph{Euler 
equations}.

Note that the NSE are still not fully understood. The fact of the matter is that 
the \emph{Clay Mathematics Institute} has devised a list of \emph{Millennium 
Problems}\footnote{\texttt{http://www.claymath.org/millennium}}, among which a 
deeper fundamental understanding of the NSE holds a reward of one million 
dollar. Because the original Navier-Stokes equations are too complex to solve, 
scientists developed solutions to specific subproblems, e.g., Euler's version of 
inviscid fluids. As an example, we give the \emph{Burgers equation}, as derived 
by Johannes Martinus Burgers \cite{BURGERS:48}, for a one-dimensional fluid in 
the form of a \emph{hyperbolic conservation law}:

\begin{equation}
\label{eq:TFM:BurgersPDE}
	\frac{\partial u}{\partial t} + u \frac{\partial u}{\partial x} = \nu \frac{\partial^{\two} u}{\partial x^{\two}},
\end{equation}

in which the $u \in \mathbb{R}$ typically represents the velocity, and $\nu$ is 
the viscosity coefficient. For inviscid fluids, $\nu = \zero$, such that 
equation \eqref{eq:TFM:BurgersPDE} corresponds to a \emph{first-order} partial 
differential equation. This type of \emph{hyperbolic PDE} is very important, as 
its solution can develop discontinuities, or more clearly stated, it can contain 
\emph{shock waves} which are of course directly relevant to the modelling of 
traffic flows. The inviscid Burgers PDE can be solved using the standard 
\emph{method of characteristics}, as will be explained in further detail in the 
next three sections.

			\subsubsection{The first-order LWR model}
			\label{sec:TFM:FirstOrderLWRModel}

Continuing the previous train of thought, we can consider traffic as an inviscid 
but compressible fluid. From this assumption, it follows that densities, mean 
speeds, and flows are defined as continuous variables, in each point in time and 
space, hence leading to the names of \emph{continuum models}, 
\emph{fluid-dynamic models}, or \emph{macroscopic models}.

The first aspect of such a fluid-dynamic description of traffic flow, consists 
of a \emph{scalar conservation law} (`scalar' because it is a first-order PDE). 
A typical derivation can be found in \cite{GARTNER:97} and \cite{JUNGEL:02}: the 
derivation is based on considering a road segment with a finite length on which 
no vehicles appear or disappear other than the ones that enter and exit it. 
After taking the infinitesimal limit (i.e., the continuum hypothesis), this will 
result in an equation that expresses the interplay between continuous densities 
and flows on a local scale. Another way of deriving the conservation law, is 
based on the use of a differentiable cumulative count function 
$\widetilde{N}(t,x)$ that represents the number of vehicles that have passed a 
certain location \cite{NEWELL:93}:

\begin{eqnarray}
	k(t,x) = - \partial \widetilde{N}(t,x) / \partial x & \text{and} & q(t,x) = \partial \widetilde{N}(t,x) / \partial t,\nonumber\\
	                                                    & \Downarrow &\nonumber\\
	\frac{\partial k(t,x)}{\partial t} = - \frac{\partial^{\two} \widetilde{N}(t,x)}{\partial t~\partial x} & \text{and} & \frac{\partial q(t,x)}{\partial x} = \frac{\partial^{\two} \widetilde{N}(t,x)}{\partial t~\partial x},\nonumber\\
	                                                    & \Downarrow &\nonumber\\
	\frac{\partial k(t,x)}{\partial t}                  & +          & \frac{\partial q(t,x)}{\partial x} = \zero,\label{eq:TFM:FirstOrderConservationLaw}
\end{eqnarray}

with the density $k$ and flow $q$ dynamically (i.e., time varying) defined over 
a single spatial dimension. Lighthill and Whitham were among the first to 
develop such a traffic flow model \cite{LIGHTHILL:55} (note that in the same 
year, Newell had constructed a theory of traffic flow at low densities 
\cite{NEWELL:96}). Crucial to their approach, was the so-called fundamental 
hypothesis, essentially stating that flow is a function of density, i.e., there 
exists a $q_{e}(k(t,x))$ equilibrium relationship \cite{MAERIVOET:05d}. 
Essentially to their theory, Lighthill and Whitham assumed that the fundamental 
hypothesis holds at all traffic densities, not just for light-density traffic 
but also for congested traffic conditions. Using this trick with the fundamental 
diagram, relates the two dependent variables in equation 
\eqref{eq:TFM:FirstOrderConservationLaw} to each other, thereby making it 
possible to solve the partial differential equation. 

One year later, in 1956, Richards independently derived the same fluid-dynamic 
model \cite{RICHARDS:56}, albeit in a slightly different form. The key 
difference, is that Richards focusses on the derivation of shock waves with 
respect to the density, whereas Lighthill and Whitham consider this more from 
the perspective of disruptions of the flow \cite{PIPES:64}. Another difference 
between both derivations, is that Richards fixed the equilibrium relation, 
whereas Lighthill and Whitham did not restrict themselves to an a priori 
definition; in Richards' paper, we can find the equation $V = a (b - D)$, with 
$V$ the space-mean speed, $D$ the density, and $a$ and $b$ fitting parameters 
\cite{RICHARDS:56}. Note that all three authors share the following same 
comment: because of the continuity assumption, the theory only holds for a large 
number of vehicles, hence the description of \emph{``long crowded roads''} in 
Lighthill and Whitham's original article.\\

\sidebar{
	Because of the nearly simultaneous and independent development of the theory, 
	the model has become known as the \emph{LWR model}, after the initials of its 
	inventors who receive the credit. In some texts, the model is also referred to 
	as the \emph{hydrodynamic model}, or the \emph{kinematic wave model} (KWM), 
	attributed to the fact that the model's solution is based on characteristics, 
	which are called kinematic waves (e.g., shock waves).
}\\

			\subsubsection{Analytical solutions of the LWR model}
			\label{sec:TFM:LWRAnalyticalSolutions}

Reconsidering equation \eqref{eq:TFM:FirstOrderConservationLaw}, taking into 
account the fundamental diagram, the conservation law is now expressed as:

\begin{equation}
\label{eq:TFM:FirstOrderPDEConservationLaw}
	k_{t} + q_{e}(k)_{x} = \zero,
\end{equation}

in which we introduced the standard differential calculus notation for PDEs. 
Recognising the fundamental relation of traffic flow theory, i.e., $q = 
k~\overline v_{s}$, equation \eqref{eq:TFM:FirstOrderPDEConservationLaw} 
\cite{MAERIVOET:05d} then becomes:

\begin{equation}
\label{eq:TFM:FirstOrderPDEBasedOnEquilibriumSpeed}
	k_{t} + (k~\overline v_{s_{e}}(k))_{x} = k_{t} + \left ( \overline v_{s_{e}}(k) + k \frac{d \overline v_{s_{e}}(k)}{dk} \right ) k_{x} = \zero.
\end{equation}

The above hyperbolic PDE, can be translated into the Burgers equation 
\eqref{eq:TFM:BurgersPDE}, using a suitable transformation to a dimensionless 
form as explained in the rigourous mathematical treatment provided by J\"ungel 
\cite{JUNGEL:02}. The conservation law 
\eqref{eq:TFM:FirstOrderPDEConservationLaw} can also be cast in a non-linear 
wave equation, using the chain rule for differentiation \cite{GARTNER:97}:

\begin{equation}
\label{eq:TFM:FirstOrderPDEIVP}
	k_{t} + \frac{dq_{e}(k)}{dk} k_{x} = \zero.
\end{equation}

Analytically solving the previous equation using the method of characteristics, 
results in shock waves that travel with speeds equal to:

\begin{equation}
	w = \frac{dq_{e}(k)}{dk},
\end{equation}

i.e., the tangent to the $q_{e}(k)$ fundamental diagram. This tangent 
corresponds to the speed $w$ of the backward propagating \emph{kinematic wave} 
\cite{MAERIVOET:05d}. As a consequence, solutions, being the characteristics, of 
equation \eqref{eq:TFM:FirstOrderPDEIVP} have the following form:

\begin{equation}
	k(t,x) = k(x - wt),
\end{equation}

with the observation that the density is constant along such a characteristic. 
Note that in order to obtain shock waves that are only decelerating, the used 
$q_{e}(k)$ fundamental diagram should be concave (a property that is often 
neglected) \cite{DELCASTILLO:95}:

\begin{equation}
\label{eq:TFM:FDConcavityCondition}
	\frac{d^{\two}q_{e}(k)}{dk^{\two}} \leq \zero.
\end{equation}

Starting from an initial condition, the problem of finding the solution to the 
conservation PDE, is also called an \emph{initial value problem} (IVP), whereby 
the solution describes how the density evolves with increasing time. The problem 
is called a \emph{generalised Riemann problem} (GRP) when we consider an 
infinitely long road with given constant initial densities up- and downstream of 
a discontinuity.

Because the method of characteristics can result in non-unique solutions, a 
trick is used to select the correct, i.e., physically relevant, one. Recall from 
equation \eqref{eq:TFM:BurgersPDE} that the right-hand side of the Burgers PDE 
contained a viscosity term $\nu$. The general principle that is adopted for 
selection of the correct solution, is based on the \emph{Oleinik entropy 
condition}, which regards the conservation law as a diffuse equation. In this 
context, the viscosity coefficient is multiplied with a small diffusion constant 
$\epsilon$. In the \emph{vanishing viscosity limit} $\epsilon \rightarrow 
\zero$, the method returns a unique, smooth, and physically relevant solution 
instead of infinitely many (weak) solutions \cite{LEVEQUE:92,NAGEL:05}. For more 
details with respect to the application of traffic flows, we refer to the 
excellent treatment given by J\"ungel \cite{JUNGEL:02}.\\

\sidebar{
	Ansorge, Bui et al., Velan and Florian later reinterpreted this entropy 
	condition, stating that it is equivalent to a \emph{driver's ride impulse}
	\cite{ANSORGE:90,BUI:92,VELAN:02}. Drivers going from free-flow to congested 
	traffic encounter a sharp shock wave, whereas drivers going in the reverse 
	direction essentially pass through all intermediate points on the fundamental 
	diagrams, i.e., the solution generates a fan of waves. It is for this latter 
	case that the `ride impulse' is relevant: it denotes the fact that stopped 
	drivers prefer to start riding again, resulting in a fan of waves.
}\\

A more intuitive explanation can also be given based on the \emph{anticipation} 
that drivers adopt when they approach a shock wave: their equilibrium speed 
$\overline v_{s_{e}}(k)$ is also a function of the change in density, e.g.:

\begin{equation}
	\overline v_{s_{e}}(k) \doteq \overline v_{s_{e}}(k) - \frac{\nu}{k} \frac{\partial k}{\partial x}.
\end{equation}

Substituting this new equilibrium relation in equation 
\eqref{eq:TFM:FirstOrderPDEBasedOnEquilibriumSpeed}, results in a right-hand 
side equal to $\nu \frac{\partial^{\two} k}{\partial x^{\two}}$. Applying the 
same methodology based on the vanishing viscosity limit of the entropy solution, 
results in the same unique solution. Because the shock waves are in fact 
mathematical discontinuities, and as such, infinitesimally small, they are 
typically `smeared out' by numerical schemes. In fact, this is just the 
equivalent of introducing an artificial viscosity (as explained earlier), which 
allows diffusion (i.e., the combined effect of dissipation and dispersion) of 
the shock waves. Note that this diffusion is a consequence of the numerical 
solution, and not necessarily corresponding to the real diffusion processes in a 
viscous fluid. This numerical smoothing helps to retain numerical stability of 
the final solution.

Whenever in the solution of the conservation equation, two of its 
characteristics intersect, the density takes on two different values (each one 
belonging to a single characteristic). As this mathematical quirk is physically 
impossible, the entropy solution states that both characteristics terminate and 
breed a \emph{shock wave}; as such, these shock waves form boundaries that 
discontinuously separate densities, flows, and space-mean speeds 
\cite{GARTNER:97}. The speed of such a shock wave is related to the following 
ratio \cite{PIPES:64}:

\begin{equation}
\label{eq:TFM:FirstOrderShockWaveSpeed}
	w_{\text{shock}} = \frac{\Delta q}{\Delta k},
\end{equation}

with $\Delta q = q_{u} - q_{d}$ and $\Delta k = k_{u} - k_{d}$ the relative 
difference in flows, respectively densities, up- and downstream of the shock 
wave.

Note that going from a low to a high density regime typically results in a shock 
wave, whereas the reverse transition is accompanied by an emanation of a 
\emph{fan of characteristics} (also called \emph{expansion}, 
\emph{acceleration}, or \emph{rarefaction waves}). In shock wave theory, the 
densities on either side of a shock are well defined (i.e., unique solutions 
exist); along the shock wave however, the density jumps discontinuously from one 
value to another. In this latter respect, equation 
\eqref{eq:TFM:FirstOrderShockWaveSpeed} is said to satisfy the 
\emph{Rankine-Hugoniot jump condition}.

The previous remarks with respect to the entropy condition, are closely related 
to the concavity of the $q_{e}(k)$ fundamental diagram, as defined by equation 
\eqref{eq:TFM:FDConcavityCondition}: for concave $q_{e}(k)$ fundamental 
diagrams, all shock waves are \emph{compression waves} going from lower to 
higher densities. However, for $q_{e}(k)$ fundamental diagrams that contain 
convex regions, application of the entropy condition can return the wrong 
solution \cite{LEVEQUE:01}. Although the mathematics of using these kinds of 
fundamental diagrams has been worked out, see for example the work of Li 
\cite{LI:03}, a unified physical interpretation is still lacking 
\cite{MAERIVOET:04e,NAGEL:05}: instead of only deceleration shock waves and 
acceleration fans, we now also have acceleration shock waves and deceleration 
fans. Finally, it is important to realise that for non-smooth $q_{e}(k)$ 
fundamental diagrams, the entropy condition is not applicable and no fans occur 
because the correct unique solution is automatically obtained \cite{VELAN:02}.

\begin{figure*}[!htb]
	\centering
	\psfrag{t}[][]{\psfragstyle{$t$}}
	\psfrag{tred}[][]{\psfragstyle{$t_{\text{red}}$}}
	\psfrag{tgreen}[][]{\psfragstyle{$t_{\text{green}}$}}
	\psfrag{x}[][]{\psfragstyle{$x$}}
	\psfrag{xlight}[][]{\psfragstyle{$x_{\text{light}}$}}
	\psfrag{k}[][]{\psfragstyle{$k$}}
	\psfrag{kc}[][]{\psfragstyle{$k_{c}$}}
	\psfrag{kj}[][]{\psfragstyle{$k_{j}$}}
	\psfrag{q}[][]{\psfragstyle{$q$}}
	\psfrag{qcap}[][]{\psfragstyle{$q_{\text{cap}}$}}
	\psfrag{1}[][]{\psfragstyle{\textbf{1}}}
	\psfrag{2}[][]{\psfragstyle{\textbf{2}}}
	\psfrag{3}[][]{\psfragstyle{\textbf{3}}}
	\psfrag{4}[][]{\psfragstyle{\textbf{4}}}
	\includegraphics[width=\textwidth]{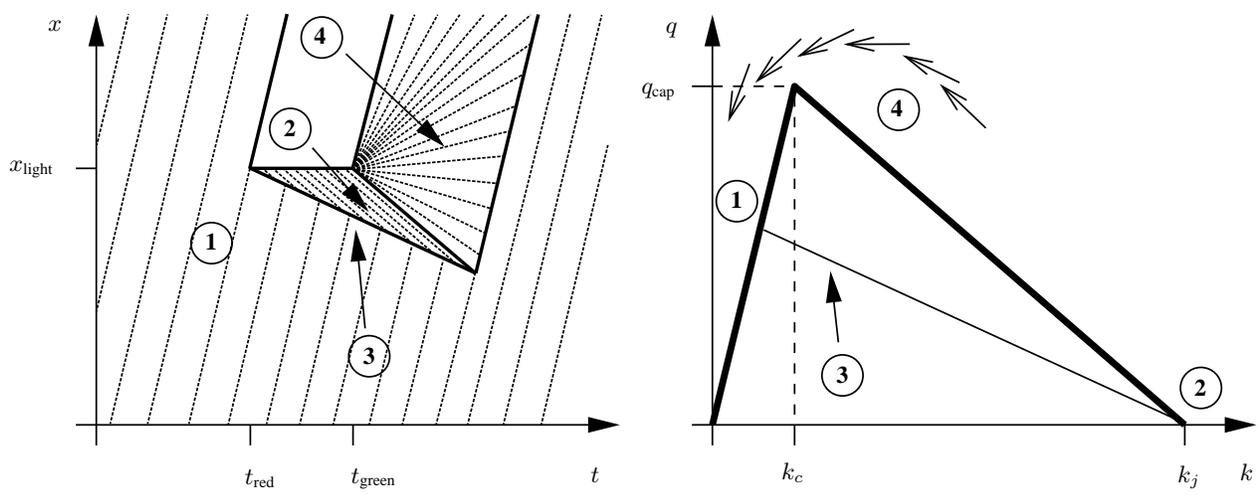}
	\caption{
		Example of an analytical solution based on the LWR first-order macroscopic 
		traffic flow model with a triangular $q_{e}(k)$ fundamental diagram. 
		\emph{Left:} a time-space diagram with a traffic light located at position 
		$x_{\text{light}}$. It is green, except during the period from 
		$t_{\text{red}}$ until $t_{\text{green}}$. The solution is visually sketched 
		by means of the characteristics that evolve during the tempo-spatial 
		evolution of the traffic flow. \emph{Right:} a triangular fundamental 
		diagram, with the initial conditions at state \textcircled{\scriptsize 
		\textbf 1}. When the traffic light is red, a queue develops in which the jam 
		density state at point \textcircled{\scriptsize \textbf 2} holds. Its 
		upstream boundary is demarcated by the shock wave \textcircled{\scriptsize 
		\textbf 3}. When the queue starts to dissipate, the solution of 
		characteristics generally becomes a fan of rarefaction waves 
		\textcircled{\scriptsize \textbf 4}.
	}
	\label{fig:TFM:LWRAnalyticalSolution}
\end{figure*}

In \figref{fig:TFM:LWRAnalyticalSolution}, we have depicted a classic example 
that is often used when illustrating the tempo-spatial evolution of a traffic 
flow at a traffic light (left part), based on the LWR first-order macroscopic 
traffic flow model with a triangular $q_{e}(k)$ fundamental diagram (right 
part). The application of the traffic flow model is visible in the time-space 
diagram to the left. A traffic light is located at position $x_{\text{light}}$; 
it is initially green, and at $t_{\text{red}}$ it turns red until 
$t_{\text{green}}$ when it switches back to green. The initial conditions at the 
road segment are located at point \textcircled{\scriptsize \textbf 1} on the 
fundamental diagram. Because all characteristics of the solution are tangential 
to the fundamental diagram, the results can be elegantly visualised when using a 
triangular diagram: except for the fan of rarefaction waves (we approximate the 
non-differentiable tip of the triangle with a smooth one, such that we can show 
the fan \textcircled{\scriptsize \textbf 4} for all didactical intents and 
purposes), only two kinematic wave speeds are possible. When the traffic light 
turns red, a queue of stopped vehicles develops. Inside this queue, the jam 
density state $k_{j}$ holds, corresponding to point \textcircled{\scriptsize 
\textbf 2} on the fundamental diagram. The upstream boundary of the queue is 
demarcated by the shock wave \textcircled{\scriptsize \textbf 3} that is formed 
by the intersections of the characteristics \textcircled{\scriptsize \textbf 1} 
and \textcircled{\scriptsize \textbf 2}. Downstream of the jam, there are no 
vehicles: because we are working with a triangular fundamental diagram, the 
characteristics are parallel to the vehicle trajectories (their speeds are equal 
to the slopes of points on the free-flow branch). The initial regime at state 
\textcircled{\scriptsize \textbf 1} and the `empty' regime downstream of the 
queue are separated from each other by a \emph{contact discontinuity} or 
\emph{slip}. When the traffic light turns green again, the queue starts to 
dissipate, whereby the solution of characteristics becomes a fan of rarefaction 
waves \textcircled{\scriptsize \textbf 4}, taking on all speeds between states 
\textcircled{\scriptsize \textbf 2} and \textcircled{\scriptsize \textbf 1} on 
the fundamental diagram. A final important aspect that can be seen from 
\figref{fig:TFM:LWRAnalyticalSolution}, is the fact that in the LWR model the 
outflow from a jam, i.e., going from a high to a low density regime, always 
proceeds via the capacity-flow regime at $q_{\text{cap}}$: so \emph{there is no 
capacity drop in the LWR model} because the outflow is always optimal.

To conclude our summary of analytical derivations, we point the reader to the 
significant work of Newell, who in 1993 cast the LWR theory in an elegant form. 
The key ideas he introduced were on the one hand the use of cumulative curves 
for deriving the conservation law, and on the other hand the use of a triangular 
$q_{e}(k)$ fundamental diagram \cite{NEWELL:93}. Due to Newell's work, traffic 
flow analysis in this respect gets very simplified, as it is now possible to 
give an exact \emph{graphical} solution to the LWR model for both free-flow and 
congested conditions \cite{NEWELL:93b}. To complete his theory, Newell also 
provided us with a means to include multi-destination flows, i.e., 
specifications of which off-ramp vehicles will use to exit the motorway 
\cite{NEWELL:93c}. Note that for the LWR model with a parabolic $q_{e}(k)$ 
fundamental diagram and piece-wise linear and piece-wise constant space and time 
boundaries, respectively, Wong and Wong recently devised an exact analytical 
solution scheme. Their method is based on the efficient tracking and fitting of 
generated and dispersed shock waves within a time-space diagram \cite{WONG:02}.

			\subsubsection{Numerical solutions of the LWR model}
			\label{sec:TFM:LWRNumericalSolutions}

Besides the previous \emph{analytic derivation} of a solution to the 
conservation law expressed as a PDE, it is also possible to treat the problem 
\emph{numerically}. By trying to find a numerical solution to the PDEs, we enter 
the field of \emph{computational fluid dynamics} (CFD). In a typical setup, the 
`fluid domain' is first discretised into adjacent cells (called a 
one-dimensional \emph{mesh}) of size $\Delta X$ (note that all cells need not to 
be equal in size), after which an \emph{iterative scheme} is used to update the 
cells' states (e.g., the density $k$ in each fluid cell) at discrete time steps 
$m~\Delta T$ with $m \in \mathbb{N}_{\zero}$. Typically, this entails 
\emph{finite difference schemes} (or in a broader context, \emph{finite element 
methods} or \emph{finite volume methods}), which replace the continuous partial 
derivative with a \emph{difference operator}, thereby transforming the 
conservation equation into a \emph{finite difference equation} (FDE). Examples 
of these difference operators are the \emph{forward difference operator} $\Delta 
f(x) = f(x + \one) - f(x)$ and the \emph{backward difference operator} $\nabla 
f(x) = f(x) - f(x - \one)$, which is not to be confused with the gradient vector 
of $f(x)$. Examples of finite difference schemes are the \emph{central scheme}, 
the \emph{Lax-Friedrichs scheme}, the \emph{downwind scheme}, the \emph{upwind 
scheme}, the \emph{MacCormack scheme}, the \emph{Lax-Wendroff scheme}, the 
\emph{Steger-Warming Flux Splitter scheme}, the \emph{Rieman-based Harten-van 
Leer-Lax and Einfeldt scheme}, \ldots For a more complete overview of these 
schemes, we refer the reader to the work of Helbing and Treiber 
\cite{HELBING:99b}, J\"ungel \cite{JUNGEL:02}, and Ngoduy et al. 
\cite{NGODUY:03}. A practical software implementation of a \emph{moving-mesh 
finite-volume solver} for the previously mentioned hyperbolic PDEs, can be found 
in van Dam's TraFlowPACK software \cite{VANDAM:02}. LeVeque also developed a 
numerical solver, called CLAWPACK, that is designed to compute numerical 
solutions to hyperbolic partial differential equations using a wave propagation 
approach \cite{LEVEQUE:03}. A central precaution for all these schemes, is the 
so-called \emph{Courant-Friedrichs-Lewy (CFL) condition} which guarantees 
numerical stability of the algorithms; for traffic flows, it has the physical 
interpretation that no vehicles are allowed to `skip' cells between consecutive 
time steps (i.e., all physical information that has an influence on the system's 
behaviour should be included):

\begin{equation}
	\Delta T \leq \frac{\Delta X}{\overline v_{\text{ff}}}.
\end{equation}

Just over a decade ago, Daganzo constructed a numerical scheme based on finite 
difference equations. It is known as the \emph{cell transmission model} (CTM), 
which solves the LWR model using a trapezoidal $q_{e}(k)$ fundamental diagram 
\cite{DAGANZO:94}. At the heart of his model lies a discretisation of the road 
into finite cells of width $\Delta X$, each containing a certain number of 
vehicles (i.e., an average cell density). When time advances, these vehicles are 
transmitted from upstream to downstream cells, taking into account the capacity 
constraints imposed by the downstream cells. The CTM converges to the LWR model 
in the limit when $\Delta X \rightarrow \zero$. In 1995, Daganzo extended the 
model to include network traffic, i.e., two-way merges and diverges, thereby 
allowing for the correct modelling of dynamic queue spill backs 
\cite{DAGANZO:95c}. He also cast the model in the context of \emph{Godunov FDE 
methods}\footnote{Sergei K. Godunov's numerical solution of PDEs is considered 
as a breakthrough in computational fluid dynamics: it provides a unique solution 
based on a stable Riemann problem \cite{GODUNOV:59,LEBACQUE:96,DAGANZO:99c}.}, 
allowing for arbitrary $q_{e}(k)$ fundamental diagrams. The exchange of vehicles 
between neighbouring cells is then governed by so-called \emph{sending} and 
\emph{receiving functions} \cite{DAGANZO:95b}. Lebacque derived a similar 
numerical scheme that performed the same functions as Daganzo's CTM, at 
approximately the same time (the debate on whomever was first is still not 
resolved) \cite{LEBACQUE:93,LEBACQUE:96}. In his derivation, he employed the 
terms \emph{demand} and \emph{supply functions} to denote the exchange of 
vehicles between cells. He also provided the means to handle general (i.e., 
multi-way) merges and diverges. Both the original cell transmission model and an 
implementation of the Godunov scheme for the LWR model with an arbitrary 
$q_{e}(k)$ fundamental diagram, were provided by Daganzo et al. in the form of a 
software package called NETCELL \cite{CAYFORD:97}. Note that, as mentioned 
earlier, numerical methods tend to smear out the shock waves; this diffusion is 
therefore a consequence of the solution methodology and not of the LWR model 
itself \cite{LOGGHE:03}.

Daganzo also developed another methodology for numerically solving the LWR 
equations, based on a \emph{variational formulation}. Rather than extending the 
existing concept of a conservation equation coupled with a vanishing viscosity 
limit, he derived a solution based on the principles behind cumulative curves. 
The initial value problem becomes well-posed, and the methodology is able to 
handle complex boundary conditions. In short, the problem is transformed into 
finding shortest paths in a network of arcs that comprise the kinematic waves; 
as a surplus, the method is computationally more efficient than traditional 
solutions based on conservation laws \cite{DAGANZO:03b,DAGANZO:05}.

Traditional cell-based numerical methods are fairly computationally intensive, 
because they have to discretise the road entirely (even in regions where there 
is no variation in density), resulting in a solution that is composed of linear 
shock waves and continuous fans (i.e., the rarefaction waves). In order to 
derive a solution that is computationally more efficient, Henn proposed to 
replace the continuous fans of rarefaction waves with a discrete set of angular 
sectors (i.e., the density now varies with discrete steps). The efficiency now 
stems from the fact that, instead of a whole array of cells, only list 
structures need to be maintained \cite{HENN:03}.

Only recently, a combination of Daganzo's CTM with a triangular $q_{e}(k)$ 
fundamental diagram and Newell's cumulative curves was constructed by Yperman et 
al, resulting in the \emph{link transmission model} (LTM). Because whole links 
can be treated at once, the LTM's computational efficiency is much higher than 
that of classic numerical solution schemes for the LWR model, whilst retaining 
the same accuracy \cite{YPERMAN}.

With respect to the applicability of the LWR model to real-life traffic flows, 
we refer the reader to two studies: the first was done by Lin and Ahanotu in the 
course of the \emph{California Partners for Advanced Transit and Highways} 
(PATH) programme (formerly known as the \emph{Program on Advanced Technology for 
the Highway}). In their work, they performed a validation for the CTM with 
respect to the formation and dissipation of queues, concluding that the most 
important first-order characteristics (correlations in measurements of free-flow 
traffic at successive detector stations, and the speed of the backward 
propagating wave under congested conditions) perform reasonable well when 
comparing them to field data \cite{LIN:95}.

A second, more thorough and critical study was done by Nagel and Nelson. In it, 
they scrutinise the LWR model, both with concave $q_{e}(k)$ fundamental diagrams 
and those with convex regions. Their main conclusion states that it remains 
difficult to judge the model's capabilities on a fair basis, largely due to the 
fact that there do not exist many real-world data sets which also contain a 
geometrical description of the local infrastructural road layout. This latter 
ingredient is a requirement for assessing whether or not an observed traffic 
breakdown is either spontaneously induced or due to the presence of an active 
bottleneck (because the LWR model constitutes a strictly deterministic model) 
\cite{NAGEL:05}.

			\subsubsection{Flavours of the LWR model}
			\label{sec:TFM:FlavoursOfTheLWRModel}

Considering this elegant first-order traffic flow model, its main advantages are 
that it is simple, and in a sense reproduces the most important features of 
traffic flows (i.e., shock waves and rarefaction waves). However, because of its 
restriction to a first-order partial differential equation, certain other 
effects, such as stop-and-go traffic waves, capacity drop and hysteresis, 
traffic flow instabilities, finite acceleration capabilities, \ldots can not be 
represented \cite{LIGHTHILL:55}. In many cases, these `deficiencies' can be 
tackled by switching to higher-order models, as will be elaborated upon in 
section \ref{sec:TFM:HigherOrderModels}. Interestingly, in their original paper, 
Lighthill and Whitham recognised the fact that drivers tend to anticipate on 
downstream conditions, changing their speed gradually when crossing shock waves. 
This in fact necessitates a diffusion term in the conservation equation that 
captures a density gradient.

Instead of using a higher-order model, traffic flow engineers can also resort to 
more sophisticated approaches, such as extensions of the first-order model. To 
conclude this section, let us give a concise overview of some of the model 
flavours that have been proposed as straightforward extensions to the seminal 
LWR model.

An interesting set of extensions launched, was created by Daganzo, dealing with 
two classes of vehicles, of which one class can use all lanes of a motorway, 
whereas the other class is restricted to a right-hand subset of these lanes. 
When the capacity of the latter vehicles in regular lanes is exceeded, a queue 
will develop in those lanes, but the former vehicles will still be able to use 
the other lanes; this is called a \emph{2-pipe} regime. Similarly, if the 
capacity of the yet freely flowing vehicles is exceeded, all lanes enter a 
queued state, which is called a \emph{1-pipe} regime. In short, interactions 
between vehicles in this and the following models are nearly always considered 
from a user equilibrium perspective \cite{DAGANZO:97d}. Daganzo et al. applied 
the theory to a case where there is a set of special lanes on which only 
priority vehicles can drive. The theory was also suited to describe congestion 
on a motorway diverge, such that the motorway itself can still be in the 
free-flow regime \cite{DAGANZO:97e}. For the special case of queue spill back at 
a motorway's off-ramp, Newell also provided a graphical solution that is based 
on the use of cumulative curves \cite{NEWELL:99}.

Continuing the previous train of thought, Daganzo provided a logical extension: 
he again considered different lanes, but now introduced two different types of 
drivers: aggressive ones (called \emph{rabbits}) and timid ones (called 
\emph{slugs}). Daganzo himself states that a correct traffic flow theory should 
involve both human psychology and lane-changing aspects, leading him to such a 
\emph{behavioural} description \cite{DAGANZO:02}. The theory was also used to 
explain the phenomenon of a capacity funnel \cite{MAERIVOET:05d}: according to 
the theory, once a capacity drop occurs, the recovery to the capacity-flow 
regime can not occur spontaneously, thereby requiring an exogeneous mechanism. 
Daganzo provides an explanation, called the \emph{pumping phenomenon}: drivers 
temporarily accept shorter time headways downstream of an on-ramp, leading to a 
`pumped state' of high-density and high-speed traffic, or in other words, a 
capacity-flow regime \cite{DAGANZO:02b}. Chung and Cassidy later provided a 
validation of the theory, by applying it to describe merge bottlenecks on 
multi-lane motorways in Toronto (Canada) and Berkeley (California). In their 
study, they introduced the concept of \emph{semi-congestion}, denoting a regime 
in which on vehicle class enters a state with a reduced mean speed, whereas the 
other vehicle class can still travel unimpeded. Their findings indicated an 
agreement between both shock wave speeds empirically observed and predicted by 
the model \cite{CHUNG:02}.

An interesting case to which the LWR theory can be applied, is the problem of 
\emph{moving bottlenecks} as stated by Gazis and Herman \cite{GAZIS:92}. 
Examples of such bottlenecks are slower trucks on the right shoulder lanes, 
which can impede upstream traffic. Newell was among the first to try to give a 
satisfactory consistent treatment of this type of bottlenecks. The trick he used 
was to translate the problem into a moving coordinate system that is travelling 
at the bottleneck's velocity. This resulted in a description of a stationary 
bottleneck, after which the classic LWR theory can be applied \cite{NEWELL:98}. 
Although the theory is sound, there exist some serious drawbacks, mainly due to 
its underlying assumptions. For example, the moving bottlenecks are assumed to 
be long convoys, and other drivers' behaviours are not affected by the 
bottlenecks' speeds; even more serious is the fact that the theory is not valid 
for very light traffic conditions, and that several strange effects are 
predicted by the theory (e.g., a bottleneck with increasing speed can result in 
a lower upstream capacity). To this end, Mu\~noz and Daganzo applied the 
previously mentioned behavioural model with rabbits and slugs to the problem of 
a moving bottleneck. Their theory performs satisfactorily and agrees well with 
empirically observed motorway features. However, because of the fact that it 
relies on the LWR model, it is not entirely valid for bottlenecks that travel at 
high speeds under light traffic conditions. In this latter case, they state that 
driver differences are much more important than the dynamics dictated by the 
kinematic model \cite{MUNOZ:02b}.

Another theory that deals with the problem of moving bottlenecks, is the one 
proposed by Daganzo and Laval: they treat moving bottlenecks as a sequence of 
consecutive \emph{fixed obstructions} that have the same capacity restraining 
effects. Despite the fact that the previous theory of Mu\~noz and Daganzo has a 
good performance, it does not easily lend itself to discretisation schemes that 
allow numerical solutions. In contrast to this, the hybrid theory (fixed 
obstructions coupled with the LWR model dynamics) of Daganzo and Laval holds 
high promise as they have shown that it can be discretised in a numerically 
stable fashion \cite{DAGANZO:03d}. As a continuation of this work, Lavel 
furthermore investigated the power of these fixed obstructions, allowing him to 
capture lane-changes as random events modelled by moving bottlenecks in a LWR 
1-pipe regime. It is suggested that (disruptive) lane changes form the main 
cause for instabilities in a traffic stream. This leads the `Berkeley school' to 
the statement that incorporating lane-change capabilities into multi-lane 
macroscopic models seems a prerequisite for observing effects such as capacity 
drops, kinematic waves of fast vehicles, \ldots \cite{LAVAL:04} In this respect, 
Jin provides a theory that explicitly takes into account to effects of lane 
changes \cite{JIN:05d}. The starting point in this model, is the presence of 
certain road areas in which traffic streams mix. The underlying assumption here 
is that all lane changes lead to the same traffic conditions in each lane: the 
crucial element in the model is that vehicles performing lane changes are 
temporarily counted twice in the density total. This new `effective density', is 
then used to transform the $q_{e}(k)$ fundamental diagram, leading to a reversed 
lambda shape. However, because the current version of the theory employs a small 
artificial constant to introduce the lane changes, we question its practical 
applicability when it comes to calibration and validation.

To conclude this overview of the first-order models, we highlight two other 
successful attempts at increasing the capabilities of the classic LWR model. A 
first important extension was made by Logghe, who derived a \emph{multi-class 
formulation}\footnote{At approximately the same time, Chanut and Buisson 
constructed a first-order model that incorporates vehicles with different 
lengths and free-flow speeds \cite{CHANUT:03}. Their model can be considered as 
a trimmed-down version of Logghe's multi-class formulation.} that allows for the 
correct modelling of heterogeneous traffic streams (e.g., preserving the FIFO 
property for interacting classes). Classes are distinguished by their maximum 
speed, vehicle length, and capacity (all intended for a triangular $q_{e}(k)$ 
fundamental diagram). A central ingredient to his theory, is the interactions 
between different user classes that reside on a road: in this respect, each 
class acts selfishly, with slower vehicles taking on the role of moving 
bottlenecks. Besides being able to construct analytical and graphical solutions, 
Logghe also provided a stable numerical scheme, as well as a complete network 
version with road inhomogeneities, and two-way merges and diverges 
\cite{LOGGHE:03}. A second extension was made by Jin; it actually deals with a 
whole plethora of extensions, in particular for inhomogeneous links (e.g., lane 
drops), merge and diverge zones, and mixed-type vehicles (i.e., having different 
$q_{e}(k)$ fundamental diagram characteristics). All these models are then 
combined in a description for a \emph{multi-commodity kinematic wave model for 
network traffic}, whereby the commodities signify vehicles taking different 
paths \cite{JIN:03}.\\

\sidebar{
	A finally important aspect that is mainly related to lane changes, is the 
	\emph{anisotropy property} of a traffic stream. This property basically states 
	that drivers are not influenced by the presence upstream vehicles. In a sense, 
	most models describing the acceleration behaviour of a vehicle, only take into 
	account the state of the vehicle directly in front. For most macroscopic 
	traffic flow models, this anisotropy constitutes a necessary ingredient. 
	However, in his original paper, Richards very subtly points out that the fact 
	of whether or not drivers only react to the conditions ahead, remains an open 
	question \cite{RICHARDS:56}. In contrast to this, Newell states that a driver 
	is only influenced by downstream conditions, leading to a natural 
	cause-and-effect relation, making the problem mathematically well-posed 
	\cite{NEWELL:93b}. Recently, Zhang stated that the anisotropy property is 
	likely to be violated in multi-lane traffic flows. His explanation is closely 
	tied to the concavity character of a $q_{e}(k)$ fundamental diagram 
	(non-concave regions can lead to characteristics that travel faster than the 
	space-mean speed of the traffic stream). He also provides an intuitive 
	reasoning based on Daganzo's rabbits and slugs, whereby \emph{tailgating} 
	vehicles induce slower downstream vehicles to `make way'. Note that for 
	single-lane traffic flows, the anisotropy property is expected to hold because 
	of the FIFO property (vehicles can not pass each other), although there are 
	exceptions in the case of some higher-order macroscopic traffic flow models 
	\cite{ZHANG:03}.
}\\

			\subsubsection{Higher-order models}
			\label{sec:TFM:HigherOrderModels}

The development of higher-order macroscopic models came as a response to the 
apparent shortcomings of the first-order LWR model. Harold Payne was among the 
first in 1971 to develop such a higher-order model \cite{PAYNE:71}. In those 
days, ramp metering\footnote{Ramp metering is an ATMS whereby a traffic light is 
placed at an on-ramp, such that traffic enters the highway from the on-ramp by 
drops. We refer the reader to the work of Bellemans \cite{BELLEMANS:03} and 
Hegyi \cite{HEGYI:04} for an overview and some recent advancements.} control 
strategies were basically an all-empirical occasion. Payne recognised the 
necessity to include dynamic models in the control of on-ramps; the celebrated 
LWR model however, was found to perform unsatisfactorily with respect to the 
modelling of real-life traffic flows. One of these shortcomings, was the model's 
inability to generate stop-and-go waves. Zhang later traced this to be a 
consequence of the model's persistent reliance on a single equilibrium curve 
(i.e., the fundamental diagrams) \cite{ZHANG:03c}. In the LWR model, drivers are 
assumed to adapt their vehicle speed \emph{instantaneously} according to the 
fundamental diagram when crossing a shock wave, a phenomenon termed the 
\emph{no-memory effect} (i.e., they encounter infinite accelerations and 
decelerations \cite{ZHANG:98}). One option that leads to a solution of the 
previously mentioned problems, is to introduce different fundamental diagrams 
for vehicles driving under different traffic conditions; this avenue was 
explored by Newell \cite{NEWELL:63} and Zhang \cite{ZHANG:99b}. Another, more 
popular type of solution was proposed by Payne (as well as by Whitham some years 
later \cite{WHITHAM:74}): they suggested to add an equation to the LWR 
conservation law \eqref{eq:TFM:FirstOrderPDEBasedOnEquilibriumSpeed} and its 
fundamental diagram\footnote{Note that Lighthill and Whitham originally proposed 
to extend the conservation law in their model with relaxation and diffusion 
terms, but the idea did not receive much thought at the time 
\cite{LIGHTHILL:55}.}. This new \emph{dynamic speed equation} was derived from 
the classical car-following theories of Gazis et al. \cite{GAZIS:59} (see also 
section \ref{sec:TFM:ClassicCarFollowingAndLaneChangingModels} for more 
details). An important aspect is this derivation, is the fact that the 
car-following model includes a \emph{reaction time}, resulting in a 
\emph{momentum equation} that relates the space-mean speed of a vehicle stream 
to its density. As a result, vehicles no longer instantaneously change their 
speed when crossing a shock wave. Payne's second-order macroscopic traffic flow 
model is now described by the following pair of PDEs, i.e., a conservation law 
and a momentum equation:

\begin{equation}
\label{eq:TFM:PayneWhithamConservationLaw}
	k_{t} + (k~\overline v_{s})_{x} = 0,
\end{equation}

\begin{equation}
\label{eq:TFM:PayneWhithamMomentumEquation}
	d \overline v_{s} = \overline v_{s_{t}} +
		\underbrace{\overline v_{s} \overline v_{s_{x}}}_{\text{convection}} =
		\underbrace{\frac{\overline v_{s_{e}}(k) - \overline v_{s}}{\tau}}_{\text{relaxation}} -
		\underbrace{\frac{c^{\two}(k)}{k} k_{x}}_{\text{anticipation}},
\end{equation}

with $\overline v_{s_{t}}$ and $\overline v_{s_{x}}$ denoting the partial 
derivatives of the space-mean speed with respect to time and space, 
respectively, $\overline v_{s_{e}}$ the traditional fundamental diagram, and 
$\tau$ the \emph{reaction time}. The function $c(k)$ corresponds to the 
model-dependent \emph{sound speed of traffic} (i.e., the typical speed of a 
backward propagating kinematic shock wave); examples of $c(k)$ are 
\cite{ZHANG:03d}:

\begin{eqnarray}
	-\sqrt{-\frac{\one}{\two \tau} \frac{d \overline v_{s_{e}}(k)}{dk}} & \quad & \text{(Payne)}\\
	-\sqrt{\frac{v}{\tau}}                                              & \quad & \text{(Whitham)}\label{eq:TFM:SoundSpeedWhitham}\\
	k~\frac{d \overline v_{s_{e}}(k)}{dk}                               & \quad & \text{(Zhang)}
\end{eqnarray}

with $v$ being a parameter in equation \eqref{eq:TFM:SoundSpeedWhitham}.

In equation \eqref{eq:TFM:PayneWhithamMomentumEquation}, the left hand side 
corresponds to the derivative of the speed, i.e., the acceleration of vehicles. 
As can be seen from the formulation, Payne identified three different aspects 
for the momentum equation: a \emph{convection} term describing how the 
space-mean speed changes due to the arrival and departure of vehicles at the 
time-space location $(t,x)$, a \emph{relaxation} term describing how vehicles 
adapt their speed to the conditions dictated by the fundamental diagram, but 
with respect to a certain reaction time (as opposed to the instantaneous 
adaption in the LWR model), and finally an \emph{anticipation} term describing 
how vehicles react to downstream traffic conditions.

In continuation of the above derivation, many other higher-order models have 
been based on the Payne-Whitham (PW) second-order traffic flow model. An example 
is the work of Phillips, who changed the reaction time $\tau$ in the relaxation 
term of equation \eqref{eq:TFM:PayneWhithamMomentumEquation} from a constant to 
a value that is dependent on the current density \cite{PHILLIPS:79}. Another 
example is due to K\"uhne, who artificially introduced a viscosity term into 
equation \eqref{eq:TFM:PayneWhithamMomentumEquation}, in order to smooth the 
shock waves \cite{KUHNE:84}. The physical role that viscosity plays in a 
vehicular traffic stream is however not entirely understood: according to Zhang, 
the viscosity reflects the resistance of drivers against sharp changes in speeds 
\cite{ZHANG:03c}. For a rather complete overview of extensions to the PW model, 
we refer the reader to the work of Helbing \cite{HELBING:01}.

			\subsubsection{Critiques on higher-order models}
			\label{sec:TFM:CritiquesOnHigherOrderModels}

Higher-order models have been successfully applied in various computer 
simulations of traffic flows, e.g., the original FREFLO implementation by Payne 
\cite{PAYNE:78}, the work of Kwon and Machalopoulos who developed KRONOS which 
is an FDE solver for a motorway corridor \cite{KWON:95}, the METANET model of 
Messmer and Papageorgiou \cite{MESSMER:90}, \ldots Despite their success, it was 
Daganzo who in 1995 published their final requiem, which stood out as an 
obituary for all higher-order models \cite{DAGANZO:95}. From a theoretical 
perspective, there were some serious \emph{physical flaws} that littered these 
second-order models. Most notably was the fact that there exist two families of 
characteristics (called \emph{Mach lines}) in the Payne-Whitham type models. On 
the one hand, there are characteristics that imply a diffusion-like behaviour, 
which under certain circumstances can lead to negative speeds at the end of a 
queue, i.e., vehicles travelling backwards. On the other hand, there are 
characteristics that have the property of travelling faster than the propagation 
of traffic flow. This latter gas-like behaviour means that vehicles can get 
influenced by upstream conditions (because information is sent along the 
characteristics), which is a clear violation of the anisotropy property for 
single-lane traffic as explained in the previous section. From a physical point 
of view, the relaxation term in equation 
\eqref{eq:TFM:PayneWhithamMomentumEquation} may even introduce a `suction 
process' because slower vehicles can get sucked along by leading faster ones 
\cite{HEGYI:01}.

Several years after these critiques, Papageorgiou responded directly to the 
comments stated in Daganzo's article \cite{PAPAGEORGIOU:98}. In his response, 
Papageorgiou put a lot of emphasis at the incapabilities of first-order traffic 
flow models for use in a traffic control strategy (e.g., ramp metering). He very 
briefly reacts to the anisotropy violation, by mentioning that in multi-lane 
traffic flows, the space-mean speeds of the different lanes are not all the 
same, leading to characteristics that are \emph{allowed} to travel faster than 
the space-mean speed of all lanes combined. With respect to negative speeds (and 
hence, negative flows), he proposes to simply include an a posteriori check that 
allows to set the negative flows equal to zero. One year later, in 1999, 
Heidemann reconsidered these higher-order models, but this time from the 
perspective of \emph{mathematical flaws}. His main argument was the fact that 
the models led to an internal inconsistency, because they ignored some aspects 
related to the conservation law \cite{HEIDEMANN:99}. However, after careful 
scrutiny, Zhang later refuted Heidemann's claims: the inconsistencies that 
plague the models are a result of the insistence on the universality of a 
conservation law and the imposing of arbitrary solutions. As a consequence, the 
Payne-Whitham type of models are mathematically consistent theories, although 
they may suffer from the aforementioned physical quirks \cite{ZHANG:03d}.\\

\sidebar{
	Note that the dynamic speed equation 
	\eqref{eq:TFM:PayneWhithamMomentumEquation}, can also be cast in another form 
	that is more closely related to a gas-kinetic analogy. With this in mind, we 
	can rewrite the momentum equation as follows \cite{HOOGENDOORN:01}:

	\begin{eqnarray}
		d \overline v_{s} & = & \overline v_{s_{t}} + \underbrace{\overline v_{s} \overline v_{s_{x}}}_{\text{transport}}\nonumber\\
		                  & = & \underbrace{\frac{\overline v_{s_{e}}(k) - \overline v_{s}}{\tau}}_{\text{relaxation}} -
			\underbrace{\frac{P_{x}}{k}}_{\text{pressure}} +
			\underbrace{\frac{\nu}{k}~\overline v_{s_{xx}}}_{\text{viscosity}},\label{eq:TFM:PayneWhithamGasKineticMomentumEquation}
	\end{eqnarray}

	with now $P$ denoting the \emph{traffic pressure} and $\nu$ the 
	\emph{kinematic traffic viscosity} (as introduced by K\"uhne \cite{KUHNE:84}). 
	The convection term has been relabelled a \emph{transport term}, describing 
	the propagation of the speed profile with the speed of the vehicles. The 
	\emph{pressure term} reflects the change in space-mean speed due to arriving 
	vehicles having different speeds, and the \emph{viscosity term} reflects 
	changes due to the `friction' between different successive vehicles. The 
	classic Payne model is obtained if we set $P = k c^{\two}(k)$ and $\nu = 
	\zero$.
}\\

In contrast to Papageorgiou's response which did not provide a definite answer, 
Aw and Rascle carefully examined the reason why the PW model exhibited the 
strange phenomena indicated in Daganzo's requiem \cite{AW:00}. The root cause of 
this behaviour can be traced back to the spatial derivative $P_{x}$ of the 
pressure term in equation \eqref{eq:TFM:PayneWhithamGasKineticMomentumEquation}. 
Their solution suggests to abandon the transport and relaxation terms, and 
replace the spatial derivative of the pressure $P$ (which is a function of the 
density $k$) with a \emph{convective (Lagrangian) derivative}, i.e., $D / Dt = 
\partial_{t} + (\overline  v_{s} \cdot \nabla) = \partial_{t} + \overline v_{s} 
\partial_{x}$, with $(\overline  v_{s} \cdot \nabla)$ called the \emph{advective 
derivative term} \cite{PRICE:05}:

\begin{equation}
\label{eq:TFM:AwRascleMomentumEquation}
	(\overline v_{s} + P(k))_{t} + \overline v_{s}~(\overline v_{s} + P(k))_{x} = 0.
\end{equation}

This new formulation allows to remedy all Daganzo's stated problems 
\cite{AW:00}. Because of the somewhat limited character of their derivation of 
equation \eqref{eq:TFM:AwRascleMomentumEquation}, Rascle add a relaxation term 
to the equation's right-hand side, and developed a numerically stable 
discretisation scheme, as well as showing convergence to the classic LWR model 
when the relaxation tends towards zero \cite{RASCLE:02}.

To end our overview of higher-order models, we illustrate two other types. The 
first model is actually a \emph{third-order model} created by Helbing. It is 
based on the two PDEs of the Payne-Whitham type models, but is extended with a 
third equation that describes the change in the \emph{variance of the speed}, 
denoted by $\Theta$ \cite{HELBING:96}. Helbing derived his equations using a 
gas-kinetic analogy, resulting in the following Navier-Stokes-like equation (it 
is typically encountered in the pressure term for $P$):

\begin{eqnarray}
\label{eq:TFM:HelbingsSpeedVariancePDE}
	\Theta_{t} + \overline v_{s} \Theta_{x} & = & \underbrace{\frac{\two (\Theta_{e}(k) - \Theta)}{\tau}}_{\text{relaxation}} +\nonumber\\
	                                        &   & \underbrace{\frac{\two}{k}~\overline v_{s_{x}} (\nu~\overline v_{s_{x}} - P)}_{\text{pressure}} +\nonumber\\
	                                        &   & \underbrace{\frac{\nu}{k}~\overline v_{s_{xx}} + \frac{\kappa}{k}~\Theta_{xx}}_{\text{viscosity}},
\end{eqnarray}

with now the equilibrium relation $\overline v_{s_{e}}(k,\Theta)$ of equation 
\eqref{eq:TFM:PayneWhithamGasKineticMomentumEquation} also depending on the 
speed variance $\Theta$. In addition to the viscosity $\nu$, the dynamic speed 
variance equation \eqref{eq:TFM:HelbingsSpeedVariancePDE} also contains an 
equilibrium relation $\Theta_{e}(k,\overline v_{s})$ for the variance of the 
speed, and $\kappa$ which is a \emph{kinetic coefficient} that is related to the 
reaction time $\tau$, the density $k$, and the speed variance $\Theta$. For $\nu 
= \kappa = \zero$, Helbing's model reduces to an inviscid Euler type model as 
explained in section \ref{sec:TFM:TheContinuumApproach} \cite{KLAR:96}. Whereas 
in the LWR model there is only one family of characteristics, and in the PW 
model there are two families, the Helbing model generates three different 
families of characteristics; this implies that small perturbations in the 
traffic flow propagate both with the traffic flow itself, as well as in upstream 
and downstream direction relative to this flow \cite{HOOGENDOORN:01}.

The second model we illustrate, is the \emph{non-equilibrium model} of Zhang. 
Because of the relaxation terms in the Payne-Whitham equations, drivers 
initially tend to `overshoot' the equilibrium speed as dictated by the 
$\overline v_{s_{e}}(k)$ fundamental diagram. It takes a certain amount of time 
for them to adapt to their speed to the new traffic conditions (i.e., a change 
in density is accompanied by a \emph{smooth} change in space-mean speed), after 
which they converge on the diagram. This latter aspect gives rise to the 
empirically observed scatter in the ($k$,$q$) phase space, leading Zhang to the 
terminology of `non-equilibrium models' because of the deviation from the 
one-dimensional equilibrium fundamental diagram \cite{ZHANG:98}.

In his model, Zhang considers equilibrium traffic to be a state in which $d 
\overline v_{s} / dt = \partial k / \partial x = \zero$. In similar spirit of 
Payne's theory, Zhang constructs his model using an equilibrium relation between 
density and space-mean speed (i.e., the fundamental diagram), a reaction time 
that allows relaxation, and an anticipation term that adjusts the space-mean 
speed to downstream traffic conditions. This results in a macroscopic model that 
contains equation \eqref{eq:TFM:PayneWhithamConservationLaw} as the conservation 
law, as well as the following momentum equation:

\begin{eqnarray}
	d \overline v_{s} & = & \overline v_{s_{t}} + \overline v_{s} \overline v_{s_{x}}\nonumber\\
	                  & = & \frac{\overline v_{s_{e}}(k) - \overline v_{s}}{\tau} -
	                        k \left ( \frac{d \overline v_{s_{e}}(k)}{dk} \right )^{\two} k_{x},
\end{eqnarray}

with the last anticipation term showing the dependence on the spatial change of 
the density. Zhang also complements the theory with a finite difference scheme 
that allows to solve the equations in a numerically stable fashion, based on an 
extension of the Godunov scheme that satisfies the entropy condition referred to 
in section \ref{sec:TFM:LWRAnalyticalSolutions} \cite{ZHANG:01c}.

Just as with the improved PW model of Aw and Rascle, this model alleviates 
Daganzo's stated problem of wrong-way travel, even though there are also two 
families of characteristics, travelling slower, respectively faster, than the 
space-mean speed of traffic. An important fact here is that for the slower 
characteristics, the associated shock waves and fans correspond perfectly to 
those of the first-order LWR model. However, the shock waves and fans associated 
with the faster family of characteristics can still violate the anisotropy 
property of traffic (although they decay exponentially), but in the end, Zhang 
questions its universal validity, stating that traffic might occasionally 
violate this principle due to the heterogeneity of a traffic stream 
\cite{ZHANG:00,ZHANG:03}. The violation of anisotropy, i.e., drivers get 
influenced by upstream traffic, is sometimes referred to as gas-like behaviour, 
because in contrast to fluid-dynamics, gas particles are not anisotropic. In an 
attempt to remove this faulty behaviour, Zhang developed yet another 
non-equilibrium model that removed the gas-like behaviour, thereby respecting 
the anisotropy property. Moreover, both families of characteristics in his model 
satisfy the condition of travelling at a lower speed than the space-mean speed 
of the traffic stream, but still keeping the one-to-one correspondence between 
the slower characteristics and those of the first-order LWR model. At present, 
it is however unclear if this new model can generate stop-and-go waves, although 
there are indications that it can because of the non-equilibrium transitions 
that can occur \cite{ZHANG:02}.\\

\sidebar{
	Despite the significant progress that has been made on the front of 
	higher-order macroscopic traffic flow models, the Berkeley school firmly holds 
	its faith in first-order models and their extensions. Its main reason is 
	because of the numerical solution schemes that are well developed and 
	understood. This is not the case for higher-order models, as these contain 
	other characteristics that complicate the finite difference schemes (because 
	information is now carried both up- and downstream, and because their 
	numerical schemes initially were flawed \cite{ZHANG:01c,LAVAL:04}). Related to 
	this critique, is the fact that in contrast to the first-order model, no 
	analytical solutions exist for the higher-order models.\\

	Another reason for sticking with first-order models, is because the school 
	believes that first-order characteristics are sufficient for the description 
	of traffic flows \cite{CASSIDY:01}. Using a triangular $q_{e}(k)$ fundamental 
	diagram that captures the most important traffic flow characteristics (i.e., 
	the free-flow speed $\overline v_{\text{ff}}$, the capacity $q_{\text{cap}}$, 
	the jam density $k_{j}$, and the backward kinematic wave speed $w$), results 
	in a further elegance of the models.
}\\

		\subsection{Mesoscopic traffic flow models}
		\label{sec:TFM:MesoscopicTFM}

The previous section dealt with macroscopic models that described traffic 
streams at an aggregated level, derived from a fluid-dynamic analogy. This 
section describes how traffic can be modelled at this aggregate level, but with 
special consideration for microscopic characteristics (e.g., driver behaviour). 
Because of the ambiguity that surrounds mesoscopic models, we first elucidate 
what is meant by the term mesoscopic (i.e., it is something between a 
macroscopic and a microscopic approach). In the sections thereafter, we zoom in 
on a derivation of mesoscopic models based on a gas-kinetic analogy. For an 
outstanding overview of gas-kinetic models, we refer the reader to the work of 
Tamp\`ere \cite{TAMPERE:04}.

			\subsubsection{The different meanings of `mesoscopic'}

Considering the amount of literature that has been generated during the last few 
decades, it seems to us that there exists no unanimous consensus as to what 
exactly constitutes mesoscopic traffic flow models. In general, there are three 
popular approaches when it comes to mesoscopic models \cite{HOOGENDOORN:01}:

\begin{itemize}
	\item \textbf{Cluster models}

	When considering vehicles driving on a road, a popular method is to group 
	nearby vehicles together with respect to one of their traffic flow 
	characteristics, e.g., their space-mean speed. Instead of having to perform 
	detailed updates of all vehicles' speeds and positions, the cluster approach 
	allows to treat these vehicles as a set of groups (called \emph{clusters}, 
	\emph{cells}, \emph{packets}, or \emph{macroparticles}); these groups are then 
	propagated downstream without the need for explicit lane-changing manoeuvres 
	(leading to the coalescing and splitting of colliding and separating groups).

	Examples of this kind of models, are the \emph{CONtinuous TRaffic Assignment 
	Model} (CONTRAM) of Leonard et al. \cite{LEONARD:89}, the work of Ben-Akiva et 
	al, called \emph{Dynamic network assignment for the Management of Information 
	to Travellers} (DynaMIT), which is based on a cell transmission model with a 
	cell of a link containing a set of vehicles with identical speeds 
	\cite{BENAKIVA:96}, the \emph{Mesoscopic Traffic Simulator} (MesoTS) of Yang, 
	which allows fast predictions of future traffic states \cite{YANG:97}, \ldots

	\item \textbf{Headway distribution models}

	This rather unknown and somewhat outdated class of models, places the emphasis 
	on the probability distributions of time headways of successive vehicles (this 
	aggregation makes them mesoscopic). Two popular examples are Buckley's 
	semi-Poisson model \cite{BUCKLEY:68}, and Branston's generalised queueing 
	model \cite{BRANSTON:76}. As clarified in the summary of Hoogendoorn and Bovy, 
	the original versions of these headway distribution models assume homogeneous 
	traffic flows and they are inadequate at describing the proper dynamics of 
	traffic flows \cite{HOOGENDOORN:01}.

	\item \textbf{Gas-kinetic models}

	The third and most important characterisation of mesoscopic models comes from 
	a gas-kinetic analogy. Because macroscopic models aim towards obeying the 
	fundamental diagram (either instantaneously as in the first-order LWR model or 
	through a relaxation process as in higher-order models), the focus there lies 
	on the generation and dissipation of shock and rarefaction waves. As a 
	consequence, more complex and non-linear dynamics can not be reproduced. To 
	remedy this, gas-kinetic models implicitly bridge the gap between microscopic 
	driver behaviour and the aggregated macroscopic modelling approach.
\end{itemize}

In the next sections, we will first give an overview of the original gas-kinetic 
model as derived by Prigogine and Herman, after which we discuss some of the 
recent successful modifications that allow for heterogeneity in the traffic 
stream (i.e., multi-class modelling), as well as the inclusion of more specific 
driver behavioural characteristics.

			\subsubsection{Mesoscopic models considered from a gas-kinetic perspective}

As opposed to the macroscopic traffic flows models that are derived from a 
conservation equation based on the Navier-Stokes equations, mesoscopic models 
can be derived from a gas-kinetic analogy. From individual driving behaviour 
(termed a microscopic approach), a macroscopic model is derived. The earliest 
model can be traced back to the work of the late Nobel laureate Ilya Prigogine, 
in cooperation with Frank Andrews and Robert Herman 
\cite{PRIGOGINE:60,PRIGOGINE:71}. A central component in their theory, is the 
concept of a phase-space density (PSD):

\begin{equation}
\label{eq:TFM:PhaseSpaceDensity}
	\widetilde{k}(t,x,\overline v_{s}) = k(t,x)~P(t,x,\overline v_{s}),
\end{equation}

in which $P(t,x,\overline v_{s})$ denotes the \emph{distribution} of the 
vehicles with space-mean speed $\overline v_{s}$ at location $x$ and time $t$; 
the concept of this distribution originated in Boltzmann's theory of gas 
dynamics. For the above density function, a kinetic conservation equation can be 
derived, looking as follows \cite{HELBING:01}:

\begin{equation}
\label{eq:TFM:PrigogineKineticConservationEquation}
	\frac{d\widetilde{k}}{dt} = \widetilde{k}_{t} + \overline v_{s}~\widetilde{k}_{x} = \left ( \widetilde{k}_{t} \right )_{\text{acc}} + \left ( \widetilde{k}_{t} \right )_{\text{int}},
\end{equation}

with now the two terms on the right hand side denoting the \emph{accelerations} 
of and \emph{interactions} between the vehicles; they are also called 
\emph{gains} and \emph{losses}, \emph{relaxation} and \emph{slowing down}, or 
\emph{continuous} and \emph{discrete} terms, respectively 
\cite{KLAR:96,TAMPERE:04}. Equation 
\eqref{eq:TFM:PrigogineKineticConservationEquation} is called the 
\emph{Prigogine-Herman kinetic model} and it actually describes three processes:

\begin{enumerate}
	\item Similar to the macroscopic conservation equation, the term $\overline 
	v_{s}~\widetilde{k}_{x}$ describes a {convective behaviour}: arriving and 
	departing vehicles cause a change in the distribution $\widetilde{k}$ of 
	vehicle speeds.

	\item The first term on the equation's right hand side, $\left ( 
	\widetilde{k}_{t} \right )_{\text{acc}}$, describes the acceleration behaviour 
	of vehicles, which is assumed to be a \emph{density-dependent relaxation} 
	process of the speed distribution $P$ of equation 
	\eqref{eq:TFM:PhaseSpaceDensity} towards some pre-specified target speed
	distribution $P_{\zero}$ (typically based on an equilibrium speed).

	\item The second term on the equation's right hand side, $\left ( 
	\widetilde{k}_{t} \right )_{\text{int}}$, describes the interactions between 
	vehicles, as fast vehicles either must slow down or overtake slower ones 
	(hence implying inherently multi-lane traffic). The decision on when to either 
	slow down or to overtake (which is assumed to be a \emph{discrete event}), is 
	governed by the probabilities $(\one - \pi)$ and $\pi$, respectively. The 
	interaction term is called a \emph{collision equation}, in analogy with the 
	physics of the Boltzmann equation (where the collision term describes the 
	scattering of the gas molecules). Because there occur joint distributions in 
	this latter equation (i.e., the probability of a faster vehicle encountering a 
	slower one), a common assumption called \emph{vehicular chaos} is used, which 
	states that vehicles' speeds are uncorrelated, hence allowing to split the 
	joint distribution.
\end{enumerate}

More than a decade later, Paveri-Fontana criticised the assumption of vehicular 
chaos in the interaction term \cite{PAVERIFONTANA:75}. He subsequently proposed 
an improved gas-kinetic model, in which he extended the phase-space density of 
equation \eqref{eq:TFM:PhaseSpaceDensity} with a dependence on the desired speed 
$v_{\text{des}}$, i.e., $\widetilde{k}(t,x,\overline v_{s},v_{\text{des}})$; in 
Prigogine's original derivation, this desired speed was incorrectly considered 
to be a property of the road, instead of being a driver-related property 
\cite{HELBING:01}.\\

\sidebar{
	An interesting property of the gas-kinetic modelling approach instigated by 
	the seminal work of Prigogine, is that for densities beyond a certain critical 
	density, Nelson and Sopasakis found that the model solutions split into two 
	distinct families. The current hypothesis surrounding this phenomenon states 
	that this corresponds to the widely observed data scatter in the empirically 
	obtained ($k$,$q$) fundamental diagrams \cite{NELSON:98}.
}\\

			\subsubsection{Improvements to the mesoscopic modelling approach}
			\label{sec:TFM:ImprovementsToTheMesoscopicModellingApproach}

Significant contributions to the gas-kinetic mesoscopic model have been 
sporadic; after the work of Paveri-Fontana, Nelson was among the first to tackle 
the computational complexity associated with the four-dimensional phase-space 
density $\widetilde{k}(t,x,\overline v_{s},v_{\text{des}})$ \cite{NELSON:95}. 
In his derivation, he reformulated the relaxation and interaction terms both as 
discrete events, based on a bimodal distribution of the vehicles' speeds (i.e., 
corresponding to stopped and moving vehicles). In contrast to the classic model 
which uses a relaxation process in the acceleration term, Nelson furthermore 
based his derivation on a microscopic behavioural model 
\cite{HELBING:01,HOOGENDOORN:01,TAMPERE:04}.

Building on the work of Nelson (which is, as he describes, just a first initial 
step towards constructing a suitable kinetic model), Wegener and Klar derived a 
kinetic model in similar spirit, based on a microscopic description of 
individual driver behaviour with respect to accelerations and lane changes. 
Attractive to their work, is the fact that they also pay attention to the 
numerical solutions of their model, with respect to the description of 
homogeneous traffic flows \cite{WEGENER:96}.

Noting that the correct reproducing of traffic flow behaviour at moderate to 
higher densities still troubled the existing mesoscopic models, Helbing et al. 
explored an interesting avenue. Not only did they capture the effect that 
vehicles require a certain finite space (leading to an Enskog- instead of a 
Boltzmann-equation), they also generalised the interaction term of equation 
\eqref{eq:TFM:PrigogineKineticConservationEquation}. This last method allowed 
them to dismiss the traditional assumption of vehicular chaos, i.e., they were 
now able to treat correlations between vehicles' speeds (which have a 
substantial influence at higher densities). The trick to obtain this behaviour, 
was to assume that drivers react to the downstream traffic conditions. This 
leads to the inclusion of non-local interaction (braking) term, and hence their 
model is referred to as the \emph{non-local gas-kinetic traffic flow model} 
\cite{HELBING:98b,HELBING:02b}. Interestingly, this non-locality can generate 
effects that are similar to the ones induced by viscosity/diffusion terms in 
macroscopic traffic flows models, causing smooth behaviour at density jumps 
\cite{HELBING:01}. The power of their model is also demonstrated as it is able 
to reproduce all traffic regimes encountered in Kerner's three-phase traffic 
theory \cite{KERNER:04}.\\

\sidebar{
	Central to some of the recently proposed models, is the step process that 
	transforms one model class into another. Starting from microscopic driver 
	behavioural principles (e.g., accelerating, braking, \ldots), a mesoscopic 
	model is deduced. This mesoscopic model can then be translated into an 
	equivalent macroscopic one by applying the \emph{method of moments}. This 
	allows to obtain PDEs that describe the dynamic evolution of the density $k$, 
	space-mean speed $\overline v_{s}$, and its variance $\Theta$ (an exception to 
	this methodology is the previously mentioned model of Wegener and Klar that 
	obtains dynamic solutions directly \cite{WEGENER:96}). As an example, Helbing 
	et al. also devised a numerical scheme for their previously discussed model. 
	It was implemented in a simulation package called MASTER \cite{HELBING:01b}.
}\\

Important progress was made by the work of Hoogendoorn et al., who extended the 
gas-kinetic traffic flow models with \emph{multiple user classes}, in the sense 
that different classes of drivers have different desired speeds. In order to 
achieve this, they replaced the traditional phase-space density with a 
\emph{multi-class phase-space density} (MUC-PSD). The kinetic conservation 
equation thus describes the tempo-spatial evolution of this MUC-PSD (i.e., the 
interactions between different user classes), after which an equivalent system 
of macroscopic model equations is derived. The generalisation power of their 
model is exemplified as the previously mentioned model of Helbing et al., which 
is just a special case, having only one class 
\cite{HOOGENDOORN:99b,HOOGENDOORN:00}. The developed multiclass gas-kinetic 
model is currently being integrated in a macroscopic simulation model for 
complete road networks, called HELENA, which allows prediction of future traffic 
states, and hence to assess the effectiveness of policy measures 
\cite{HOOGENDOORN:02}.

Recently, Waldeer derived a kinetic model that is based on the description of a 
driver's acceleration behaviour (as opposed to his \emph{observed} speed 
behaviour). This novel approach attempts to alleviate the unrealistic jumps in 
speeds that are typically encountered in kinetic models. To this end, Waldeer 
extends the phase-space density even further, including a vehicle's acceleration 
in addition to its position, speed, and desired speed (leading to an even more 
complex system). Because now the acceleration is updated discretely, the speed 
will change continuously as a result \cite{WALDEER:04}. Furthermore, Waldeer 
provided a numerical scheme for solving his model, by employing a Monte Carlo 
technique that is frequently used in non-equilibrium gas-kinetic theory 
\cite{WALDEER:04b}.

To end this overview of gas-kinetic models, we mention the important work of 
Tamp\`ere, who significantly extended the previous modelling approaches 
\cite{TAMPERE:04,TAMPERE:04b}. In his work, he used the \emph{generalised 
phase-space density} (as derived by Hoogendoorn \cite{HOOGENDOORN:99b}), which 
incorporates a dependency on the \emph{traffic state} $S$ (e.g., encompassing 
vehicles' speeds and their desired speeds). As it is an increasingly recognised 
fact that a complete traffic flow model should contain elements which describe 
the human behaviour (see for example the comments made by Daganzo 
\cite{DAGANZO:02,DAGANZO:02b}), Tamp\`ere proposes to include a driver's 
\emph{activation level}. His \emph{human-kinetic model} (HKM) is, just like that 
of Helbing et al. and Hoogendoorn, able to reproduce all known traffic regimes. 
Because of the dependency of the PSD on a behavioural parameter (i.e., the 
activation level that describes a driver's awareness of the governing traffic 
conditions), the model is well-suited to evaluate the applicability of 
\emph{advanced driver assistance systems} (ADAS). As another illustrating 
example, the phenomenon of a capacity funnel can be realistically explained and 
reproduced \cite{TAMPERE:03}. However, despite the progress related to 
incorporating human behaviour into mathematical models for traffic flows, 
Tamp\`ere argues that most of the work can currently not be validated because 
there is no appropriate data yet available.

		\subsection{Microscopic traffic flow models}

Having discussed both mesoscopic and macroscopic traffic flows models, we now 
arrive at the other end of the spectrum where the microscopic models reside. 
Whereas the former describe traffic operations on an aggregate scale, the latter 
kind is based on the explicit consideration of the \emph{interactions between 
individual vehicles} within a traffic stream. The models typically employ 
characteristics such as vehicle lengths, speeds, accelerations, and time and 
space headways, vehicle and engine capabilities, as well as some rudimentary 
human characteristics that describe the driving behaviour.

The material in this section is organised as follows: we first introduce the 
classic car-following (and lane-changing) models as well as some of their modern 
successors, after which we discuss the optimal velocity model, then introduce 
the more human behaviourally psycho-physiological spacing models, which are 
subsequently followed by a brief description of traffic cellular automata 
models. After some words on models based on queueing theory, the section 
concludes with a concise overview of some of the (commercially) available 
microscopic traffic flow simulators, as well as some of the issues that are 
related to the calibration and validation of microscopic traffic flow models.

More detailed information with respect to microscopic models (more specifically, 
car-following models), can be found in the book of May \cite{MAY:90}, the 
overview of Rothery \cite{GARTNER:97}, the work of Ahmed \cite{AHMED:99}, and 
the overview of Brackstone and McDonald \cite{BRACKSTONE:00}.

			\subsubsection{Classic car-following and lane-changing models}
			\label{sec:TFM:ClassicCarFollowingAndLaneChangingModels}

Probably the most widely known class of microscopic traffic flow models is the 
so-called family of \emph{car-following} or \emph{follow-the-leader} models. One 
of the oldest `models' in this case, is the one due to Reuschel 
\cite{REUSCHEL:50}, Pipes \cite{PIPES:53}, and Forbes et al. \cite{FORBES:58}. It 
is probably best known as the \emph{``two-second rule''} taught in driving 
schools everywhere\footnote{Note that, in his article, Pipes actually stated his 
safe-distance rule as keeping at least a space gap equal to a vehicle length for 
every 15 km/h of speed you are travelling at \cite{PIPES:53,HOOGENDOORN:01}.} . 
An earlier example of this line of reasoning is the work of Herrey and Herrey, 
who specified a \emph{safe driving distance} that also included the distance 
needed to come to a full stop \cite{HERREY:45}.

It still remains astonishing that the seemingly daunting and complex task that 
encompasses driving a vehicle, can be executed with such relative ease and 
little exercise, as is testified by the many millions of kilometres that are 
driven each year. In spite of this remark, the first mathematical car-following 
models that have been developed, were based on a description of the interaction 
between two neighbouring vehicles in a traffic stream, i.e., a follower and its 
leader. In this section, we historically sketch the development of car-following 
theories, as they evolved from conclusions about early experiments into more 
sophisticated models.

The above mentioned model was originally formulated as the following 
\emph{ordinary differential equation} (ODE) for single-lane traffic:

\begin{equation}
\label{eq:TFM:PipesCarFollowingODE}
	\frac{dv_{i}(t)}{dt} = \frac{v_{i + \one}(t) - v_{i}(t)}{T},
\end{equation}

with $v_{i}(t)$ and $v_{i + \one}(t)$ the speeds of the following, respectively 
leading, vehicle at time $t$, and $T$ a relaxation parameter. For the above 
case, the underlying assumption/justification is that vehicle $i$ (the follower) 
tries to achieve the speed $v_{i + \one}(t)$ of vehicle $i + \one$ (its leader), 
whilst taking a certain relaxation time $T$ into account.

As equation \eqref{eq:TFM:PipesCarFollowingODE} describes a stable system, 
Chandler et al. were among the first to include an explicit \emph{reaction time} 
$\tau$ into the model (e.g., $\tau = $1.5~s), leading to destabilisation of 
vehicle platoons \cite{CHANDLER:58}. This reaction encompasses both a 
\emph{perception-reaction time} (PRT), i.e., the driver sees an event occurring 
(for example the brake lights of the leading vehicle), as well as a 
\emph{movement time} (MT), i.e., the driver needs to take action by applying 
pressure to the vehicle's brake pedal \cite{GARTNER:97}. Introducing this 
behaviour, resulted in what is called a \emph{stimulus-response model}, whereby 
the right-hand side of equation \eqref{eq:TFM:PipesCarFollowingODE} describes 
the stimulus and the left-hand side the response (the response is frequently 
identified as the acceleration, i.e., the actions a driver takes by pushing the 
acceleration or brake pedal). The relaxation parameter is then reciprocally 
reformulated as the sensitivity to the stimulus, i.e., $\lambda = T^{-\one}$, 
resulting in the following expression:

\begin{eqnarray}
	\text{response}             & = & \text{sensitivity} \times \text{stimulus}\nonumber\\
	                            &   & \nonumber\\
	\frac{dv_{i}(t + \tau)}{dt} & = & \lambda~(v_{i + \one}(t) - v_{i}(t))\label{eq:TFM:GeneralStimulusResponseRelation}.
\end{eqnarray}

Additional to this theoretical work, there were also some early controlled 
car-following experiments, e.g., the ones done by Kometani and Sasaki, who add a 
non-zero acceleration term to the right-hand side of the stimulus-response 
relation, in order to describe collision-free driving based on a safety distance 
\cite{KOMETANI:58,KOMETANI:61}.

Equation \eqref{eq:TFM:GeneralStimulusResponseRelation} is called a 
\emph{delayed differential equation} (DDE), which, in this case, is known to 
behave in an unstable manner, even resulting in collisions under certain initial 
conditions. Gazis et al. remedied this situation by making the stimulus $\lambda$ 
dependent on the distance, i.e., the space gap $g_{s_{i}}$ between both vehicles 
\cite{GAZIS:59}:

\begin{equation}
\label{eq:TFM:GazisCarFollowingModel}
	\frac{dv_{i}(t + \tau)}{dt} = \lambda~\frac{v_{i + \one}(t) - v_{i}(t)}{x_{i + \one}(t) - x_{i}(t)}.
\end{equation}

Further advancements to this car-following model were made by Edie, who 
introduced the current speed of the following vehicle \cite{EDIE:61}. Gazis et 
al, forming the club of people working at General Motors' research laboratories, 
generalised the above set of models into what is called the \emph{General Motors 
non-linear model} or the \emph{Gazis-Herman-Rothery (GHR) model} 
\cite{GAZIS:61}:

\begin{equation}
\label{eq:TFM:GeneralMotorsCarFollowingModel}
	\frac{dv_{i}(t + \tau)}{dt} = \lambda~v_{i}^{m}(t)\frac{v_{i + \one}(t) - v_{i}(t)}{(x_{i + \one}(t) - x_{i}(t))^{l}},
\end{equation}

with now $\lambda$, $l$, and $m$ model parameters (in the early days, the model 
was also called the \emph{L\&M model} \cite{GAZIS:02}). For a good overview of 
the different combinations of parameters attributed to the resulting models, we 
refer the reader to the book of May \cite{MAY:90}, and the work of Ahmed 
\cite{AHMED:99}.

A recent extension to the classic car-following theory, is the work of Treiber 
and Helbing, who developed the \emph{intelligent driver model} (IDM). Its 
governing equation is the following \cite{TREIBER:99,TREIBER:00,TREIBER:01}:

\begin{equation}
	\frac{dv_{i}}{dt} = a_{\text{max}} \left \lbrack \underbrace{\one - \left ( \frac{v_{i}}{v_{\text{des}}} \right )^{\delta}}_{\text{acceleration}} - \underbrace{\left ( \frac{g_{s}^{*}(v_{i},\Delta v_{i})}{g_{s_{i}}} \right )^{\two}}_{\text{deceleration}} \right \rbrack,
\end{equation}

with $a_{\text{max}}$ the maximum acceleration, $v_{\text{des}}$ the vehicles' 
desired speed, and $\Delta v_{i}$ the speed difference with the leading vehicle 
(we have dropped the dependencies on time $t$ for the sake of visual clarity). 
The first terms within the brackets denote the tendency of a vehicle to 
accelerate on a free road, whereas the last term is used to allow braking in 
order to avoid a collision (the effective desired space gap 
$g_{s}^{*}(v_{i},\Delta v_{i})$ is based on the vehicle's speed, its relative 
speed with respect to its leader, a comfortable maximum deceleration, a desired 
time headway, and a jam space gap). The finer qualities of the IDM are that it 
elegantly generalises most existing car-following models, and that it has an 
explicit link with the non-local gas-kinetic mesoscopic model discussed in 
section \ref{sec:TFM:ImprovementsToTheMesoscopicModellingApproach} 
\cite{TREIBER:00}. It is furthermore quite capable of generating all known 
traffic regimes. Based on the IDM, Treiber et al. also constructed the 
\emph{human driver model} (HDM), which includes a finite reaction time, 
estimation errors, temporal and spatial anticipation, and adaptation to the 
global traffic situation \cite{TREIBER:05}.

Similar to the work of Kometani and Sasaki, Gipps proposed a car-following model 
based on a safe braking distance, leading to collision-free dynamics 
\cite{GIPPS:81}. The model is interesting because no differential equations are 
involved (i.e., the speeds are computed directly from one discrete time step to 
another), and because it can capture underestimation and overreactions of 
drivers, which can lead to traffic flow instabilities. In similar spirit of 
Gipps' work, Krau\ss~developed a model that is based on assumptions about 
general properties of traffic flows, as well as typical acceleration and 
deceleration capabilities of vehicles. Fundamental to his approach, is that all 
vehicles strive for collision-free driving, resulting in a model that has the 
ability to generalise most known car-following models 
\cite{KRAUSS:97,KRAUSS:98}.

Another example of a recently proposed car-following model, is the `simple' 
model of Newell, who formulates his theory in terms of vehicle trajectories 
whereby the trajectory of a following vehicle is essentially the same as that of 
its leader\footnote{A similar model was proposed earlier by Helly 
\cite{HELLY:61,GERLOUGH:64,NEWELL:02b}.}. Remarkable properties are that the 
model has no driver reaction time, and that it corresponds to the first-order 
macroscopic LWR traffic flow model with a triangular $q_{e}(k)$ fundamental 
diagram (see section \ref{sec:TFM:LWRAnalyticalSolutions}) \cite{NEWELL:02b}. 
The model furthermore also agrees quite well with empirical observations made at 
a signallised intersection, which support the model and consequently also the 
first-order macroscopic LWR model \cite{AHN:04}.

As a final example, we briefly illustrate Zhang's car-following theory which is 
based on a multi-phase vehicular traffic flow. This means that the model is able 
to reproduce both the capacity drop and hysteresis phenomena, because his theory 
is based on the asymmetry between acceleration and deceleration characteristics 
of vehicles \cite{MAERIVOET:05d}. The model also holds a generalisation 
strength, as it is possible to derive all other classic car-following models 
\cite{ZHANG:05}.

With respect to the stability of the car-following models, there exist two 
criteria, i.e., \emph{local} and \emph{asymptotic stability} (also called 
\emph{string stability}). The former describes how initial disturbances in the 
behaviour of a leading vehicle affect a following vehicle, whereas the latter is 
used to denote the stability of a \emph{platoon of following vehicles}. By such 
a stable platoon it is then meant that initial finite disturbances exponentially 
die out along the platoon. Early experiments by Herman et al. already considered 
these criteria for both real-life as for the developed mathematical 
car-following models \cite{HERMAN:59}.

\begin{figure*}[!htb]
	\centering
	\includegraphics[width=0.32\textwidth]{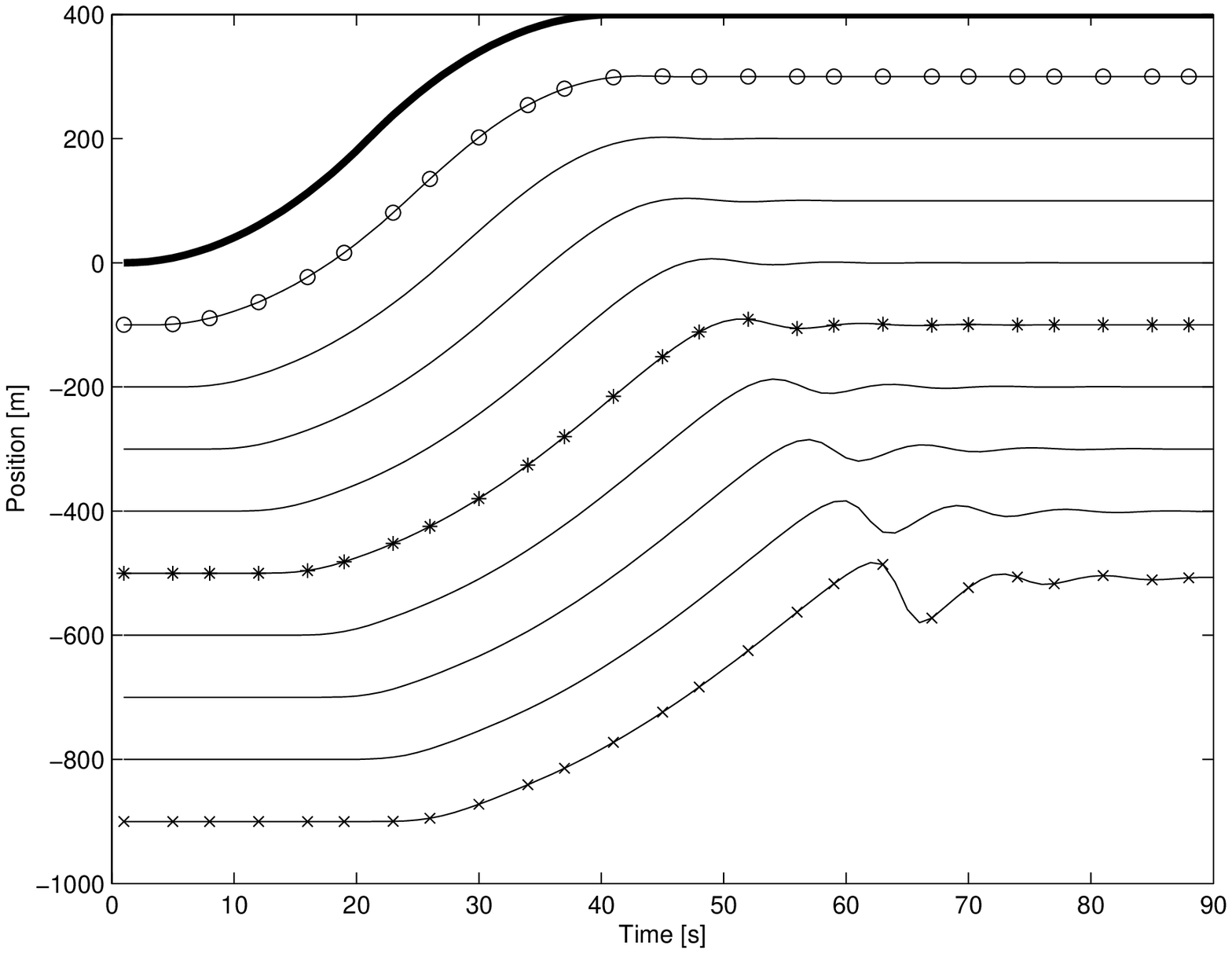}
	\includegraphics[width=0.32\textwidth]{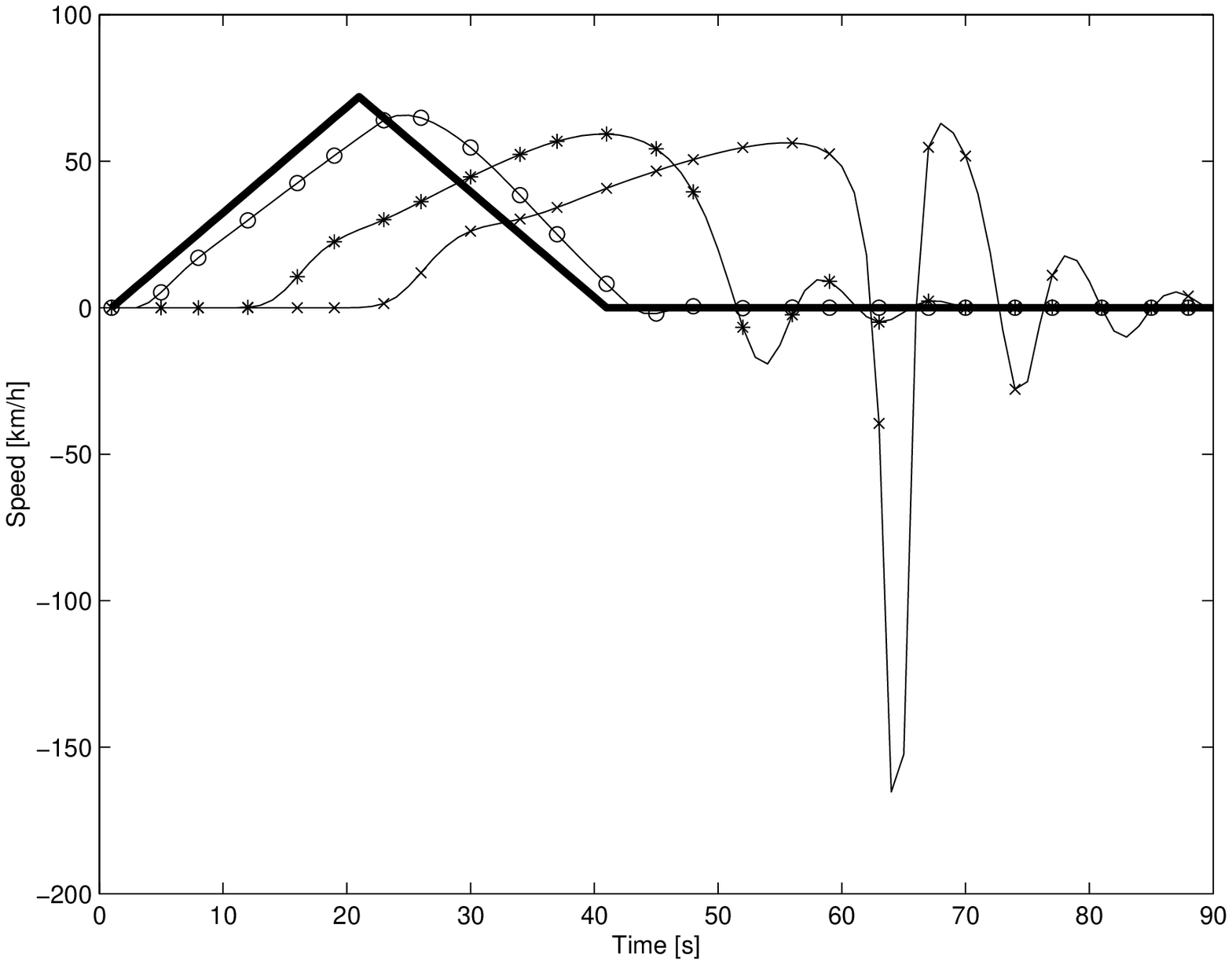}
	\includegraphics[width=0.32\textwidth]{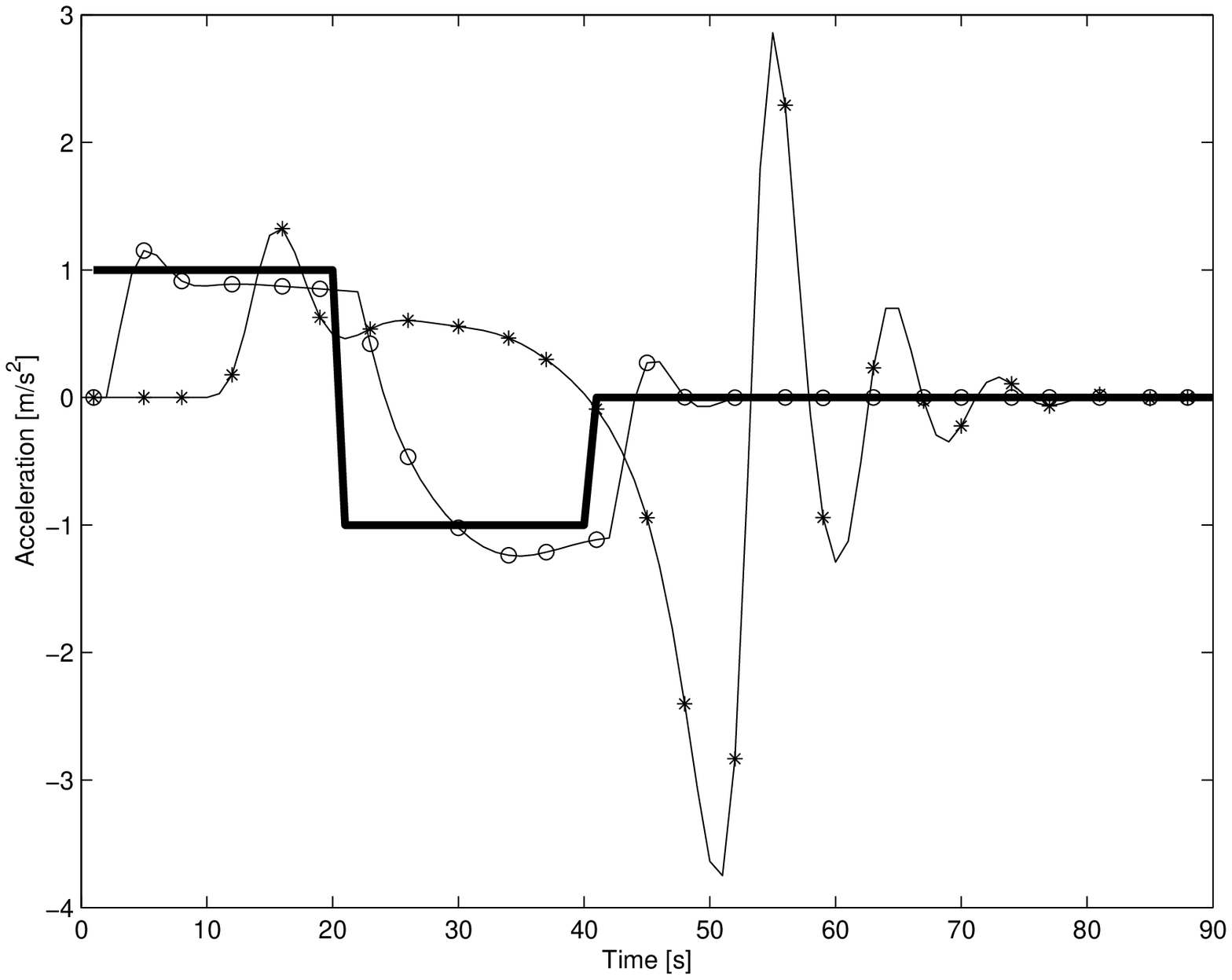}
	\caption{
		An example based on Gazis et al.'s car-following model of equation 
		\eqref{eq:TFM:GazisCarFollowingModel}, indicating an asymptotically unstable 
		platoon of 10 vehicles. \emph{Left:} the time-space trajectories of all ten 
		vehicles (the leading vehicle is shown with a thick solid line). We can see 
		an instability occurring at the 6\theth ($\star$) vehicle, growing severely 
		such that the last vehicle even has to drive backwards at a high speed of 
		-150~km/h. \emph{Middle:} the speeds of the first, the 2\thend ($\circ$), 
		the 6\theth ($\star$), and the last vehicle ($\times$). Note the 
		oscillations and negative values for the speeds of the vehicles at the end 
		of the platoon. \emph{Right:} the accelerations of the first, the 2\thend 
		($\circ$), and the 6\theth ($\star$) vehicle (example based on 
		\cite{IMMERS:98}).
	}
	\label{fig:TFM:CarFollowingStability}
\end{figure*}

As an example, we graphically illustrate in 
\figref{fig:TFM:CarFollowingStability} the asymptotic stability of a platoon of 
some 10 identical vehicles. We have used the simple car-following model of Gazis 
et al. of equation \eqref{eq:TFM:GazisCarFollowingModel} to describe how a 
following vehicle changes his acceleration, based on the speed difference and 
space gap with its direct leader in the platoon (the sensitivity $\lambda$ was 
set to 5000~m/s${^\two}$ with a reaction time $\tau = \one$~s). The left part of 
the figure shows all the vehicles' positions, whereas the middle and right parts 
show the speeds and accelerations of the 2\thend, the 5\theth, and the 10\theth 
vehicle respectively. We can see that all vehicles are initially at rest 
(homogeneously spaced), after which the leading vehicle applies an acceleration 
of 1~m/s$^{\two}$, decelerates with -1~m/s$^{\two}$, and then comes to a full 
stop. As can be seen, the first 4 following vehicles mimic the leader's 
behaviour rather well, but from the 5\theth following vehicle on, an instability 
starts to form (note that all following vehicles suffer from oscillations in 
their acceleration behaviour). This instability grows and leads to very large 
accelerations for the last vehicle, which even momentarily reaches a negative 
speed of some -150~km/h; this is clearly unrealistic (the vehicles shouldn't be 
driving backwards on the road), indicating that the specified car-following 
model is unsuitable to capture the realistic behaviour of drivers under these 
circumstances.

In continuation of this small excerpt on stability, we refer the reader to the 
work of Zhang and Jarrett who analytically and numerically derive the general 
stability conditions (in function of the reaction time and the sensitivity to 
the stimulus) for the previously mentioned classic car-following models 
\cite{ZHANG:97}, the work of Holland who derives general stability conditions 
and validates them with empirical data containing non-identical drivers (i.e., 
aggressive and timid ones); central to Holland's work is the source for 
instability with respect to a breakdown of a traffic flow. He relates this event 
to a so-called anticipation time that describes the duration for a wave 
containing an instability to travel to the current driver \cite{HOLLAND:98}. 
Finally, we mention that stability analysis is of paramount importance for e.g., 
automated vehicle technologies (`smart cars') such as \emph{intelligent} or 
\emph{adaptive cruise control} (ICC/ACC), as in for example the platooning 
experiments in the PATH project where a platoon of vehicles autonomously drives 
close to each other at high speeds.

To conclude this section, we shed some light on the typical mechanisms behind 
lane-changing models. With respect to microscopic models for multi-lane traffic, 
it is a frequent approximation to only take lateral movements \emph{between} 
neighbouring lanes into account (as opposed to the \emph{within-lane} lateral 
dynamics of a vehicle). In such cases, a vehicle changes a lane based on an 
incentive: these lane changes can then be classified as being 
\emph{discretionary} (e.g., to overtake a slower vehicle), or \emph{mandatory} 
(e.g., to take an off-ramp). When a vehicle (i.e., driver) has decided to 
perform a lane change, a check is made on whether or not it is physically 
possible to merge in to the adjacent lane (note this lane changing process also 
describes vehicles turning at street intersections). This latter process is 
called the \emph{gap acceptance} behaviour: if there is no such possibility (as 
it is frequently the case in dense traffic), a driver may initiate at 
\emph{forced merging}, in which case the following vehicle in the target lane 
might have to \emph{yield}. This interaction between forced merging and yielding 
can be frequently observed at on-ramps where heavy duty vehicles enter the 
motorway. Although it seems intuitive that there is an asymmetry between the 
frontal and backward space gaps in the target lane (i.e., the former is usually 
smaller than the latter due to the human behaviour associated with forced 
merging and yielding), there is in our opinion nevertheless not enough empirical 
data available to calibrate the microscopic models that describe lane-changing 
(see for example the work of Ahmed \cite{AHMED:99}). One way to obtain a correct 
behaviour is to use a kind of a black box approach, in which for example the 
downstream capacity of a motorway section is used as a measure for calibrating 
the interactions (i.e., lane changes) between vehicles in a traffic stream. Note 
that as technology advances, new detailed data sets are constructed. An example 
is the work of Hoogendoorn et al. who use a \emph{remote sensing} technique to 
capture vehicle trajectories based on aerial filming of driving behaviour under 
congested conditions \cite{HOOGENDOORN:03}.

			\subsubsection{Optimal velocity models}
			\label{sec:TFM:OptimalVelocityModels}

Closely related to the previously discussed classic car-following models, are 
the so-called \emph{optimal velocity models} (OVM) of Newell and Bando et al. 
Whereas the previous car-following models mostly describe the behaviour of a 
vehicle that is following a leader, the OVMs modify the acceleration mechanism, 
such that a vehicle's desired speed is selected on the basis of its space 
headway, instead of only considering the speed of the leading vehicle 
\cite{HELBING:01}. Newell was the first to suggest such an approach, using an 
equilibrium relation for the desired speed as a function of its space headway 
(e.g., the $\overline v_{s}(\overline h_{s})$ fundamental diagram 
\cite{NEWELL:63b}.

Bando et al. later improved this model, resulting in the following equation that 
describes a vehicle's acceleration behaviour \cite{BANDO:95}:

\begin{equation}
\label{eq:TFM:OVMModel}
	\frac{dv_{i}(t)}{dt} = \alpha \left ( V(h_{s_{i}}(t)) - v_{i}(t) \right ),
\end{equation}

in which $V()$ is called the \emph{optimal velocity function} (OVF). The 
difference between this desired speed, associated with the driver's current 
space headway, and the vehicle's current speed, is corrected with an 
acceleration $\alpha V()$, with now $\alpha$ a coefficient expressing the 
sensitivity of a driver. Specification of the optimal velocity function 
(typically a sigmoid function such as $\tanh$) is done such that it is zero for 
$h_{s_{i}} \rightarrow \zero$, and bounded to $v_{\text{max}}$ for $h_{s_{i}} 
\rightarrow +\infty$; this latter condition means that the model is able to 
describe the acceleration of vehicles without the explicit need for a leader as 
in the previous car-following models.

Interestingly, the OVM requires, in contrast to the classic car-following 
models, no need for a reaction time in order to obtain spontaneous clustering of 
vehicles \cite{KRAUSS:98}. Unfortunately, the model is not always free of 
collisions, and can result in unrealistically large accelerations 
\cite{NAGEL:03}.

			\subsubsection{Psycho-physiological spacing models}

Instead of using continuous changes in space gaps and relative speeds, it was 
already recognised in the early sixties that drivers respond to certain 
\emph{perception thresholds} \cite{BRACKSTONE:00}. For example, a leading 
vehicle that is looming in front of a follower, will be perceived as having 
approximately the same small size for a large duration, but once the space gap 
has shrunk to a certain size, the size of the looming vehicle will suddenly seem 
a lot bigger (i.e., like crossing a threshold), inducing the following vehicle 
to either slow down or overtake.

The underlying thresholds with respect to speeds, speed differences, and space 
gaps, were cast into a model by the work of Wiedemann et al. 
\cite{WIEDEMANN:74}. In this respect, the models are called 
\emph{psycho-physiological spacing models}, and although they seem quite 
successful in explaining the traffic dynamics from a behavioural point of view 
(even lane-change dynamics can be included based on suitable perception 
thresholds), calibration of the models has nevertheless been a difficult issue 
\cite{BRACKSTONE:00}.

			\subsubsection{Traffic cellular automata models}
			\label{sec:TFM:TrafficCellularAutomata}

In the field of traffic flow modelling, microscopic traffic simulation has 
always been regarded as a time consuming, complex process involving detailed 
models that describe the behaviour of individual vehicles. Approximately a 
decade ago, however, new microscopic models were being developed, based on the 
\emph{cellular automata} programming paradigm from \emph{statistical physics}. 
The main advantage was an \emph{efficient and fast performance} when used in 
computer simulations, due to their rather low accuracy on a microscopic scale. 
These so-called \emph{traffic cellular automata} (TCA) are dynamical systems 
that are discrete in nature, in the sense that time advances with discrete steps 
and space is coarse-grained (e.g., the road is discretised into cells of 7.5 
metres wide, each cell being empty or containing a vehicle). This 
coarse-graininess is fundamentally different from the usual microscopic models, 
which adopt a semi-continuous space, formed by the usage of IEEE floating-point 
numbers. TCA models are very flexible and powerful, in that they are also able 
to capture all previously mentioned basic phenomena that occur in traffic flows 
\cite{BARLOVIC:99,CHOWDHURY:00}. In a larger setting, these models describe 
\emph{self-driven, many-particle systems, operating far from equilibrium}. And 
in contrast to strictly gaseous analogies, the particles in these systems are 
intelligent and able to learn from past experience, thereby opening the door to 
the incorporation of behavioural and psychological aspects 
\cite{CHOWDHURY:99b,WOLF:99,HELBING:01}.

Not only in the field of vehicular traffic flow modelling, but also in other 
fields such as \emph{pedestrian behaviour}, \emph{escape} and \emph{panic 
dynamics}, \ldots the cellular automata approach proved to be quite useful. It 
is now feasible to simulate large systems containing many `intelligent 
particles', such that is it possible to observe their interactions, collective 
behaviour, self-organisation, \ldots 
\cite{IMMERS:98b,VANZUYLEN:99,HELBING:99,HELBING:01,NAGEL:02b,NAGEL:02c,CHOWDHURY:04}

			\subsubsection{Models based on queueing theory}

In this final section dealing with types of microscopic traffic flow models, we 
briefly summarise some of the models that are based on the paradigm of queueing 
theory. Early applications of queueing theory to the field of transportation 
engineering are mostly related to descriptions of the behaviour signallised and 
unsignallised intersections, overtaking on two-lane roads with opposing traffic, 
\ldots \cite{CLEVELAND:64}. Another more theoretically oriented application can 
be traced back to the work of Newell, who gives a nice summary of the 
mathematical details related to the \emph{practical} application of the 
methodology. Newell was one of the few people who directly questioned the 
usefulness of cleverly devising a lot of methods and solutions, whereby 
corresponding problems remained absent \cite{NEWELL:82}. In his later work, 
Newell reintroduced the concept of arrival and departure functions (i.e., the 
cumulative curves as briefly described in section 
\ref{sec:TFM:FirstOrderLWRModel}), giving an analytical but still highly 
intuitive method for solving traffic flow problems, and drawing parallels with 
the well-known and studied first-order macroscopic LWR model 
\cite{NEWELL:93,NEWELL:93b,NEWELL:93c}.

During the mid-nineties, Heidemann developed several queueing-based traffic flow 
models, of which the most powerful version deals with non-stationary conditions 
and is able to model the capacity drop and hysteresis phenomena, as well as 
providing an explanation for the wide scatter observed in empirical fundamental 
diagrams \cite{HEIDEMANN:01,MAERIVOET:05d}.

Central to the approach in this field, is the partitioning of a road into equal 
pieces of width $\one / k_{\text{jam}}$. Each of these pieces is then considered 
as a \emph{service station} operating with a \emph{service rate} $\mu = 
k_{\text{jam}} \cdot \overline v_{\text{ff}}$. Equivalently, vehicles arrive at 
each service station with an \emph{arrival rate} $\lambda = k \cdot \overline 
v_{\text{ff}}$, with the assumption that $k$ is the prevailing density and that 
traffic can flow unimpeded in the free-flow traffic regime. When vehicles enter 
the motorway, they can get stuck inside the queues, thereby reducing the 
space-mean speed in the system. Different \emph{queueing policies} can be 
specified in the form of service and arrival distributions. In queueing theory, 
the \emph{Kendall notation} is adopted, whereby a system is described as $A/S/m$ 
with $A$ the arrival distribution, $S$ the service distribution, and $m$ the 
number of servers (i.e., service stations). Typical forms are the $M/M/\one$ 
queues that have an exponentially distributed arrival time, exponentially 
distributed service time, and one server (with an infinite buffer).

Recently, Van Woensel extended the existing queueing models for traffic flows, 
leading to e.g., analytical derivations of fundamental diagrams based on $G/G/m$ 
queues that have general distributions for the arrival and service rates with 
multiple servers \cite{VANWOENSEL:03}. The methodology also includes queues with 
finite buffers, and has been applied to the estimation of emissions, although we 
question the validity of this latter approach (which is essentially based on a 
one-dimensional fundamental diagram) as we believe dynamic models are necessary, 
e.g., to capture transients in traffic flows \cite{VANDAELE:00}.

Queue-based models were also used to describe large-scale traffic systems, e.g., 
complete countries, as was mentioned in section 
\ref{sec:TFM:CritiqueOnTripBasedApproaches} \cite{CETIN:03}. In that section, we 
already mentioned that queues with finite buffer capacities are to be preferred 
in order to correctly model queue spill back. However, with respect to a proper 
description of traffic flow phenomena, some of the problems can not be so easily 
solved, e.g., the speed of a backward propagating kinematic shock wave. Take for 
example vehicles queued behind each other at a traffic light: once the light 
turns green, the first-order macroscopic LWR model correctly shows the dispersal 
of this queue. In a queue-based model however, once a vehicle exits the front of 
the queue, all vehicles simultaneously and instantly move up one place, thus the 
kinematic wave propagates backwards at an infinite speed~!

To conclude this short summary on queueing models, we mention the work of 
J\'ulvez and Boel, who present a similar approach, based on the use of 
\emph{Petri nets\footnote{Petri nets (invented in the sixties by Carl Adam 
Petri) are a formalism for describing discrete systems \cite{PETRI:62}; they 
consist of directed graphs of `transitions' and `places', with arcs forming the 
connections between them. Places can contain `tokens', which can be `consumed' 
when a transition `fires'.}}. Their work allows them to construct complete urban 
networks, based on the joining together of short sections, with continuous Petri 
nets for the propagation of traffic flows, and discrete Petri nets for the 
description of the traffic lights \cite{BASILE:04,JULVEZ:05}.

			\subsubsection{Microscopic traffic flow simulators}

In continuation of the previous sections that gave an overview of the different 
types of existing microscopic traffic flow models, this section introduces some 
of the computer implementations that have been built around these models. In 
most cases, the computer simulators incorporate the car-following and 
lane-changing processes as submodels, as opposed to strategic and operational 
modules that work at a higher-level layer (i.e., route choice, \ldots).

\begin{figure}[!htb]
	\centering
	\includegraphics[width=\figurewidth]{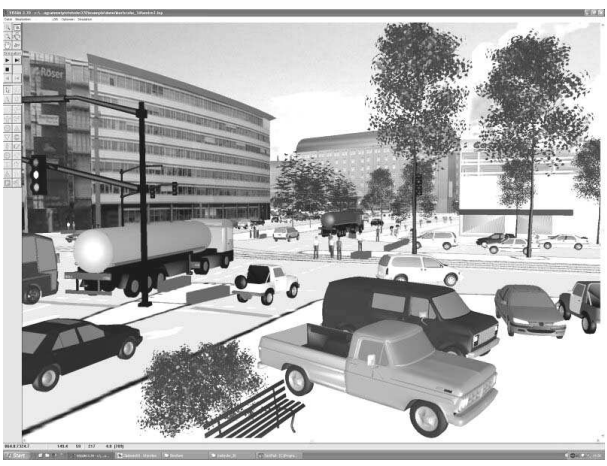}
	\caption{
		A screenshot of the VISSIM microscopic traffic flow simulator, showing a 
		detailed virtual environment containing trees, buildings, pedestrians, 
		\ldots (image reproduced after \cite{PTV:05}).
	}
	\label{fig:TFM:VISSIMScreenshot}
\end{figure}

Whereas most microscopic traffic simulators allow to build a road network, 
specify travel demands (e.g., by means of OD tables), there was quite some 
effort spent over the last decade, in order to achieve a qualitative 
visualisation (e.g., complete virtual environments with trees, buildings, 
pedestrians, bicycles, \ldots An example of such a virtual environment is shown 
in \figref{fig:TFM:VISSIMScreenshot}, which is based on VISSIM's visualisation 
module. Note that in our opinion, the usefulness of these virtual scenes should 
not be underestimated, as in some cases a project's approval might hinge on a 
good visual representation of the results. It is one thing for policy makers to 
judge the effects of replacing a signallised intersection with a roundabout, 
based on a report of the observed downstream flows of each intersection arm, but 
it gives a whole other feeling when they are able to \emph{see} how the traffic 
streams will interact~! Even in the early sixties, it was recognised that a 
visual representation of the underlying traffic flow process, was an undeniable 
fact for promoting its acceptance among traffic engineers \cite{GERLOUGH:64}. 
With respect to this latest comment, Lieberman even states that \emph{``There is 
a need to view vehicle animation displays, to gain an understanding of how the 
system is behaving, in order to explain why the resulting statistics were 
produced''} \cite{GARTNER:97}.

Quite a large amount of microscopic traffic flow models have been developed, in 
most cases starting from a research tool, and --- by the law of profit --- 
naturally evolving into full-blown commercial packages, including e.g., dynamic 
traffic assignment and other transportation planning features. Note the sad 
observation that this commercialisation inherently tends to obscure the 
underlying models. In such cases, privacy concerns, company policies, and 
project contracts and agreements prohibit a total disclosure of the mathematical 
details involved. Some of these computer models are listed here. For starters, 
the \emph{Generic Environment for TRaffic Analysis and Modeling} (GETRAM) 
couples the multi-modal traffic assignment model EMME/2 to the \emph{Advanced 
Interactive Microscopic Simulator for Urban and Non-Urban Networks} (AIMSUN2) 
model \cite{BARCELO:02c}. Next, the \emph{Parallel microscopic traffic 
simulator} (Paramics), initially developed at the Edinburgh Parallel Computing 
Centre, but afterwards bought by Quadstone \cite{CAMERON:94,QUADSTONE:00}. 
Subsequently, Yang developed a \emph{MIcroscopic Traffic flow SIMulator} 
(MITSIM) \cite{YANG:97}, and Maerivoet constructed a \emph{MIcroscopic TRAffic 
flow SIMulator} (Mitrasim \emph{2000}) which was mostly based on and influenced 
by MITSIM's and Paramics' dynamic behaviour \cite{MAERIVOET:01}. Two further 
examples are the Open Source Software (OSS) package called \emph{Simulation of 
Urban MObility} (SUMO), developed at the Deutsches Zentrum f\"ur Luft- und 
Raumfahrt \cite{KRAJZEWICZ:04}, and the VISSIM programme developed by the German 
PTV group \cite{PTV:05}. In addition, there is the \emph{INteractive DYnamic 
traffic assignment} (INDY) model \cite{MALONE:03}, and the INTEGRATION software 
package developed by Van Aerde et al. This latter simulator deserves a special 
mention: it is microscopic in nature, but the speeds of the vehicles that are 
propagated through the network, are based on a macroscopic $\overline 
v_{s_{e}}(\overline h_{s})$ fundamental diagram for each link 
\cite{VANAERDE:96}. Finally, we mention the \emph{TRansportation ANalysis and 
SIMulation System} (TRANSIMS) project \cite{NAGEL:98b}, \ldots

An extensive overview of all existing microscopic traffic flow simulators until 
1998 is provided by the \emph{Simulation Modelling Applied to Road Transport 
European Scheme Tests}, or better known as the SMARTEST report 
\cite{ALGERS:98}.\\

\sidebar{
	When using one of these microscopic simulators, it is important to understand 
	the assumptions and limitations inherent to the implemented models, in order 
	to judge the results objectively. Indeed, as with any model, the question on 
	whether some observed behaviour arises due to the implemented model, or as a 
	result of the imposed boundary conditions, should always be asked, understood, 
	and answered.
}\\

			\subsubsection{Calibration and validation issues}

Due to the sometimes large amount of parameters typically involved in 
microscopic traffic flow models, their computational complexity is often a 
significant disadvantage when compared to meso- or macroscopic models (although 
there are some exceptions, e.g., the traffic cellular automata models of section 
\ref{sec:TFM:TrafficCellularAutomata}). From the point of view of model 
calibration and validation, this poses an interesting conundrum, as in many 
cases not all parameters are equally influential on the results (thus requiring 
some sensitivity analyses). In this sense, microscopic models contain a real 
danger of purporting to convey a sort of fake accuracy. Different parameter 
combinations can lead to the same phenomenological effects, leaving us pondering 
as to what exactly is causing the observed behaviour \cite{TAMPERE:04}. As there 
is no clear road map on how to calibrate microscopic traffic flow models, we 
here give a small sample of some of the numerous attempts that have been made.

There is the work of Jayakrishan and Sahraoui who distinguish between 
calibration in the conceptual (i.e., at the level of the underlying mathematical 
model) and operational (i.e., within the global context of the study) phases; 
they apply their operational methodology to both PARAMICS (micro) and DYNASMART 
(macro), using the \emph{California Freeway Performance Measurement System} 
(PeMS) database from the PATH project to feed and couple both models 
\cite{JAYAKRISHAN:00}.

Based on a publicly available data set of a one-lay road corridor of six 
kilometres long (the data contained detailed cumulative curves), Brockfeld et 
al. systematically tested the predicted travel times of some ten well-known 
microscopic traffic flow models. As a result of a non-linear optimisation 
process to calibrate the models, they found that the intelligent driver model 
and the cell-transmission model perform the best (i.e., below an error rate of 
17\%), due to the fact that these models require the least amount of parameters 
(there were even some models such as the Gipps-based ones that had \emph{hidden} 
parameters). Their final conclusion is noteworthy, as they state that 
\emph{``creating a new model is often done, however calibrating this model to 
reality is a formidable task, which explains why there currently are more models 
than results about them''} \cite{BROCKFELD:03}.

Related to the previous study, Hourdakis et al. present an automated systematic 
calibration methodology based on an optimisation process, applied to the AIMSUN2 
simulator. The data used for the calibration procedure stem from a twenty 
kilometres long motorway in Minneapolis, Minnesota. The process first involves a 
calibration of the global model parameters (i.e., to get the macroscopic flows 
and speeds correct), after which the local parameters are dealt with (i.e., ramp 
metering setups, et cetera). In their results, Hourdakis et al. state an obtained 
average correlation coefficient of 0.961 for manual calibration (the results for 
the automated calibration are similar), which is quite high (they mostly explain 
this due to the data's high level of detail, as well as the quality of the 
simulator software) \cite{HOURDAKIS:03}.

Recently, Chu et al. extended the systematic, multi-stage calibration approach 
for the PARAMICS simulator. Based on data of a highly congested six kilometres 
long corridor network in the city of Irvine, Orange County, California, they 
first calibrate the driving behaviour models, then the route choice model, after 
which estimation and fine-tuning of the OD tables is done. Despite the good 
reproduction of travel times, their calibration methodology was done manually, 
and an automated optimisation procedure remains future work \cite{CHU:04}.\\

Other examples of calibration of microscopic traffic flow models, include the 
work of Dowling et al., who give an extensive account on the application of 
commercially available simulation tools to typically encountered traffic 
engineering problems \cite{DOWLING:02}, and the work of Mahanti which is 
primarily based on the correct representation of OD tables \cite{MAHANTI:04}.\\

\sidebar{
	To end this section, we state some important principles that are --- in our 
	opinion --- related to a correct calibration methodology. First and foremost, 
	we believe that all traffic flow models (whether they are macro-, meso- or 
	microscopic in nature), should be able to accurately reproduce and predict the 
	encountered delays, queue lengths, and other macroscopic \emph{first-order 
	characteristics} (i.e., the kinematic wave speed, a correct and realistic road 
	capacity, \ldots). One way to test this is the use of cumulative curves, as 
	they provide an elegant way to automatically perform a good calibration. It is 
	for example possible to consider the difference between observed and simulated 
	curves, and then use a Kolmogorov-Smirnov goodness-of-fit statistical test to 
	decide on whether the difference is statistically significant, or if it is 
	just a Brownian motion with a zero mean. Only when these first-order effects 
	can be correctly reproduced, the next step can be to consider 
	\emph{second-order effects} such as waves of stop-and-go traffic, 
	oscillations, \ldots
}\\

Furthermore, it is important to take into account the \emph{spatial nature} of 
the study area, i.e., a detailed description of the road infrastructure, with 
bottleneck locations as well as up- and downstream boundary conditions. With 
respect to the model that is created within the computer, it is paramount to 
know how the model behaves on both the link as well as the node level. Because 
the models are most of the time working with fairly homogeneous road links 
(e.g., constant elevations, no road curvature, \ldots), it might be necessary to 
allow for small deviations from (or fixes to) reality (e.g. inserting extra 
intermediate nodes in the network in order to artificially obtain bottlenecks).

		\subsection{Submicroscopic traffic flow models}

As the level of modelling detail is increased, we enter the realm of 
submicroscopic models. Traditional microscopic models describe vehicles as 
single operating units, putting emphasis on the interactions between different 
(successive) vehicles. In addition to this, submicroscopic models push the 
boundaries even further, giving detailed descriptions of a vehicle's inner 
workings. This typically entails modelling of the \emph{physical 
characteristics} such as engine performance, detailed gearbox operations, 
acceleration, braking, and steering manoeuvres, \ldots Complementary to the 
functioning of a vehicle's physical components, submicroscopic models can also 
describe a \emph{human driver's decision taking process} in much more detail 
than is usually done. Some examples of submicroscopic models are:

\begin{itemize}
	\item van Arem's \emph{Microscopic model for Simulation of Intelligent Cruise 
	Control} MIXIC: it contains a driver model (for deciding on and executing of 
	lane changes, car-following behaviour, and the application of intelligent (or 
	adaptive) cruise control -- ICC/ACC) and a vehicle model (dealing with the 
	engine, the transmission, road friction, aerodynamic, rolling, and slope 
	resistance) \cite{VANAREM:97}.

	\item In similar spirit, Minderhoud has developed the \emph{Simulation model 
	of Motorways with Next generation vehicles} (Simone); this model focusses on 
	intelligent driver support systems, such as ICC/ACC, platoon driving, 
	centralised control of vehicles, et cetera. In contrast to most other 
	(sub)microscopic models, Simone explicitly allows for rear-end collisions to 
	occur under certain parameter combinations. As there is a close coupling 
	between driver behaviour related parameters and those of the simulation, these 
	collision dynamics enable the modeller to find realistic values (or ranges) 
	for these parameters \cite{MINDERHOUD:99}.

	\item Ludmann's \emph{Program for the dEvelopment of Longitudinal micrOscopic 
	traffic Processes in a Systemrelevant environment} (PELOPS), is akin to the 
	previous two models. It is however more technologically oriented with respect 
	to the car-following behaviour of vehicles, aiming at merging both a driver's 
	perceptions and decisions, the car's handling, and the surrounding traffic 
	conditions. At the core of the model, there are four modules that respectively 
	describe vehicle routing in a road network, human decision taking (i.e., 
	car-following, tactical decisions with respect to lane-changing, \ldots), 
	vehicle handling (i.e., a driver's physical acts of steering, accelerating and 
	braking, \ldots), and finally a module that describes physical vehicle 
	characteristics (traction on elevations, engine capabilities, exhaust gas
	modelling, \ldots) \cite{LUDMANN:98,EHMANNS:00}.
\end{itemize}

To conclude this section, we like to mention an often scientifically neglected 
area of research, namely the popular field of simulation in the \emph{computer 
gaming industry}. Over the last couple of decades, numerous arcade-style 
\emph{racing simulations} have been developed, allowing a \emph{player} to be 
completely immersed in a three-dimensional virtual world in which racing at high 
speeds is paramount. Examples of these kinds of programmes are the highly 
addictive world of \emph{Formula 1} racing, street racing in city environments, 
off-road rally races, \ldots The underlying submicroscopic models in these 
games, have over the course of several years been evolved to incorporate all 
sorts of physical effects. Friction characteristics (e.g., pavement versus 
asphalt), road elevation, wet conditions, air drag and wind resistance 
(including effects such as slip streaming and downforce), car weight depending 
on fuel consumption, tyre wear, \ldots have had influences on what we commonly 
refer to as \emph{car handling}, i.e., realistic behaviour with respect to car 
acceleration, braking, and steering. Thanks to the increasing computational 
power of desktop computers, graphics cards, as well as dedicated gaming consoles 
(e.g., Microsoft's \emph{Xbox}, Sony's \emph{PlayStation}, Nintendo's 
\emph{GameCube}, \ldots), the path to a whole plethora of extra realistic 
effects has been paved: skidding, under- and oversteering, sun glare, overly 
realistic collision dynamics (in our opinion, this is where the arcade sensation 
plays a major role), \ldots

		\subsection{The debate between microscopic and macroscopic models}
		\label{sec:TFM:MicroscopicAndMacroscopicDebate}

Deciding which class of models, i.e., microscopic, or macroscopic (and we also 
include the mesoscopic models), is the correct one to formulate traffic flow 
problems, has been a debate among traffic engineers ever since the late fifties. 
Although the debate was not as intense as say, the one between first- and 
higher-order macroscopic traffic flow models (see section 
\ref{sec:TFM:CritiquesOnHigherOrderModels} for more details), it nevertheless 
sparkled some interesting issues. As is nearly always the case, the true answer 
to the above question depends on the kind of problem one is interested in 
solving \cite{GAZIS:02}.

In the beginning years of traffic flow engineering, a bridge was formed between 
the microscopic General Motors car-following model of equation 
\eqref{eq:TFM:GeneralMotorsCarFollowingModel}, and the Greenberg macroscopic 
model \cite{GREENBERG:59,GAZIS:61}. This proved to be quite a significant 
breakthrough, as it was now possible to obtain all known steady-state 
macroscopic fundamental diagrams, by integrating the car-following equation with 
suitably chosen parameter values \cite{GAZIS:02}. A recent example of this kind 
of linking, was done by Treiber and Helbing, who provided a \emph{micro-macro 
link} between their non-local gas-kinetic mesoscopic model (see section 
\ref{sec:TFM:ImprovementsToTheMesoscopicModellingApproach}) and the intelligent 
driver model (see section 
\ref{sec:TFM:ClassicCarFollowingAndLaneChangingModels}) 
\cite{HELBING:98b,HELBING:02b}.

Besides this explicit translating of microscopic into macroscopic (mesoscopic) 
models and vice versa, it is also possible to develop \emph{hybrid models} that 
couple macroscopically modelled road links to microscopically modelled ones. 
Examples include the work of Magne et al., who develop a hybrid simulator that 
couples a METANET-like second-order macroscopic traffic flow model with the 
\emph{SImulation TRAfic} (SITRA-B+) microscopic traffic flow model. Special 
attention is given to the interfaces between macroscopically and microscopically 
modelled road segments; each macroscopic time iteration in the simulator, is 
accompanied by a number of microscopic iterations \cite{MAGNE:00}. In similar 
spirit, the work of Bourrel and Henn links macroscopic representations of 
traffic flows to microscopic ones, using interfaces that describe the 
transitions between them. As an application of their methodology, they describe 
the translation between the first-order macroscopic LWR model and a vehicle 
representation of this model (based on trajectories) \cite{BOURREL:02}. Another 
avenue was pursued by the Wilco \cite{BURGHOUT:04} and Wilco et al. 
\cite{BURGHOUT:05}, who developed an integration framework between the MITSIMLab 
microscopic model and the Mezzo mesoscopic model. By building upon a mesoscopic 
approach, the strength of their work lies in the fact that no aggregation and 
disaggregation of flows needs to be performed.

	\section{Conclusions}

The material elaborated upon in this paper, spanned a broad range going from 
transportation planning models that operate on a high level, to traffic flow 
models that explicitly describe the physical propagation of traffic flows.

As explained in the introduction, we feel there is a frequent confusion among 
traffic engineers and policy makers when it comes to transportation planning 
models and the role that traffic flow models play therein. To this day, many 
transportation planning bureaus continue to use static tools for evaluating 
policy decisions, whereas the need for dynamic models is getting more and more 
pronounced \cite{MAERIVOET:04g}.

Even after more than sixty years of traffic flow modelling, the debate on what 
is the correct modelling approach remains highly active. On the transportation 
planning side, many agencies still primarily focus on the traditional four-step 
model (4SM), because it is the best intuitively understood approach. In contrast 
to this, activity-based modelling (ABM) is gaining momentum, although it remains 
a rather obscure discipline to many people. At the basis of this scrutiny 
towards the ABM, lies the absence of a generally accepted framework such as the 
one of the 4SM. It is tempting to translate the ABM approach to the 4SM, by 
which e.g., the ABM's synthetic population generation (including activity 
generation, household choices and scheduling) corresponds to the 4SM's 
production and attraction, distribution, and modal split (or to discrete choice 
theory in a broader setting), thereby generating (time dependent) OD tables. 
Similarly, the ABM's agent simulation can be seen as an implementation of the 
4SM's traffic assignment. However, it remains difficult to gain insight into 
this kind of direct translation and the resulting travel behaviour, although the 
ABM's scientific field is continuously in a state of flux thanks to the 
increasing computational power.

On the traffic flow modelling side, the debate on whether or not to use 
macro-/meso- or microscopic models still continues to spawn many intriguing 
discussions. Despite the respective criticisms, it is widely agreed upon that 
modelling driver behaviour entails complex human-human, human-vehicle, and 
vehicle-vehicle interactions. These call for interdisciplinary research, drawing 
from fields such as mathematics, physics, and engineering, as well as sociology 
and psychology (see e.g., the overview of Helbing and Nagel \cite{HELBING:04}).

\appendix
%

\section{Glossary of terms}
\label{appendix:Glossary}

	\subsection{Acronyms and abbreviations}

\begin{tabular}{ll}
	4SM            & four step model\\
	AADT           & annual average daily traffic\\
	ABM            & activity-based modelling\\
	ACC            & adaptive cruise control\\
	ACF            & average cost function\\
\end{tabular}

\begin{tabular}{ll}
	ADAS           & advanced driver assistance systems\\
	AIMSUN2        & Advanced Interactive Microscopic\\
	               & Simulator for Urban and Non-Urban\\
	               & Networks\\
	AMICI          & Advanced Multi-agent Information and\\
	               & Control for Integrated multi-class traffic\\
	               & networks\\
	AON            & all-or-nothing\\
	ASDA           & Automatische StauDynamikAnalyse\\
	ASEP           & asymmetric simple exclusion process\\
	ATIS           & advanced traveller information systems\\
	ATMS           & advanced traffic management systems\\
	BCA            & Burgers cellular automaton\\
	BJH            & Benjamin, Johnso, and Hui\\
	BJH-TCA        & Benjamin-Johnson-Hui traffic cellular\\
	               & automaton\\
	BL-TCA         & brake-light traffic cellular automaton\\
	BML            & Biham, Middleton, and Levine\\
	BML-TCA        & Biham-Middleton-Levine traffic cellular\\
	               & automaton\\
	BMW            & Beckmann, McGuire, and Winsten\\
	BPR            & Bureau of Public Roads\\
	CA             & cellular automaton\\
	CA-184         & Wolfram's cellular automaton rule 184\\
	CAD            & computer aided design\\
	CBD            & central business district\\
	CFD            & computational fluid dynamics\\
	CFL            & Courant-Friedrichs-Lewy\\
	ChSch-TCA      & Chowdhury-Schadschneider traffic\\
	               & cellular automaton\\
	CLO            & camera Linkeroever\\
	CML            & coupled map lattice\\
	CONTRAM        & CONtinuous TRaffic Assignment\\
	               & Model\\
	COMF           & car-oriented mean-field theory\\
	CPM            & computational process models\\
	CTM            & cell transmission model\\
	DDE            & delayed differential equation\\
	DFI-TCA        & deterministic Fukui-Ishibashi traffic\\
	               & cellular automaton\\
	DGP            & dissolving general pattern\\
	DLC            & discretionary lane change\\	
	DLD            & double inductive loop detector\\
	DNL            & dynamic network loading\\
	DRIP           & dynamic route information panel\\
	DTA            & dynamic traffic assignment\\
	DTC            & dynamic traffic control\\
	DTM            & dynamic traffic management\\
	DUE            & deterministic user equilibrium\\
	DynaMIT        & Dynamic network assignment for the\\
	               & Management of Information to\\
	               & Travellers\\
	DYNASMART      & DYnamic Network Assignment-\\
	               & Simulation Model for Advanced\\
	               & Roadway Telematics\\
	ECA            & elementary cellular automaton\\
	EP             & expanded congested pattern\\
	ER-TCA         & Emmerich-Rank traffic cellular\\
	               & automaton\\
	FCD            & floating car data\\
\end{tabular}

\begin{tabular}{ll}
	FDE            & finite difference equation\\
	FIFO           & first-in, first-out\\
	FOTO           & Forecasting of Traffic Objects\\
	GETRAM         & Generic Environment for TRaffic\\
	               & Analysis and Modeling\\
	GHR            & Gazis-Herman-Rothery\\
	GIS            & geographical information systems\\
	GNSS           & Global Navigation Satellite System\\
	               & (e.g., Europe's Galileo)\\
	GoE            & Garden of Eden state\\
	GP             & general pattern\\
	GPRS           & General Packet Radio Service\\
	GPS            & Global Positioning System\\
	               & (e.g., USA's NAVSTAR)\\
	GRP            & generalised Riemann problem\\
	GSM            & Groupe Sp\'eciale Mobile\\
	GSMC           & Global System for Mobile\\
	               & Communications\\
	HAPP           & household activity pattern problem\\
	HCM            & Highway Capacity Manual\\
	HCT            & homogeneously congested traffic\\
	HDM            & human driver model\\
	HKM            & human-kinetic model\\
	HRB            & Highway Research Board\\
	HS-TCA         & Helbing-Schreckenberg traffic cellular\\
	               & automaton\\
	ICC            & intelligent cruise control\\
	IDM            & intelligent driver model\\
	INDY           & INteractive DYnamic traffic assignment\\
	ITS            & intelligent transportation systems\\
	IVP            & initial value problem\\
	JDK            & Java\trademark Development Kit\\
	KKT            & Karush-Kuhn-Tucker\\
	KKW-TCA        & Kerner-Klenov-Wolf traffic cellular\\
	               & automaton\\
	KWM            & kinematic wave model\\
	LGA            & lattice gas automaton\\
	LOD            & level of detail\\
	LOS            & level of service\\
	LSP            & localised synchronised-flow pattern\\
	LTM            & link transmission model\\
	LWR            & Lighthill, Whitham, and Richards\\
	MADT           & monthly average daily traffic\\
	MC-STCA        & multi-cell stochastic traffic cellular\\
	               & automaton\\
	MesoTS         & Mesoscopic Traffic Simulator\\
	MFT            & mean-field theory\\
	MITRASIM       & MIcroscopic TRAffic flow SIMulator\\
	MITSIM         & MIcroscopic Traffic flow SIMulator\\
	MIXIC          & Microscopic model for Simulation of\\
	               & Intelligent Cruise Control\\
	MLC            & mandatory lane change\\	
	               & moving localised cluster\\	
	MOE            & measure of effectiveness\\
	MPA            & matrix-product ansatz\\
	MPCF           & marginal private cost function\\
	MSA            & method of successive averages\\
	MSCF           & marginal social cost function\\
	MSP            & moving synchronised-flow pattern\\
\end{tabular}

\begin{tabular}{ll}
	MT             & movement time\\
	MUC-PSD        & multi-class phase-space density\\
	NaSch          & Nagel and Schreckenberg\\
	NAVSTAR        & Navigation Satellite Timing and Ranging\\
	NCCA           & number conserving cellular automaton\\
	NSE            & Navier-Stokes equations\\
	OCT            & oscillatory congested traffic\\
	OD             & origin-destination\\
	ODE            & ordinary differential equation\\
	OSS            & Open Source Software\\
	OVF            & optimal velocity function\\
	OVM            & optimal velocity model\\
	Paramics       & Parallel microscopic traffic simulator\\
	PATH           & California Partners for Advanced Transit\\
	               & and Highways\\
	               & Program on Advanced Technology for\\
	               & the Highway\\
	PCE            & passenger car equivalent\\
	PCU            & passenger car unit\\
	PDE            & partial differential equation\\
	PELOPS         & Program for the dEvelopment of\\
	               & Longitudinal micrOscopic traffic\\
	               & Processes in a Systemrelevant\\
	               & environment\\
	PeMS           & California Freeway Performance\\
	               & Measurement System\\
	PHF            & peak hour factor\\
	PLC            & pinned localised cluster\\
	pMFT           & paradisiacal mean-field theory\\
	PRT            & perception-reaction time\\
	PSD            & phase-space density\\
	PW             & Payne-Whitham\\
	QoS            & quality of service\\
	SFI-TCA        & stochastic Fukui-Ishibashi traffic\\
	               & cellular automaton\\
	Simone         & Simulation model of Motorways with\\
	               & Next generation vehicles\\
	SLD            & single inductive loop detector\\
	SMARTEST       & Simulation Modelling Applied to Road\\
	               & Transport European Scheme Tests\\
	SMS            & space-mean speed\\
	SOC            & self-organised criticality\\
	SOMF           & site-oriented mean-field theory\\
	SP             & synchronised-flow pattern\\
	SSEP           & symmetric simple exclusion process\\
	STA            & static traffic assignment\\
	STCA           & stochastic traffic cellular automaton\\
	STCA-CC        & stochastic traffic cellular automaton\\
	               & with cruise control\\
	SUE            & stochastic user equilibrium\\
	SUMO           & Simulation of Urban MObility\\
	T$^{\two}$-TCA & Takayasu-Takayasu traffic cellular\\
	               & automaton\\
	TASEP          & totally asymmetric simple exclusion\\
	               & process\\
	TCA            & traffic cellular automaton\\
	TDF            & travel demand function\\
	TMC            & Traffic Message Channel\\
	TMS            & time-mean speed\\
\end{tabular}

\begin{tabular}{ll}
	TOCA           & time-oriented traffic cellular\\
	               & automaton\\
	TRANSIMS       & TRansportation ANalysis and SIMulation\\
	               & System\\
	TRB            & Transportation Research Board\\
	TSG            & triggered stop-and-go traffic\\
	UDM            & ultra-discretisation method\\
	UMTS           & Universal Mobile Telecommunications\\
	               & System\\
	VDR-TCA        & velocity-dependent randomisation traffic\\
	               & cellular automaton\\
	VDT            & total vehicle distance travelled\\
	VHT            & total vehicle hours travelled\\
	VMS            & variable message sign\\
	VMT            & total vehicle miles travelled\\
	VOT            & value of time\\
	WSP            & widening synchronised-flow pattern\\
	WYA            & whole year analysis\\
\end{tabular}

	\subsection{List of symbols}

\renewcommand{\arraystretch}{1.2}

\begin{tabular}{ll}
	$a_{\text{max}}$                     & the maximum acceleration in the IDM\\
	$c(k)$                               & the sound speed of traffic\\
	$C(q)$                               & the economical cost associated with the\\
	                                     & travel demand $q$\\
	$\Delta f(x)$                        & the forward difference operator applied\\
	                                     & to $f(x)$\\
	$\Delta k$                           & the difference in density up- and\\
	                                     & downstream of a shock wave\\
	$\Delta q$                           & the difference in flow up- and\\
	                                     & downstream of a shock wave\\
	$\Delta X$                           & the width of a cell in a numerical\\
	                                     & discretisation scheme\\
	$\Delta T$                           & the size of a time step in a numerical\\
	                                     & discretisation scheme\\
	$D_{j}$                              & a destination zone $j$\\
	$\epsilon$                           & a small diffusion constant for the\\
	                                     & viscosity $\nu$\\
	$g_{s}^{*}(v_{i},\Delta v_{i})$      & the effective desired space gap in\\
	                                     & the IDM\\
	$\kappa$                             & a kinetic coefficient related to $\tau$, $k$,\\
	                                     & and $\Theta$\\
	$k_{t}$                              & the partial derivative of $k(t,x)$ with\\
	                                     & respect to time\\
	$k_{x}$                              & the partial derivative of $k(t,x)$ with\\
	                                     & respect to space\\
	$\widetilde{k}(t,x,\overline v_{s})$ & the phase-space density at $(t,x)$\\
	                                     & associated with SMS $\overline v_{s}$\\
	$\widetilde{k}_{t}$                  & the partial derivative of $\widetilde{k}(t,x)$ with\\
	                                     & respect to time\\
\end{tabular}

\begin{tabular}{ll}
	$\widetilde{k}_{x}$                  & the partial derivative of $\widetilde{k}(t,x)$ with\\
	                                     & respect to space\\
	$\lambda$                            & the sensitivity to the stimulus in a\\
	                                     & car-following model\\
	                                     & the arrival rate at a server in queueing\\
	                                     & theory\\
	$\mu$                                & the service rate of a server in queueing\\
	                                     & theory\\
	$\nabla f(x)$                        & the backward difference operator applied\\
	                                     & to $f(x)$\\
	                                     & the gradient vector of $f(x)$\\
	$\nu$                                & the kinematic traffic viscosity coefficient\\
	$O_{i}$                              & an origin zone $i$\\
	$\pi$                                & the probability of overtaking (as opposed\\
	                                     & to slowing down)\\
	$P$                                  & the traffic pressure\\
	$P_{x}$                              & the partial derivative of the traffic pressure\\
	                                     & with respect to space\\
	$P(t,x,\overline v_{s})$             & the distribution of the vehicles with SMS\\
	                                     & $\overline v_{s}$ at $(t,x)$\\
	$q_{\text{pc}}$                      & the practical capacity\\
	$q_{\text{so}}$                      & travel demand associated with a system\\
	                                     & optimum\\
	$q_{\text{ue}}$                      & travel demand associated with a user\\
	                                     & equilibrium\\	
	$S$                                  & a traffic state in the human-kinetic model\\
	$\tau$                               & a driver's reaction time\\
	$\Theta$                             & the variance of the speed\\
	$\Theta_{e}(k,\overline v_{s})$      & an equilibrium relation between the speed\\
	                                     & variance, the density, and the SMS\\
	$T$                                  & a travel time\\
	                                     & a relaxation parameter (in Pipes' car-\\
	                                     & following model)\\
	$T_{\text{ff}}$                      & a travel time under free-flow conditions\\
	$u$                                  & the velocity (in the context of a Navier-\\
	                                     & Stokes fluid)\\
	$v_{\text{des}}$                     & the desired speed of drivers\\
	$\overline v_{s_{t}}$                & the partial derivative of the space-mean\\
	                                     & speed with respect to time\\
	$\overline v_{s_{x}}$                & the partial derivative of the space-mean\\
	                                     & speed with respect to space\\
	$\overline v_{s_{e}}(k,\Theta)$      & an equilibrium relation between the SMS,\\
	                                     & the density, and the speed variance\\
	$V()$                                & the optimal velocity function\\
	$w_{\text{shock}}$                   & the speed of a shock wave\\
\end{tabular}

\section*{Acknowledgements}

Dr. Bart De Moor is a full professor at the Katholieke Universiteit Leuven, Belgium.
\noindent
Our research is supported by:
\textbf{Research Council KUL}: GOA AMBioRICS, several PhD/post\-doc
\& fellow grants,
\textbf{Flemish Government}:
\textbf{FWO}: PhD/post\-doc grants, projects, G.0407.02 (support vector machines),
G.\-0197.02 (power islands), G.0141.03 (identification and cryptography), G.0491.03
(control for intensive care glycemia), G.0120.\-03 (QIT), G.0452.04 (new quantum algorithms),
G.0499.04 (statistics), G.0211.05 (Nonlinear), research communities (ICCoS, ANMMM, MLDM),
\textbf{IWT}: PhD Grants, GBOU (McKnow),
\textbf{Belgian Federal Science Policy Office}: IUAP P5/\-22 (`Dynamical Systems and
Control: Computation, Identification and Modelling', 2002-2006), PODO-II (CP/40:
TMS and Sustainability),
\textbf{EU}: FP5-Quprodis, ERNSI,
\textbf{Contract Research/agreements}: ISMC/IPCOS, Data4s,TML, Elia, LMS,
Mastercard.

\bibliography{paper}

\end{document}